\newif\iflongversion
\longversiontrue

\iflongversion
\documentclass[12pt]{article}
\usepackage{fullpage}
\usepackage{times}
\usepackage{doi}

\usepackage{graphicx}
\usepackage{fancyhdr}
\else
\documentclass[runningheads,envcountsame]{llncs}
\fi

\usepackage{graphicx}
\usepackage{hyperref}

\usepackage{xcolor}
\usepackage{paralist}
\usepackage{wrapfig}

\usepackage{etoolbox}
\BeforeBeginEnvironment{wrapfigure}{\setlength{\intextsep}{0pt}}

\usepackage{bm}
\usepackage[bbsets,Dfprime,setrelation]{math}

\usepackage{wasysym}
\usepackage[prefixflatinterpret,bracketmodalinterpret,fixformat,silentconst,sidenotecalculus,longseqcontext,seqinsist]{logic}
\usepackage[prefixflatinterpret,bracketmodalinterpret,fixformat,silentconst,differentialdL,simplenames]{dL}
\def\leftrule{L}%
\def\rightrule{R}%
\newcommand{\bebecomes}{\mathrel{::=}}
\newcommand{\alternative}{~|~}

\usepackage[proofatend,proofonly]{endproof2}

\newcommand{\fvarA}{\phi}
\newcommand{\fvarB}{\psi}
\newcommand{\rfvar}{P}
\newcommand{\invvar}{I}
\newcommand{\rcfvar}{C}
\newcommand{\rgvar}{G}
\newcommand{\rrfvar}{R}
\newcommand{\rsfvar}{S}
\newcommand{\rsfvarhat}{\widetilde{S}}
\newcommand{\rtfvar}{T}

\newsavebox{\Lightningval}%
\sbox{\Lightningval}{\mbox{\lightning}}

\newsavebox{\Rval}%
\sbox{\Rval}{$\scriptstyle\mathbb{R}$}
\relax

  \newdimen\linferenceRulehskipamount%
  \newdimen\lcalculuscollectionvskipamount%
\renewcommand{\linferenceRuleNameSeparation}{\hspace{4pt}}

\definecolor{vblue}{rgb}{.1,.15,.62}
\definecolor{vgray}{rgb}{.35,.35,.35}
\renewcommand*{\irrulename}[1]{\text{\textcolor{vgray}{\upshape#1}}}%

\renewcommand{\axkey}[1]{#1}
\renewcommand*{\lie}[3][]
{\mathcal{L}_{#2}^{\ifthenelse{\equal{#1}{}}{}{^{(#1)}}}(#3)}
\newcommand*{\lied}[3][]{\overset{\bm .}{#3}\ifthenelse{\equal{#1}{}}{}{^{(#1)}}}
\newcommand{\cmp}{\succcurlyeq}

\newcommand{\pmc}{\preccurlyeq}

\newcommand{\bdr}[1]{\partial #1}

\usepackage{prettyref}
\newcommand{\rref}[2][]{\prettyref{#2}}
\newrefformat{sec}{Section\,\ref{#1}}
\newrefformat{subsec}{Section\,\ref{#1}}
\newrefformat{def}{Def.\,\ref{#1}}
\newrefformat{thm}{Theorem\,\ref{#1}}
\newrefformat{prop}{Proposition\,\ref{#1}}
\newrefformat{rem}{Remark\,\ref{#1}}
\newrefformat{lem}{Lemma\,\ref{#1}}
\newrefformat{cor}{Corollary\,\ref{#1}}
\newrefformat{ex}{Example\,\ref{#1}}
\newrefformat{cex}{Counterexample\,\ref{#1}}
\newrefformat{tab}{Table\,\ref{#1}}
\newrefformat{fig}{Fig.\,\ref{#1}}
\newrefformat{case}{case\,\ref{#1}}
\newrefformat{foot}{Footnote\,\ref{#1}}
\newrefformat{app}{Appendix\,\ref{#1}}
\newenvironment{proofsketch}[1][TODO]{\proof[Proof Sketch (\rref{#1})]}{\endproof}

\iflongversion
\newtheorem{theorem}{Theorem}
\newtheorem{lemma}[theorem]{Lemma}

\newtheorem{corollary}[theorem]{Corollary}
\newtheorem{conjecture}[theorem]{Conjecture}

\theoremstyle{remark}
\newtheorem{example}{Example}
\newtheorem{counterexample}[example]{Counterexample}{\itshape}{}
\else
\spnewtheorem{counterexample}[example]{Counterexample}{\itshape}{}
\fi

\definecolor{highlightred}{rgb}{.7, 0.0, 0.0}
\iflongversion
\usepackage{ulem}
\newcommand{\highlight}[1]{\textcolor{vblue}{\uline{#1}}}
\else
\newcommand{\highlight}[1]{\textcolor{vblue}{\underline{#1}}}
\fi

\newcommand{\ptermA}{p}
\newcommand{\ptermB}{q}
\renewcommand{\allvars}{\mathbb{V}}
\newcommand{\States}{\mathbb{S}}

\newcommand{\I}{\dLint[const=I,state=\omega]}

\usepackage{ upgreek }
\newcommand{\solvar}{\boldsymbol\upvarphi}
\newcommand{\timevar}{t}
\newcommand{\boundedf}{B}
\newcommand{\constt}[1]{#1()}

\newcommand{\exlinear}{\ensuremath{\alpha_l}}
\newcommand{\exnonlinear}{\ensuremath{\alpha_n}}
\newcommand{\lnorm}[1]{{{\norm{#1}}_{\infty}}}

\newcommand{\bigchi}{\ensuremath{\mathcal{X}}}

\newcommand{\footnotesizeoff}{}%
\makeatletter
\newcommand{\oset}[3][0ex]{%
  \mathrel{\mathop{#3}\limits^{
    \vbox to#1{\kern-2\ex@
    \hbox{$\scriptstyle#2$}\vss}}}}
\makeatother

\newcommand{\osetf}[2]{\oset[1.6ex]{#1}{#2}}

\iflongversion
\newrefformat{app}{Appendix\,\ref{#1}}

\newcommand{\thereport}[1]{#1}
\else
\newrefformat{app}{\cite{DBLP:journals/corr/abs-1904-07984}}

\newcommand{\thereport}[1]{the report #1}
\fi

\newcommand{\openset}{\mathcal{O}}

\begin{document}
\title{An Axiomatic Approach to Liveness\\ for Differential Equations}
\iflongversion
\author{Yong Kiam Tan \and
Andr\'e Platzer
\thanks{
  Computer Science Department, Carnegie Mellon University, Pittsburgh, USA
  {\{yongkiat$|$aplatzer\}@cs.cmu.edu}
  }
}
\date{}
\else
\author{Yong Kiam Tan \orcidID{0000-0001-7033-2463} \and
Andr\'e Platzer \orcidID{0000-0001-7238-5710}
}%
\authorrunning{Tan Y.K., Platzer A.}

\institute{
Computer Science Department, Carnegie Mellon University, Pittsburgh, USA\\
\email{\{yongkiat|aplatzer\}@cs.cmu.edu}
}
\fi

\maketitle              %
\iflongversion
\thispagestyle{empty}
\fi

\begin{abstract}
This paper presents an approach for deductive liveness verification for ordinary differential equations (ODEs) with differential dynamic logic.
Numerous subtleties complicate the generalization of well-known discrete liveness verification techniques, such as loop variants, to the continuous setting.
For example, ODE solutions may blow up in finite time or their progress towards the goal may converge to zero.
Our approach handles these subtleties by successively refining ODE liveness properties using ODE invariance properties which have a well-understood deductive proof theory.
This approach is widely applicable: we survey several liveness arguments in the literature and derive them all as special instances of our axiomatic refinement approach.
We also correct several soundness errors in the surveyed arguments, which further highlights the subtlety of ODE liveness reasoning and the utility of our deductive approach.
The library of common refinement steps identified through our approach enables both the sound development and justification of new ODE liveness proof rules from our axioms.

\iflongversion
\noindent
\textbf{Keywords:} {differential equations, liveness, differential dynamic logic}
\else
\keywords{differential equations, liveness, differential dynamic logic}
\fi
\end{abstract}
\section{Introduction}
\label{sec:introduction}
Hybrid systems are mathematical models describing discrete and continuous dynamics, and interactions thereof~\cite{DoyenFPP18}.
This flexibility makes them natural models of cyber-physical systems (CPSs) which feature interactions between discrete computational control and continuous real world physics \cite{10.2307/j.ctt17kkb0d,Platzer18}.
Formal verification of hybrid systems is of significant practical interest because the CPSs they model frequently operate in safety-critical settings.
Verifying properties of the continuous dynamics is a key aspect of any such endeavor.

This paper focuses on deductive liveness verification for continuous dynamics described by ordinary differential equations (ODEs).
We work with differential dynamic logic (\dL)~\cite{DBLP:conf/lics/Platzer12a,DBLP:journals/jar/Platzer17,Platzer18}, a logic for \emph{deductive verification} of hybrid systems, which compositionally lifts our results to the hybrid systems setting.
Methods for proving liveness in the discrete setting are well-known: loop variants show that discrete loops eventually reach a desired goal, while temporal logic is used to specify and study liveness properties in concurrent and infinitary settings~\cite{DBLP:books/daglib/0077033,DBLP:journals/toplas/OwickiL82}.
In the continuous setting, \emph{liveness} for an ODE means that its solutions eventually enter a desired goal region in finite time without leaving the domain of allowed (or safe) states.\footnote{This property has also been called, e.g., \emph{eventuality}~\cite{DBLP:journals/siamco/PrajnaR07,DBLP:conf/fm/SogokonJ15} and \emph{reachability}~\cite{DBLP:conf/emsoft/TalyT10}.
To minimize ambiguity, this paper refers to the property as \emph{liveness}, with a precise formal definition in~\rref{sec:background}.
Other advanced notions of liveness for ODEs are discussed in~\rref{sec:relatedwork}, although their formal deduction is left for future work.}
Deduction of such ODE liveness properties is hampered by several difficulties:
\begin{inparaenum}[\it i)]
\item solutions of ODEs may converge towards a goal without ever reaching it,
\item solutions of (non-linear) ODEs may blow up in finite time leaving insufficient time for the desired goal to be reached, and
\item the goal may be reachable but only by leaving the domain constraint.
\end{inparaenum}
In contrast, \emph{invariance} properties for ODEs are better understood~\cite{DBLP:conf/tacas/GhorbalP14,DBLP:conf/emsoft/LiuZZ11} and have a complete \dL axiomatization~\cite{DBLP:conf/lics/PlatzerT18}.
Motivated by the aforementioned difficulties, we present \dL axioms enabling step-by-step refinement of ODE liveness properties with a sequence of ODE invariance properties.
This brings the full deductive power of \dL's ODE invariance proof rules to bear on liveness proofs.
Our approach is a general framework for understanding ODE liveness arguments.
We use it to survey several arguments from the literature and derive them all as (corrected) \dL proof rules, see \rref{tab:survey}.
This logical presentation has two key benefits:
\begin{itemize}
\item The proof rules are \emph{derived} from sound axioms of \dL, guaranteeing their correctness.
Many of the surveyed arguments contain subtle soundness errors, see~\rref{tab:survey}.
These errors do not diminish the surveyed work.
Rather, they emphasize the need for an axiomatic, uniform way of presenting and analyzing ODE liveness arguments rather than ad hoc approaches.
\item The approach identifies common refinement steps that form a basis for the surveyed liveness arguments. %
This library of building blocks enables sound development and justification of new ODE liveness proof rules, e.g., by generalizing individual refinement steps or by exploring different combinations of those steps.
Corollaries~\ref{cor:higherdv},~\ref{cor:boundedandcompact}, and~\ref{cor:combination} are examples of new ODE liveness proof rules that can be derived and justified using our uniform approach.
\end{itemize}

\iflongversion
All proofs are in Appendix~\ref{app:proofcalc} and~\ref{app:proofs}. Counterexamples for the soundness errors in~\rref{tab:survey} are given in~\rref{app:counterexamples}.
\else
All proofs are in the companion report \rref{app:}, together with counterexamples for the soundness errors listed in~\rref{tab:survey}.
\fi

\begin{table}[t]
\centering
\caption{Surveyed ODE liveness arguments with our corrections \highlight{highlighted in blue}. The referenced corollaries are our corresponding (corrected) derived proof rules.}
\label{tab:survey}
\setlength{\tabcolsep}{4pt}
\begin{tabular}{c|l l | l l}
\hline
Source                                & \multicolumn{2}{l|}{Without Domain Constraints} & \multicolumn{2}{l}{With Domain Constraints} \\ \hline
\cite{DBLP:journals/logcom/Platzer10}   & OK                         & (Cor.~\ref{cor:atomicdvcmp})  & \highlight{if open/closed, initially false} & (Cor.~\ref{cor:atomicdvcmpQ})   \\
\cite{DBLP:conf/hybrid/PrajnaR05,DBLP:journals/siamco/PrajnaR07} & \multicolumn{2}{l|}{\cite[Remark 3.6]{DBLP:journals/siamco/PrajnaR07} \highlight{is incorrect}} & \highlight{if conditions checked globally} & (Cor.~\ref{cor:prq})     \\
\cite{DBLP:journals/siamco/RatschanS10} &  \highlight{if compact} & (Cor.~\ref{cor:rs}) & \highlight{if  compact} & (Cor.~\ref{cor:rsq}) \\
\cite{DBLP:conf/fm/SogokonJ15}          & OK                      & (Cor.~\ref{cor:SP}) &  OK                     & (Cor.~\ref{cor:SPQ}) \\
\cite{DBLP:conf/emsoft/TalyT10}         &  \highlight{if globally Lipschitz} & (Cor.~\ref{cor:tt}) & \highlight{if globally Lipschitz} & (Cor.~\ref{cor:ttq}) \\
\end{tabular}
\end{table}
\section{Background}
\label{sec:background}

This section reviews the syntax and semantics of \dL, focusing on its continuous fragment which has a complete axiomatization for ODE invariants~\cite{DBLP:conf/lics/PlatzerT18}.
Full presentations of \dL, including its discrete fragment, are available elsewhere~\cite{DBLP:journals/jar/Platzer17,Platzer18}.

\subsection{Syntax}
\label{subsec:syntax}

\iflongversion
\begin{wrapfigure}[33]{r}{0.35\textwidth}
\else
\begin{wrapfigure}[31]{r}{0.35\textwidth}
\fi
\centering
\includegraphics[width=0.33\textwidth]{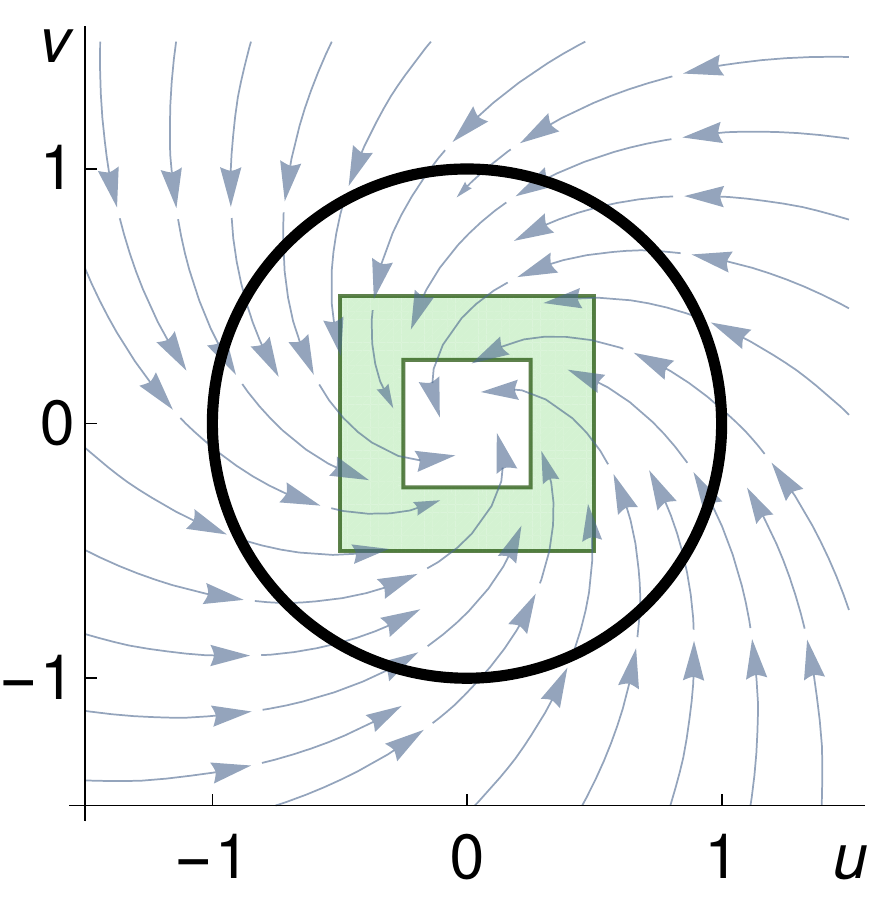}
\includegraphics[width=0.33\textwidth]{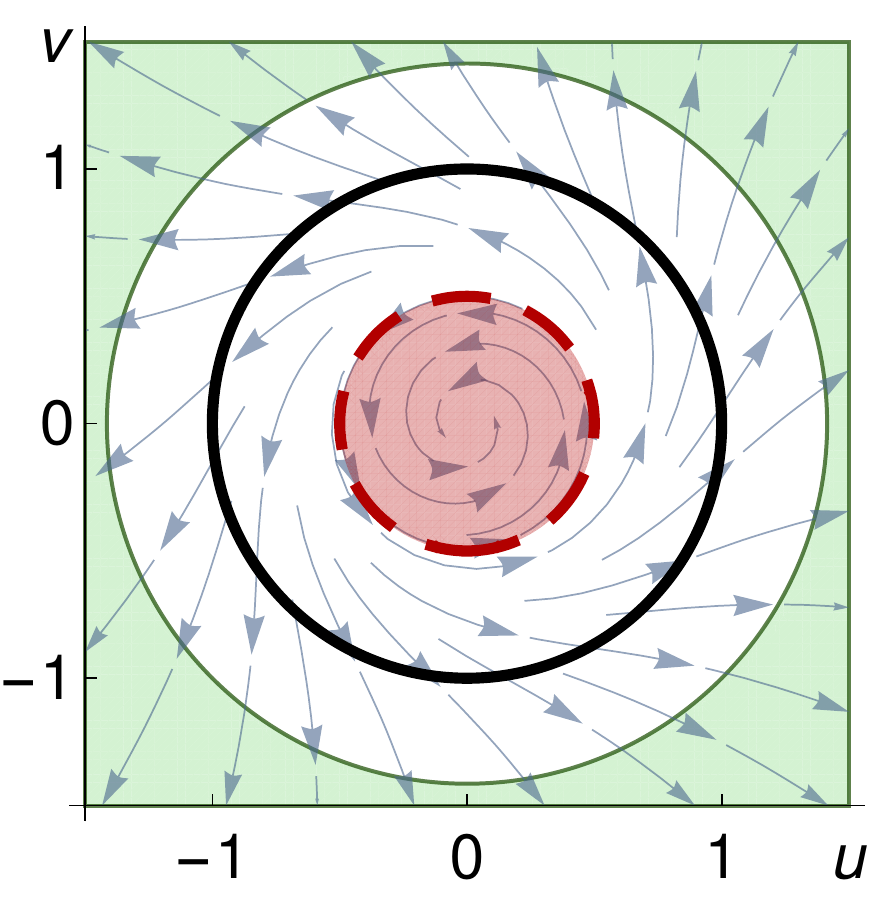}
\caption{Visualization of $\exlinear$ (above) and $\exnonlinear$ (below).
Solutions of $\exlinear$ globally spiral towards the origin.
In contrast, solutions of $\exnonlinear$ spiral inwards within the inner red disk (dashed boundary), but spiral outwards otherwise.
For both ODEs, solutions starting on the black unit circle eventually enter their respective shaded green goal regions.}
\label{fig:odeexamples}
\end{wrapfigure}

The grammar of \dL terms is as follows, where $v \in \allvars$ is a variable and $c \in \rationals$ is a rational constant.
These terms are polynomials over the set of variables $\allvars$:
\[
	\ptermA,\ptermB~\bebecomes~v \alternative c \alternative \ptermA + \ptermB \alternative \ptermA \cdot \ptermB
\]

The grammar of \dL formulas is as follows, where $\sim {\in}~\{=,\neq,\geq,>,\leq,<\}$ is a comparison operator and $\alpha$ is a hybrid program:
\begin{align*}
  \fvarA,\fvarB~\bebecomes&~\overbrace{\ptermA \sim \ptermB \alternative \fvarA \land \fvarB \alternative \fvarA \lor \fvarB \alternative \lnot{\fvarA} \alternative \lforall{v}{\fvarA} \alternative \lexists{v}{\fvarA}}^{\text{First-order formulas of real arithmetic }\rfvar,\ivr}\\
 &\alternative \dbox{\alpha}{\fvarA} \alternative\ddiamond{\alpha}{\fvarA}
\end{align*}

The notation $\ptermA \cmp \ptermB$ (resp.~$\pmc$) is used when the comparison operator can be either $\geq$ or $>$ (resp.~$\leq$ or $<$).
Other standard logical connectives, e.g., $\limply,\lbisubjunct$, are definable as in classical logic.
Formulas not containing the modalities $\dibox{\cdot},\didia{\cdot}$ are formulas of first-order real arithmetic and are written as $\rfvar,\ivr$.
The box ($\dbox{\alpha}{\fvarA}$) and diamond ($\ddiamond{\alpha}{\fvarA}$) modality formulas express dynamic properties of the hybrid program $\alpha$.
We focus on \emph{continuous} programs, where $\alpha$ is given by a system of ODEs $\pevolvein{\D{x}=\genDE{x}}{\ivr}$.
Here, $\D{x}=\genDE{x}$ is an $n$-dimensional system of differential equations, $\D{x_1}=f_1(x), \dots, \D{x_n}=f_n(x)$, over variables $x = (x_1,\dots,x_n)$, where the LHS $\D{x_i}$ is the time derivative of $x_i$ and the RHS $f_i(x)$ is a polynomial over variables $x$.
The domain constraint $\ivr$ specifies the set of states in which the ODE is allowed to evolve continuously.
When there is no domain constraint, i.e., $\ivr$ is the formula $\ltrue$, the ODE is written as $\D{x}=\genDE{x}$.

Two running example ODEs are visualized in~\rref{fig:odeexamples} with directional arrows corresponding to their RHS evaluated at points on the plane.
The first ODE, $\exlinear \mnodefequiv \D{u}=-v-u,\D{v}=u-v$, is \emph{linear} because its RHS depends linearly on $u,v$.
The second ODE, $\exnonlinear \mnodefequiv \D{u}=-v-u(\frac{1}{4}-u^2-v^2),\D{v}=u-v(\frac{1}{4}-u^2-v^2)$, is \emph{non-linear}.
The non-linearity of $\exnonlinear$ results in more complex behavior for its solutions, e.g., the difference in spiraling behavior shown in~\rref{fig:odeexamples}.
In fact, solutions of $\exnonlinear$ blow up in finite time iff they start outside the disk characterized by $u^2+v^2 \leq \frac{1}{4}$.
Finite time blow up is impossible for linear ODEs like $\exlinear$~\cite{Chicone2006,Walter1998}.

When terms (or formulas) appear in contexts involving ODEs $\D{x}=\genDE{x}$, it is sometimes necessary to restrict the set of free variables they are allowed to mention.
These restrictions are always stated explicitly and are also indicated as arguments\footnote{This understanding of variable dependencies is made precise using function and predicate symbols in \dL's uniform substitution calculus~\cite{DBLP:journals/jar/Platzer17}.} to terms (or formulas), e.g., $\ptermA()$ means the term $\ptermA$ does not mention any of $x_1,\dots,x_n$ free, while $\rfvar(x)$ means the formula $\rfvar$ may mention all of them.%
\subsection{Semantics}
\label{subsec:semantics}

States $\iget[state]{\I} : \allvars \to \reals$ assign real values to each variable in $\allvars$; the set of all states is written $\States$.
The semantics of polynomial term $\ptermA$ in state $\iget[state]{\I} \in \States$ is the real value $\ivaluation{\I}{\ptermA}$ of the corresponding polynomial function evaluated at $\iget[state]{\I}$.
The semantics of formula $\fvarA$ is the set of states $\imodel{\I}{\fvarA} \subseteq \States$ in which that formula is true.
The semantics of first-order logical connectives are defined as usual, e.g., $\imodel{\I}{\fvarA \land \fvarB} = \imodel{\I}{\fvarA} \cap \imodel{\I}{\fvarB}$.

For ODEs, the semantics of the modal operators is defined directly as follows.\footnote{The semantics of \dL formulas is defined compositionally elsewhere~\cite{DBLP:journals/jar/Platzer17,Platzer18}.}
Let $\iget[state]{\I} \in \States$ and $\solvar : [0, T) \to \States$ (for some $0<T\leq\infty$), be the unique, right-maximal solution~\cite{Chicone2006,Walter1998} to the ODE $\D{x}=\genDE{x}$ with initial value $\solvar(0)=\iget[state]{\I}$:
\begin{align*}
 \m{\imodels{\I}{\dbox{\pevolvein{\D{x}=\genDE{x}}{\ivr} }{\fvarA}}}~\text{iff}~&\text{for all}~0 \leq \tau < T~\text{where}~\solvar(\zeta)\,{\in}\,\imodel{\I}{\ivr}~\text{for all}~0 \leq \zeta \leq \tau\text{:}\\
&\solvar(\tau) \in \imodel{\I}{\fvarA} \\
 \m{\imodels{\I}{\ddiamond{\pevolvein{\D{x}=\genDE{x}}{\ivr}}{\fvarA}}}~\text{iff}~&\text{there exists}~0 \leq \tau < T~\text{such that:}\\
&\solvar(\tau) \in \imodel{\I}{\fvarA}~\text{and}~\solvar(\zeta) \in \imodel{\I}{\ivr}~\text{for all}~0 \leq \zeta \leq \tau
\end{align*}

Informally, $\dbox{\pevolvein{\D{x}=\genDE{x}}{\ivr}}{\fvarA}$ is true in initial state $\iget[state]{\I}$ if \emph{all} states reached by following the ODE from $\iget[state]{\I}$ while remaining in the domain constraint $\ivr$ satisfy postcondition $\fvarA$.
Dually, the \emph{liveness} property $\ddiamond{\pevolvein{\D{x}=\genDE{x}}{\ivr}}{\fvarA}$ is true in initial state $\iget[state]{\I}$ if \emph{some} state which satisfies the postcondition $\fvarA$ is eventually reached in \emph{finite} time by following the ODE from $\iget[state]{\I}$ while staying in domain constraint $\ivr$.
For the running example,~\rref{fig:odeexamples} suggests that formulas\footnote{Here, $\lnorm{(u,v)}$ denotes the $L^\infty$ norm. The inequality $\lnorm{(u,v)} \leq \frac{1}{2}$ is expressible in first-order real arithmetic as $u^2 \leq \frac{1}{4} \land v^2 \leq \frac{1}{4}$ (similarly for $\frac{1}{4} \leq \lnorm{(u,v)}$).}
$\ddiamond{\exlinear}{\big(\frac{1}{4} \leq \lnorm{(u,v)} \leq \frac{1}{2}\big)}$ and
$\ddiamond{\exnonlinear}{u^2 + v^2 \geq 2}$ are true for initial states $\iget[state]{\I}$ on the unit circle.
These liveness properties are rigorously proved in Examples~\ref{ex:exlinproof} and~\ref{ex:exnonlinproof} respectively.

Variables $y \in \allvars \setminus \{x\}$ not occurring on the LHS of ODE $\D{x}=\genDE{x}$ remain constant along solutions $\solvar : [0, T) \to \States$ of the ODE, with $\solvar(\tau)(y) = \solvar(0)(y)$ for all $\tau \in [0,T)$.
Since only the values of $x=(x_1,\dots,x_n)$ change along the solution $\solvar$ it may also be viewed geometrically as a trajectory in $\reals^n$, dependent on the initial values of the constant \emph{parameters} $y$.
Similarly, the value of terms and formulas depends only on the values of their free variables~\cite{DBLP:journals/jar/Platzer17}.
Thus, terms (or formulas) whose free variables are all parameters for $\D{x}=\genDE{x}$ also have constant (truth) values along solutions of the ODE.
For formulas $\fvarA$ that only mention free variables $x$, $\imodel{\I}{\fvarA}$ can also be viewed geometrically as a subset of $\reals^n$.
Such a formula is said to \emph{characterize} a (topologically) open (resp. closed, bounded, compact) set with respect to variables $x$ iff the set $\imodel{\I}{\fvarA} \subseteq \reals^n$ is topologically open (resp. closed, bounded, compact) with respect to the Euclidean topology.
These topological conditions are used as side conditions for some of the axioms and proof rules in this paper.
In~\thereport{\rref{app:proofcalc}}, a more general definition of these side conditions is given for formulas $\fvarA$ that mention parameters $y$.
\iflongversion
These side conditions are decidable~\cite{Bochnak1998} when $\fvarA$ is a formula of first-order real arithmetic and there are simple syntactic criteria for checking if they hold (\rref{app:proofcalc}).
\else
These side conditions are decidable~\cite{Bochnak1998} when $\fvarA$ is a formula of first-order real arithmetic and there are simple syntactic criteria for checking if they hold~\rref{app:proofcalc}.
\fi

Formula $\fvarA$ is valid iff $\imodel{\I}{\fvarA} = \States$, i.e., $\fvarA$ is true in all states.
In particular, if the formula \(\invvar \limply \dbox{\pevolvein{\D{x}=\genDE{x}}{\ivr}}{\invvar}\) is valid, the formula $\invvar$ is an \emph{invariant} of the ODE $\pevolvein{\D{x}=\genDE{x}}{\ivr}$.
Unfolding the semantics, this means that from any initial state $\iget[state]{\I}$ satisfying $\invvar$, all states reached by the solution of the ODE $\D{x}=\genDE{x}$ from $\iget[state]{\I}$ while staying in the domain constraint $\ivr$ satisfy $\invvar$.

\subsection{Proof Calculus}
\label{subsec:proofcalculus}
\irlabel{qear|\usebox{\Rval}}
\irlabel{notr|$\lnot$\rightrule}
\irlabel{notl|$\lnot$\leftrule}
\irlabel{orr|$\lor$\rightrule}
\irlabel{orl|$\lor$\leftrule}
\irlabel{andr|$\land$\rightrule}
\irlabel{andl|$\land$\leftrule}
\irlabel{implyr|$\limply$\rightrule}
\irlabel{implyl|$\limply$\leftrule}
\irlabel{equivr|$\lbisubjunct$\rightrule}
\irlabel{equivl|$\lbisubjunct$\leftrule}
\irlabel{id|id}
\irlabel{cut|cut}
\irlabel{weakenr|W\rightrule}
\irlabel{weakenl|W\leftrule}
\irlabel{existsr|$\exists$\rightrule}
\irlabel{existsrinst|$\exists$\rightrule}
\irlabel{alll|$\forall$\leftrule}
\irlabel{alllinst|$\forall$\leftrule}
\irlabel{allr|$\forall$\rightrule}
\irlabel{existsl|$\exists$\leftrule}
\irlabel{iallr|i$\forall$}
\irlabel{iexistsr|i$\exists$}

All derivations are presented in a classical sequent calculus with usual rules for manipulating logical connectives and sequents.
The semantics of \emph{sequent} \(\lsequent{\Gamma}{\fvarA}\) is equivalent to the formula \((\landfold_{\fvarB \in\Gamma} \fvarB) \limply \fvarA\) and a sequent is valid iff its corresponding formula is valid.
Completed branches in a sequent proof are marked with $\lclose$.
First-order real arithmetic is decidable~\cite{Bochnak1998} so we assume such a decision procedure and label proof steps with \irref{qear} when they follow from real arithmetic.
An axiom (schema) is \emph{sound} iff all instances of the axiom are valid.
Proof rules are \emph{sound} iff validity of all premises (above the rule bar) entails validity of the conclusion (below the rule bar).
Axioms and proof rules are \emph{derivable} if they can be deduced from sound \dL axioms and proof rules.
Soundness of the base \dL axiomatization ensures that derived axioms and proof rules are sound~\cite{DBLP:journals/jar/Platzer17,Platzer18,DBLP:conf/lics/PlatzerT18}.

The \dL proof calculus (briefly recalled below) is \emph{complete} for ODE invariants~\cite{DBLP:conf/lics/PlatzerT18}, i.e., any true ODE invariant expressible in first-order real arithmetic can be proved in the calculus.
The proof rule~\irref{dIcmp} (below) uses the \emph{Lie derivative} of polynomial $\ptermA$ with respect to the ODE $\D{x}=\genDE{x}$, which is defined as $\lie[]{\genDE{x}}{\ptermA} \mdefeq \sum_{x_i\in x} \Dp[x_i]{\ptermA} f_i(x)$. Higher Lie derivatives $\lied[i]{\genDE{x}}{\ptermA}{}$ are defined inductively:
$ \lied[0]{\genDE{x}}{\ptermA}{} \mdefeq p, \lied[i+1]{\genDE{x}}{\ptermA}{} \mdefeq \lie{\genDE{x}}{\lied[i]{\genDE{x}}{\ptermA}{}}{}, \lied[]{\genDE{x}}{\ptermA} \mdefeq \lied[1]{\genDE{x}}{\ptermA}{}$.
Syntactically, Lie derivatives $\lied[i]{\genDE{x}}{\ptermA}$ are polynomials in the term language.
They are provably definable in \dL using differentials~\cite{DBLP:journals/jar/Platzer17}.
Semantically, the value of Lie derivative $\lied[]{\genDE{x}}{\ptermA}$ is equal to the time derivative of the value of $\ptermA$ along solution $\solvar$ of the ODE $\D{x}=\genDE{x}$.

\begin{lemma}[Axioms and proof rules of \dL~\cite{DBLP:journals/jar/Platzer17,Platzer18,DBLP:conf/lics/PlatzerT18}]
\label{lem:dlaxioms}
The following are sound axioms and proof rules of \dL.\\
\begin{calculuscollection}
\begin{calculus}
\cinferenceRule[diamond|$\didia{\cdot}$]{diamond axiom}
{\linferenceRule[equiv]
  {\lnot\dbox{\alpha}{\lnot{\rfvar}}}
  {\axkey{\ddiamond{\alpha}{\rfvar}}}
}
{}
\end{calculus}
\quad\quad\quad\quad
\begin{calculus}
\cinferenceRule[K|K]{K axiom / modal modus ponens} %
{\linferenceRule[impl]
  {\dbox{\alpha}{(\rrfvar \limply \rfvar)}}
  {(\dbox{\alpha}{\rrfvar}\limply\axkey{\dbox{\alpha}{\rfvar}})}
}{}
\end{calculus}\\
\begin{calculus}
\dinferenceRule[dIcmp|dI$_\cmp$]{}
{\linferenceRule
  {\lsequent{\ivr}{\lied[]{\genDE{x}}{\ptermA}\geq\lied[]{\genDE{x}}{\ptermB}}
  }
  {\lsequent{\Gamma,\ptermA \cmp \ptermB }{\dbox{\pevolvein{\D{x}=\genDE{x}}{\ivr}}{\ptermA \cmp \ptermB}} }
  \quad
}{where $\cmp$ is either $\geq$ or $>$}
\dinferenceRule[dC|dC]{}
{\linferenceRule
  {\lsequent{\Gamma}{\dbox{\pevolvein{\D{x}=\genDE{x}}{\ivr}}{\rcfvar}} \quad
   \lsequent{\Gamma}{\dbox{\pevolvein{\D{x}=\genDE{x}}{\ivr \land \rcfvar}}{\rfvar}}
  }
  {\lsequent{\Gamma}{\dbox{\pevolvein{\D{x}=\genDE{x}}{\ivr}}{\rfvar}}}
}{}
\end{calculus}\\
\begin{calculus}
\dinferenceRule[dW|dW]{}
{\linferenceRule
  {\lsequent{\ivr}{\rfvar}}
  {\lsequent{\Gamma}{\dbox{\pevolvein{\D{x}=\genDE{x}}{\ivr}}{\rfvar}}}
}{}
\dinferenceRule[MbW|M${\dibox{'}}$]{}
{\linferenceRule
  {\lsequent{\ivr,\rrfvar}{\rfvar} \quad \lsequent{\Gamma}{\dbox{\pevolvein{\D{x}=\genDE{x}}{\ivr}}{\rrfvar}}}
  {\lsequent{\Gamma}{\dbox{\pevolvein{\D{x}=\genDE{x}}{\ivr}}{\rfvar}}}
}{}
\end{calculus}
\iflongversion
\qquad
\else
\hfill
\fi
\begin{calculus}
\dinferenceRule[dGt|dGt]{diff ghost clock}
{\linferenceRule
  {\lsequent{\Gamma,\timevar=0} {\ddiamond{\pevolvein{\D{x}=\genDE{x},\D{\timevar}=1}{\ivr}}{\rfvar}}}
  {\lsequent{\Gamma}{\ddiamond{\pevolvein{\D{x}=\genDE{x}}{\ivr}}{\rfvar}}}
}{}
\dinferenceRule[MdW|M${\didia{'}}$]{}
{\linferenceRule
  {\lsequent{\ivr,\rrfvar}{\rfvar} \quad \lsequent{\Gamma}{\ddiamond{\pevolvein{\D{x}=\genDE{x}}{\ivr}}{\rrfvar}}}
  {\lsequent{\Gamma}{\ddiamond{\pevolvein{\D{x}=\genDE{x}}{\ivr}}{\rfvar}}}
}{}
\end{calculus}
\end{calculuscollection}
\end{lemma}

Axiom~\irref{diamond} expresses the duality between the box and diamond modalities.
It is used to switch between the two in proofs and to dualize axioms between the box and diamond modalities.
Axiom~\irref{K} is the modus ponens principle for the box modality.
Differential invariants~\irref{dIcmp} says that if the Lie derivatives obey the inequality $\lied[]{\genDE{x}}{\ptermA} \geq \lied[]{\genDE{x}}{\ptermB}$, then $\ptermA \cmp \ptermB$ is an invariant of the ODE.
Differential cuts~\irref{dC} says that if we can separately prove that formula $\rcfvar$ is always satisfied along the solution, then $\rcfvar$ may be assumed in the domain constraint when proving the same for formula $\rfvar$.
In the box modality, solutions are restricted to stay in the domain constraint $\ivr$; differential weakening~\irref{dW} says that postcondition $\rfvar$ is always satisfied along solutions if it is already implied by the domain constraint.
Liveness arguments are often based on analyzing the duration that solutions of the ODE are followed.
Rule~\irref{dGt} is a special instance of the more general differential ghosts rule~\cite{DBLP:journals/jar/Platzer17,Platzer18,DBLP:conf/lics/PlatzerT18} which allows \emph{new} auxiliary variables to be introduced for the purposes of proof.
It augments the ODE $\D{x}=\genDE{x}$ with an additional differential equation, $\D{\timevar}=1$, so that the (fresh) variable $\timevar$, with initial value $\timevar=0$, tracks the progress of time.
Using \irref{dW+K+diamond}, the final two monotonicity proof rules~\irref{MbW+MdW} for differential equations are derivable.
They strengthen the postcondition from $\rfvar$ to $\rrfvar$, assuming domain constraint $\ivr$, for the box and diamond modalities respectively.

Throughout this paper, we present proof rules, e.g.,~\irref{dW}, that discard all assumptions $\Gamma$ on initial states when moving from conclusion to the premises.
Intuitively, this is necessary for soundness because the premises of these rules internalize reasoning that happens \emph{along solutions} of the ODE $\pevolvein{\D{x}=\genDE{x}}{\ivr}$ rather than in the initial state.
On the other hand, the truth value of constant assumptions $\constt{\rfvar}$ do not change along solutions, so they can be soundly kept across rule applications~\cite{Platzer18}.
These additional constant contexts are useful when working with assumptions on symbolic parameters e.g., $\constt{v} > 0$ to represent a (constant) positive velocity.

\section{Liveness via Box Refinements}
\label{sec:livenessaxioms}
Suppose we already know an initial liveness property $\ddiamond{\pevolvein{\D{x}=\genDE{x}}{\ivr_0}}{\rfvar_0}$ for the ODE $\D{x}=\genDE{x}$.
How could this be used to prove a desired liveness property $\ddiamond{\pevolvein{\D{x}=\genDE{x}}{\ivr}}{\rfvar}$ for that ODE?
Logically, this amounts to proving:
\begin{equation}
\ddiamond{\pevolvein{\D{x}=\genDE{x}}{\ivr_0}}{\rfvar_0} \limply \ddiamond{\pevolvein{\D{x}=\genDE{x}}{\ivr}}{\rfvar}
\label{eq:refinementimpl}
\end{equation}

Proving implication~\rref{eq:refinementimpl} \emph{refines} the initial liveness property to the desired one.
Our approach is built on refinement axioms that conclude such implications from box modality formulas.
The following are two basic derived refinement axioms:

\begin{lemma}[Diamond refinement axioms]
\label{lem:diarefaxioms}
The following $\didia{\cdot}$ refinement axioms are derivable in \dL.\\
\begin{calculuscollection}
\begin{calculus}
\dinferenceRule[dDR|DR${\didia{\cdot}}$]{}
{\linferenceRule[impl]
  {\dbox{\pevolvein{\D{x}=\genDE{x}}{\rrfvar}}{\ivr}}
  {\big( \ddiamond{\pevolvein{\D{x}=\genDE{x}}{\rrfvar}}{\rfvar} \limply \axkey{\ddiamond{\pevolvein{\D{x}=\genDE{x}}{\ivr}}{\rfvar}}\big)}
}{}

\dinferenceRule[Prog|K${\didia{\&}}$]{}
{
\linferenceRule[impl]
  {\dbox{\pevolvein{\D{x}=\genDE{x}}{\ivr \land \lnot{\rfvar}}}{\lnot{\rgvar}}}
  {\big(\ddiamond{\pevolvein{\D{x}=\genDE{x}}{\ivr}}{\rgvar} \limply \axkey{\ddiamond{\pevolvein{\D{x}=\genDE{x}}{\ivr}}{\rfvar}}\big)}
}{}
\end{calculus}
\end{calculuscollection}
\end{lemma}
\begin{proofatend}
Axiom~\irref{Prog} is derived as follows starting with \irref{diamond+notl+notr}, which turns the diamond modalities in the antecedent and succedent into box modality formulas. A \irref{dC} step using the right antecedent completes the proof.
{\footnotesizeoff%
\begin{sequentdeduction}[array]
\linfer[diamond+notl+notr]{
\linfer[dC]{
  \lclose
}
  {\lsequent{\dbox{\pevolvein{\D{x}=\genDE{x}}{\ivr \land \lnot{\rfvar}}}{\lnot{\rgvar}}, \dbox{\pevolvein{\D{x}=\genDE{x}}{\ivr}}{\lnot{\rfvar}}} {\dbox{\pevolvein{\D{x}=\genDE{x}}{\ivr}}{\lnot{\rgvar}}}}
}
  {\lsequent{\dbox{\pevolvein{\D{x}=\genDE{x}}{\ivr \land \lnot{\rfvar}}}{\lnot{\rgvar}}, \ddiamond{\pevolvein{\D{x}=\genDE{x}}{\ivr}}{\rgvar}} {\ddiamond{\pevolvein{\D{x}=\genDE{x}}{\ivr}}{\rfvar}}}
\end{sequentdeduction}
}%
Axiom~\irref{dDR} similarly derives from axiom~\irref{DMP} with~\irref{diamond}~\cite{DBLP:conf/lics/PlatzerT18}.
\end{proofatend}

In axiom \irref{Prog}, formula $\dbox{\pevolvein{\D{x}=\genDE{x}}{\ivr \land \lnot{\rfvar}}}{\lnot{\rgvar}}$ says the solution cannot get to $\rgvar$ before getting to $\rfvar$ as $\rgvar$ never happens while $\lnot{\rfvar}$ holds.
In axiom~\irref{dDR}, formula $\dbox{\pevolvein{\D{x}=\genDE{x}}{\rrfvar}}{\ivr}$ says that the ODE solution never leaves $\ivr$ while staying in $\rrfvar$, so the solution getting to $\rfvar$ within $\rrfvar$ implies that it also gets to $\rfvar$ within $\ivr$.
These axioms prove implication~\rref{eq:refinementimpl} in just one refinement step.
Logical implication is transitive though, so we can also chain a longer sequence of such steps to prove implication~\rref{eq:refinementimpl}.
This is shown in~\rref{eq:refinementchain}, with neighboring implications informally chained together for illustration:
{\footnotesizeoff
\begin{align}
\ddiamond{\pevolvein{\D{x}=\genDE{x}}{\ivr_0}}{\rfvar_0} &\overbrace{\limply}^{\hidewidth \irref{dDR}~\text{with}~\dbox{\pevolvein{\D{x}=\genDE{x}}{\ivr_1}}{\ivr_0} \quad\quad\;\; \hidewidth} \ddiamond{\pevolvein{\D{x}=\genDE{x}}{\ivr_1}}{\rfvar_0} \overbrace{\limply}^{\hidewidth \;\;\quad\quad \irref{Prog}~\text{with}~\dbox{\pevolvein{\D{x}=\genDE{x}}{\ivr_1 \land \lnot{P_1}}}{\lnot{P_0}} \hidewidth} \ddiamond{\pevolvein{\D{x}=\genDE{x}}{\ivr_1}}{\rfvar_1} \nonumber \\
&~\limply \cdots \limply \ddiamond{\pevolvein{\D{x}=\genDE{x}}{\ivr}}{\rfvar}
\label{eq:refinementchain}
\end{align}
}%

The chain of refinements~\rref{eq:refinementchain} proves the desired implication~\rref{eq:refinementimpl}, but to formally conclude the liveness property $\ddiamond{\pevolvein{\D{x}=\genDE{x}}{\ivr}}{\rfvar}$, we still need to prove the hypothesis $\ddiamond{\pevolvein{\D{x}=\genDE{x}}{\ivr_0}}{\rfvar_0}$ on the left of the implication.
The following axioms provide a means of formally establishing such an initial liveness property:

\begin{lemma}[Existence axioms]
\label{lem:diainitaxioms}
The following existence axioms are sound.
In both axioms, $\constt{\ptermA}$ is constant for the ODE $\D{x}=\genDE{x},\D{\timevar}=1$.
In axiom~\irref{GEx}, the ODE $\D{x}=\genDE{x}$ is globally Lipschitz continuous.
In axiom~\irref{BEx}, the formula $\boundedf(x)$ characterizes a bounded set over variables $x$.\\
\begin{calculuscollection}
\begin{calculus}
\cinferenceRule[GEx|GEx]{}
{
  \axkey{\ddiamond{\pevolve{\D{x}=\genDE{x},\D{\timevar}=1}}{\timevar > \constt{\ptermA}}}
}{}

\cinferenceRule[BEx|BEx]{}
{
  \axkey{\ddiamond{\pevolve{\D{x}=\genDE{x},\D{\timevar}=1}}{(\lnot{\boundedf(x)} \lor \timevar > \constt{\ptermA})}}
}{}
\end{calculus}
\end{calculuscollection}
\end{lemma}
\begin{proofatend}
Let $\iget[state]{\I} \in \States$ and $\solvar : [0, T) \to \States, 0<T\leq\infty$ be the unique, right-maximal solution~\cite{Chicone2006,Walter1998} to the ODE $\D{x}=\genDE{x},\D{\timevar}=1$ with initial value $\solvar(0)=\iget[state]{\I}$.
Let $\timevar_0 \mnodefeq \iget[state]{\I}(t)$ be the initial value of the time variable $\timevar$.
Since this time variable obeys the ODE $\D{\timevar}=1$, by uniqueness, its value along solution $\solvar$ is given by $\solvar(\zeta)(\timevar) = \timevar_0 + \zeta$ for $\zeta \in [0,T)$.
Further, let $\ptermA_0 \mnodefeq \ivaluation{\I}{\constt{\ptermA}}$ be the initial value of term $\constt{\ptermA}$.
Since $\constt{\ptermA}$ is constant for the ODE $\D{x}=\genDE{x}$, its value along solution $\solvar$ remains constant at $\ptermA_0$.

\begin{itemize}
\item For axiom~\irref{GEx}, the ODE $\D{x}=\genDE{x},\D{\timevar}=1$ is globally Lipschitz continuous in $x$. By~\cite[\S10.VII]{Walter1998}, this implies that solutions exist for all time, i.e., $T = \infty$ for the right-maximal solution $\solvar$.
Thus, for any $\zeta > \ptermA_0 - \timevar_0$ where $\zeta \in [0,\infty)$, $\solvar(\zeta)(\timevar) = \timevar_0 + \zeta > \ptermA_0$, which yields the desired conclusion.

\item For axiom~\irref{BEx}, if $T = \infty$, then the conclusion follows by a similar argument to~\irref{GEx} with the right disjunct of~\irref{BEx} satisfied at some time $\zeta$ along the right-maximal solution.
Otherwise, $0 < T < \infty$ and the solution $\solvar$ only exists on the finite time interval $[0,T)$.
By~\cite[Theorem 1.4]{Chicone2006}, $|\solvar(\zeta)|$ is unbounded, approaching $\infty$ as $\zeta$ approaches $T$ (where $|\cdot|$ is the Euclidean norm).
The value of $\timevar$ is bounded above by the constant $\timevar_0 + T \in \reals$ and all parameters $y \in \allvars \setminus \{x\}$ have constant real values along $\solvar$.
Therefore, the only way for $|\solvar(\zeta)|$ to approach $\infty$ is for $|\solvar(\zeta)(x)|$ to approach $\infty$ as $\zeta$ approaches $T$.

Let $\gamma \in \reals^r$ be the real (constant) value of the parameters $y = (y_1,\dots,y_r)$ along the solution $\solvar$.
By assumption, the set characterized by formula $B(x)$ for these parameters is bounded with $\beta \mnodefeq \imodel{\I}{B(x)}_\gamma \subseteq \reals^n$.
It must be the case that $\solvar(\zeta)(x) \notin \beta$ for some $\zeta \in [0,T)$.
Otherwise, $|\solvar(\zeta)(x)|$ is bounded for all $\zeta \in [0,T)$ which is a contradiction.
The conclusion follows since the left disjunct of~\irref{BEx} is satisfied at time $\zeta \in [0,T)$ where $\solvar(\zeta)(x) \notin \beta$.
\qedhere
\end{itemize}
\end{proofatend}

Axioms~\irref{GEx+BEx} are stated for ODEs with an explicit time variable $\timevar$, where $\D{x}=\genDE{x}$ does not mention $\timevar$.
Within proofs, these axioms can be accessed after using rule~\irref{dGt} to add a fresh time variable $\timevar$.
Solutions of globally Lipschitz ODEs exist for all time so axiom~\irref{GEx} says that along such solutions, the value of time variable $\timevar$ eventually exceeds that of the constant term $\ptermA()$.\footnote{It is important for soundness that $\ptermA()$ is constant for the ODE, e.g., instances of axiom~\irref{GEx} with postcondition $\timevar > 2\timevar$ are clearly not valid.}
This global Lipschitz continuity condition is satisfied e.g., by $\exlinear$, and more generally by linear ODEs of the form $\D{x}=Ax$, where $A$ is a matrix of (constant) parameters~\cite{Chicone2006}.
Global Lipschitz continuity is a strong requirement that does not hold even for simple non-linear ODEs like $\exnonlinear$, which only have short-lived solutions (see~\rref{fig:odeexamples}).
This phenomenon, where the right-maximal ODE solution $\solvar$ is only defined on a finite time interval $[0,T)$ with $T < \infty$, is known as \emph{finite time blow up of solutions}~\cite{Chicone2006}.
Axiom~\irref{BEx} removes the global Lipschitz continuity requirement but weakens the postcondition to say that solutions must either exist for sufficient duration or blow up and leave the \emph{bounded} set characterized by formula $\boundedf(x)$.

Refinement with axiom~\irref{dDR} requires proving the formula $\dbox{\pevolvein{\D{x}=\genDE{x}}{\rrfvar}}{\ivr}$.
Na\"ively, we might expect that adding $\lnot{\rfvar}$ to the domain constraint should also work, i.e., the solution only needs to be in $\ivr$ while it has not yet gotten to $\rfvar$:
\[
\cinferenceRule[badaxiom|DR$\didia{\cdot}$\usebox{\Lightningval}]{}
{
\linferenceRule[impl]
  {\dbox{\pevolvein{\D{x}=\genDE{x}}{\rrfvar \land \lnot{\rfvar}}}{\ivr} }
  {\big(\ddiamond{\pevolvein{\D{x}=\genDE{x}}{\rrfvar}}{\rfvar} \limply \axkey{\ddiamond{\pevolvein{\D{x}=\genDE{x}}{\ivr}}{\rfvar}}\big)}
}{}
\]

This conjectured axiom is unsound (indicated by $\mbox{\lightning}$) as the solution could sneak out of $\ivr$ when it crosses from $\lnot{\rfvar}$ into $\rfvar$.
In continuous settings, the language of topology makes precise what this means.
The following topological refinement axioms soundly restrict what happens at the crossover point:

\begin{lemma}[Topological refinement axioms]
\label{lem:diatopaxioms}
The following topological $\didia{\cdot}$ refinement axioms are sound.
In axiom~\irref{CORef}, $\rfvar,\ivr$ either both characterize topologically open or both characterize topologically closed sets over variables $x$.\\
\begin{calculuscollection}
\begin{calculus}
\cinferenceRule[CORef|COR]{}
{
\linferenceRule[impl]
  {\lnot{\rfvar} \land \dbox{\pevolvein{\D{x}=\genDE{x}}{\rrfvar \land \lnot{\rfvar}}}{\ivr} }
  {\big(\ddiamond{\pevolvein{\D{x}=\genDE{x}}{\rrfvar}}{\rfvar} \limply \axkey{\ddiamond{\pevolvein{\D{x}=\genDE{x}}{\ivr}}{\rfvar}}\big)}
}{}

\cinferenceRule[SARef|SAR]{}
{
\linferenceRule[impl]
  {\dbox{\pevolvein{\D{x}=\genDE{x}}{\rrfvar \land \lnot{(\rfvar \land \ivr)}}}{\ivr}}
  {\big( \ddiamond{\pevolvein{\D{x}=\genDE{x}}{\rrfvar}}{\rfvar} \limply \axkey{\ddiamond{\pevolvein{\D{x}=\genDE{x}}{\ivr}}{\rfvar}}\big)}
}{}
\end{calculus}
\end{calculuscollection}
\end{lemma}
\begin{proofatend}
Let $\iget[state]{\I} \in \States$ and $\solvar : [0, T) \to \States, 0<T\leq\infty$ be the unique, right-maximal solution~\cite{Chicone2006,Walter1998} to the ODE $\D{x}=\genDE{x}$ with initial value $\solvar(0)=\iget[state]{\I}$.
By definition, $\solvar$ is differentiable, and therefore continuous.
This proof uses the fact that the inverse image of an open set for a continuous function is open~\cite[Theorem 4.8]{MR0385023}.
In particular, if $\solvar(t) \in \openset$ for some time $0 \leq t < T$ and open set $\openset$, then the inverse image of a sufficiently small open ball $\openset_\varepsilon \subseteq \openset$ centered at $\solvar(t)$ is open.
Thus, if $t>0$,\footnote{In case $t=0$, the time interval in~\rref{eq:openball} is truncated to the left with $\solvar(\zeta) \in \openset~\text{for all}~0 \leq \zeta < t+\varepsilon$.}, then $\solvar$ stays in the open set $\openset$ for an open time interval about $t$, i.e., for some $\varepsilon > 0$:
\begin{align}
\solvar(\zeta) \in \openset~\text{for all}~t-\varepsilon \leq \zeta \leq t+\varepsilon
\label{eq:openball}
\end{align}

For the soundness proof of both~\irref{CORef} and~\irref{SARef}, assume that $\iget[state]{\I} \in \imodel{\I}{\ddiamond{\pevolvein{\D{x}=\genDE{x}}{\rrfvar}}{\rfvar}}$, i.e., there is time $\tau \in [0,T)$ such that $\solvar(\tau) \in \imodel{\I}{\rfvar}$ and $\solvar(\zeta) \in \imodel{\I}{\rrfvar}$ for all $0 \leq \zeta \leq \tau$.

\begin{itemize}
\item For axiom~\irref{CORef}, suppose further that~$\iget[state]{\I} \in \imodel{\I}{\lnot{\rfvar} \land \dbox{\pevolvein{\D{x}=\genDE{x}}{\rrfvar \land \lnot{\rfvar}}}{\ivr}}$.
Consider the set $\{t~|~\solvar(\zeta) \notin \imodel{\I}{\rfvar}~\text{for all}~0 \leq \zeta \leq t \}$ which is non-empty since $\iget[state]{\I} = \solvar(0) \notin \imodel{\I}{\rfvar}$.
This set has a supremum $t$ with $0 \leq t \leq \tau$ and $\solvar(\zeta) \notin \imodel{\I}{\rfvar}$ for all $0 \leq \zeta < t$.

\begin{itemize}
\item Suppose $\rfvar,\ivr$ both characterize topologically closed sets.
Since $\rfvar$ characterizes a topologically closed set, if $\solvar(t) \notin \imodel{\I}{\rfvar}$, then by~\rref{eq:openball}, $t$ is not the supremum, which is a contradiction.
Thus, $\solvar(t) \in \imodel{\I}{\rfvar}$ and $0 < t$.
Hence, $\solvar(\zeta) \in \imodel{\I}{\rrfvar \land \lnot{\rfvar}}$ for all $0 \leq \zeta < t$, which, together with the assumption $\iget[state]{\I} \in \imodel{\I}{\dbox{\pevolvein{\D{x}=\genDE{x}}{\rrfvar \land \lnot{\rfvar}}}{\ivr}}$ implies $\solvar(\zeta) \in \imodel{\I}{\ivr}$ for all $0 \leq \zeta < t$.
Since $\ivr$ characterizes a topologically closed set, this implies $\solvar(t) \in \imodel{\I}{\ivr}$ since~\rref{eq:openball} again yields a contradiction otherwise.
Thus, $\iget[state]{\I} \in \imodel{\I}{\ddiamond{\pevolvein{\D{x}=\genDE{x}}{\ivr}}{\rfvar}}$ because $\solvar(t) \in \imodel{\I}{\rfvar}$ and $\solvar(\zeta) \in \imodel{\I}{\ivr}$ for all $0 \leq \zeta \leq t$.

\item Suppose $\rfvar,\ivr$ both characterize topologically open sets.
Then, $\solvar(t) \notin \imodel{\I}{\rfvar}$ because otherwise, by~\rref{eq:openball}, $t$ is not the supremum.
Note that $t < \tau$ and $\solvar(\zeta) \in \imodel{\I}{\rrfvar \land \lnot{\rfvar}}$ for all $0 \leq \zeta \leq t$, which, together with the assumption $\iget[state]{\I} \in \imodel{\I}{\dbox{\pevolvein{\D{x}=\genDE{x}}{\rrfvar \land \lnot{\rfvar}}}{\ivr}}$ implies $\solvar(\zeta) \in \imodel{\I}{\ivr}$ for all $0 \leq \zeta \leq t$.
Since $\ivr$ characterizes a topologically open set, by~\rref{eq:openball}, there exists $\varepsilon > 0$ where $t+\varepsilon < \tau$ such that $\solvar(t+\zeta) \in \imodel{\I}{\ivr}$ for all $0 \leq \zeta \leq \varepsilon$.
By definition of the supremum, for every such $\varepsilon > 0$, there exists $\solvar(t+\zeta) \in \imodel{\I}{\rfvar}$ for some $\zeta$ where $0 < \zeta \leq \varepsilon$.
This yields the desired conclusion.
\end{itemize}

\item For axiom~\irref{SARef}, assume that
\begin{equation}
\iget[state]{\I} \in \imodel{\I}{\dbox{\pevolvein{\D{x}=\genDE{x}}{\rrfvar \land \lnot{(\rfvar \land \ivr)}}}{\ivr}}
\label{eq:sarefassum}
\end{equation}
If $\iget[state]{\I} \in \imodel{\I}{\rfvar \land \ivr}$, then $\iget[state]{\I} \in \ddiamond{\pevolvein{\D{x}=\genDE{x}}{\ivr}}{\rfvar}$ trivially by following the solution $\solvar$ for duration $0$.
Thus, assume $\iget[state]{\I} \notin \imodel{\I}{\rfvar \land \ivr}$.
From~\rref{eq:sarefassum}, $\iget[state]{\I} \in \imodel{\I}{\ivr}$ which further implies $\iget[state]{\I} \notin \imodel{\I}{\rfvar}$.
Consider the set of times $\{t~|~\solvar(\zeta) \notin \imodel{\I}{\rfvar}~\text{for all}~0 \leq \zeta \leq t \}$ which is non-empty since $\iget[state]{\I} = \solvar(0) \notin \imodel{\I}{\rfvar}$.
This set has a supremum $t$ with $0 \leq t \leq \tau$ and $\solvar(\zeta) \notin \imodel{\I}{\rfvar}$ for all $0 \leq \zeta < t$.
Thus, $\solvar(\zeta) \in \imodel{\I}{\rrfvar \land \lnot{(\rfvar \land \ivr)}}$ for all $0 \leq \zeta < t$.
By~\rref{eq:sarefassum}, $\solvar(\zeta) \in \imodel{\I}{\ivr}$ for all $0 \leq \zeta < t$.
Classically, either $\solvar(t) \in \imodel{\I}{\rfvar}$ or $\solvar(t) \notin \imodel{\I}{\rfvar}$.

\begin{itemize}
\item Suppose $\solvar(t) \in \imodel{\I}{\rfvar}$, then if $\solvar(t) \in \imodel{\I}{\ivr}$, $\solvar(\zeta) \in \imodel{\I}{\ivr}$ for all $0 \leq \zeta \leq t$ and so by definition, $\iget[state]{\I} \in \imodel{\I}{\ddiamond{\pevolvein{\D{x}=\genDE{x}}{\ivr}}{\rfvar}}$.
On the other hand, if $\solvar(t) \notin \imodel{\I}{\ivr}$, then $\solvar(\zeta) \in \imodel{\I}{\rrfvar \land \lnot{(\rfvar \land \ivr)}}$ for all $0 \leq \zeta \leq t$, so from~\rref{eq:sarefassum}, $\solvar(t) \in \imodel{\I}{\ivr}$, which yields a contradiction.

If the formula $\rfvar$ is further assumed to characterize a closed set, this sub-case (with $\solvar(t) \in \imodel{\I}{\rfvar}$) is the only possibility because $\solvar(\zeta) \notin \imodel{\I}{\rfvar}$ for all $0 \leq \zeta < t$ which implies $\solvar(t) \in \imodel{\I}{\rfvar}$ by~\rref{eq:openball}.

\item Suppose $\solvar(t) \notin \imodel{\I}{\rfvar}$, then $t < \tau$ and $\solvar(\zeta) \in \imodel{\I}{\rrfvar \land \lnot{(\rfvar \land \ivr)}}$ for all $0 \leq \zeta \leq t$, so from~\rref{eq:sarefassum}, $\solvar(t) \in \imodel{\I}{\ivr}$.
Since $\ivr$ is a formula of first-order real arithmetic, solutions of polynomial ODEs either locally progress into the set characterized by $\ivr$ or $\lnot{\ivr}$~\cite{DBLP:conf/lics/PlatzerT18,DBLP:conf/fm/SogokonJ15}.\footnote{This property is specific to sets characterized by first-order formulas of real arithmetic and is not necessarily true for arbitrary sets and ODEs.}
In particular, there exists $\varepsilon > 0$, where $t + \varepsilon < \tau$, such that either \textcircled{1} $\solvar(t+\zeta) \in \imodel{\I}{\ivr}$ for all $0 < \zeta \leq \varepsilon$ or \textcircled{2} $\solvar(t+\zeta) \notin \imodel{\I}{\ivr}$ for all $0 < \zeta \leq \varepsilon$.
By definition of the supremum, for every such $\varepsilon$ there exists $\solvar(t+\zeta) \in \imodel{\I}{\rfvar}$ for some $\zeta$ where $0 < \zeta \leq \varepsilon$.

In case \textcircled{1}, since $\solvar(t+\zeta) \in \imodel{\I}{\rfvar}$ and $\solvar(\nu) \in \imodel{\I}{\ivr}$ for all $0 \leq \nu \leq t+\zeta$, we have $\iget[state]{\I} \in \imodel{\I}{\ddiamond{\pevolvein{\D{x}=\genDE{x}}{\ivr}}{\rfvar}}$.
If the formula $\ivr$ is further assumed to characterize an open set, this sub-case (\textcircled{1}) is the only possibility, even if $\ivr$ is not a formula of first-order real arithmetic, because $\solvar(t) \in \imodel{\I}{\ivr}$ implies $\solvar$ continues to satisfy $\ivr$ for some time interval to the right of $t$ by~\rref{eq:openball}.

In case \textcircled{2}, observe that $\solvar(\nu) \in \imodel{\I}{\rrfvar \land \lnot{(\rfvar \land \ivr)}}$ for all $0 \leq \nu\leq t+\zeta$, from~\rref{eq:sarefassum}, $\solvar(t+\zeta) \in \imodel{\I}{\ivr}$, which yields a contradiction.
\qedhere
\end{itemize}

\end{itemize}
\end{proofatend}

Axiom~\irref{CORef} is the more informative topological refinement axiom.
Like the (unsound) axiom candidate~\irref{badaxiom}, it allows formula $\lnot{\rfvar}$ to be assumed in the domain constraint when proving the box refinement.
For soundness though, axiom~\irref{CORef} has additional topological side conditions on formulas $\rfvar,\ivr$ so it can only be used when these conditions are met.
Axiom~\irref{SARef} applies more generally but only assumes the less informative formula $\lnot{(\rfvar \land \ivr)}$ in the domain constraint for the box modality formula in the refinement.
Its proof crucially relies on $\ivr$ being a formula of real arithmetic so that the set it characterizes has tame topological behavior~\cite{Bochnak1998}, see the proof in~\thereport{\rref{app:proofs}} for more details.\footnote{By topological considerations similar to~\irref{CORef}, axiom~\irref{SARef} is also sound if it requires that the formula $\rfvar$ (or resp. $\ivr$) characterizes a topologically closed (resp. open) set over the ODE variables $x$.
These additional cases are also proved in~\thereport{\rref{app:proofs}} without relying on the fact that $\ivr$ is a formula of real arithmetic.}

\section{Liveness Without Domain Constraints}
\label{sec:nodomconstraint}
This section presents proof rules for liveness properties of ODEs $\D{x}=\genDE{x}$ without domain constraints, i.e., where $\ivr$ is the formula $\ltrue$.
Errors and omissions in the surveyed techniques are \highlight{highlighted in blue}.

\subsection{Differential Variants}
A fundamental technique for verifying liveness of discrete loops is the identification of a loop variant, i.e., a quantity that decreases monotonically across each loop iteration.
Differential variants~\cite{DBLP:journals/logcom/Platzer10} are their continuous analog:

\begin{corollary}[Atomic differential variants~\cite{DBLP:journals/logcom/Platzer10}]
\label{cor:atomicdvcmp}
The following proof rules (where $\cmp$ is either $\geq$ or $>$) are derivable in \dL.
Terms $\constt{\varepsilon},\constt{\ptermA_0}$ are constant for ODE $\D{x}=\genDE{x},\D{\timevar}=1$.
In rule~\irref{dVcmp}, $\D{x}=\genDE{x}$ is globally Lipschitz continuous.\\
\begin{calculuscollection}
\begin{calculus}
\dinferenceRule[dVcmpA|dV$_\cmp^*$]{}
{\linferenceRule
  { \lsequent{\lnot{(\ptermA \cmp 0)}}{\lied[]{\genDE{x}}{\ptermA}\geq \constt{\varepsilon}}
  }
  {\lsequent{\Gamma,\ptermA{=}\constt{\ptermA_0},\timevar{=}0,\ddiamond{\pevolve{\D{x}=\genDE{x},\D{\timevar}=1}}{\big(\constt{\ptermA_0} {+} \constt{\varepsilon} \timevar {>} 0\big)}}{\ddiamond{\pevolve{\D{x}=\genDE{x},\D{\timevar}=1}}{\ptermA \cmp 0}} }
}{}

\dinferenceRule[dVcmp|dV$_\cmp$]{}
{\linferenceRule
  { \lsequent{\lnot{(\ptermA \cmp 0)}}{\lied[]{\genDE{x}}{\ptermA}\geq \constt{\varepsilon}}
  }
  {\lsequent{\Gamma, \constt{\varepsilon} > 0}{\ddiamond{\pevolve{\D{x}=\genDE{x}}}{\ptermA \cmp 0}} }
}{}
\end{calculus}
\end{calculuscollection}
\end{corollary}
\begin{proofsketch}[app:proofs]
Rule~\irref{dVcmpA} derives by using axiom \irref{Prog} with the choice of formula $\rgvar \mnodefequiv \constt{\ptermA_0} {+} \constt{\varepsilon} \timevar {>} 0$:

\begin{minipage}[b]{0.05\textwidth}
\end{minipage}\hfill
\begin{minipage}[b]{0.95\textwidth}
{\footnotesizeoff%
\begin{sequentdeduction}
\linfer[Prog]{
  \lsequent{\Gamma,\ptermA {=} \constt{\ptermA_0},\timevar{=}0}{\dbox{\pevolvein{\D{x}=\genDE{x},\D{\timevar}=1}{\lnot{(\ptermA \cmp 0)}}}{\constt{\ptermA_0} {+} \constt{\varepsilon} \timevar \leq 0}}
}
  {\lsequent{\Gamma,\ptermA {=} \constt{\ptermA_0},\timevar{=}0,\ddiamond{\pevolve{\D{x}=\genDE{x},\D{\timevar}=1}}{\big(\constt{\ptermA_0} {+} \constt{\varepsilon} \timevar {>} 0\big)}}{\ddiamond{\pevolve{\D{x}=\genDE{x},\D{\timevar}=1}}{\ptermA {\cmp} 0}}}
\end{sequentdeduction}
}%
\end{minipage}

Monotonicity \irref{MbW} strengthens the postcondition to $\ptermA \geq \constt{\ptermA_0} + \constt{\varepsilon} \timevar$ with the domain constraint $\lnot{(\ptermA \cmp 0)}$.
A subsequent use of \irref{dIcmp} completes the derivation:
{\footnotesizeoff%
\begin{sequentdeduction}[array]
  \linfer[MbW]{
  \linfer[dIcmp]{
    \lsequent{\lnot{(\ptermA \cmp 0)}}{\lied[]{\genDE{x}}{\ptermA}\geq \constt{\varepsilon}}
  }
    {\lsequent{\Gamma, \ptermA = \constt{\ptermA_0}, \timevar=0}{\dbox{\pevolvein{\D{x}=\genDE{x},\D{\timevar}=1}{\lnot{(\ptermA \cmp 0)}}}{\ptermA \geq \constt{\ptermA_0} + \constt{\varepsilon} \timevar}}}
  }
  {\lsequent{\Gamma,\ptermA = \constt{\ptermA_0}, \timevar=0}{\dbox{\pevolvein{\D{x}=\genDE{x},\D{\timevar}=1}{\lnot{(\ptermA \cmp 0)}}}{\constt{\ptermA_0} + \constt{\varepsilon} \timevar \leq 0}}}
\end{sequentdeduction}
}%

Rule~\irref{dVcmp} is derived in~\thereport{\rref{app:proofs}} as a corollary of rule~\irref{dVcmpA}.
It uses the global existence axiom~\irref{GEx} and rule~\irref{dGt} to introduce the time variable.
\end{proofsketch}
\begin{proofatend}
The derivation of~\irref{dVcmp} starts by introducing fresh variables $\ptermA_0, i$ representing the initial values of $\ptermA$ and the multiplicative inverse of $\varepsilon()$ respectively using arithmetic cuts (\irref{cut+qear}) and Skolemizing (\irref{existsl}).
It then uses \irref{dGt} to introduce a fresh time variable to the system of differential equations:
{\footnotesizeoff%
\begin{sequentdeduction}[array]
  \linfer[cut+qear]{
  \linfer[existsl]{
  \linfer[dGt]{
    \lsequent{\Gamma, \constt{\varepsilon} > 0, \ptermA=\ptermA_0, i\varepsilon() = 1, \timevar=0}{\ddiamond{\pevolve{\D{x}=\genDE{x},\D{\timevar}=1}}{\ptermA \cmp 0}}
  }
    {\lsequent{\Gamma, \constt{\varepsilon} > 0, \ptermA=\ptermA_0, i\varepsilon() = 1}{\ddiamond{\pevolve{\D{x}=\genDE{x}}}{\ptermA \cmp 0}}}
  }
  {\lsequent{\Gamma, \constt{\varepsilon} > 0, \lexists{\ptermA_0}{\ptermA=\ptermA_0}, \lexists{i}{i\varepsilon() = 1}}{\ddiamond{\pevolve{\D{x}=\genDE{x}}}{\ptermA \cmp 0}}}
  }
  {\lsequent{\Gamma, \constt{\varepsilon} > 0}{\ddiamond{\pevolve{\D{x}=\genDE{x}}}{\ptermA \cmp 0}} }
\end{sequentdeduction}
}%

Next, an initial liveness assumption $\ddiamond{\pevolve{\D{x}=\genDE{x},\D{\timevar}=1}}{\ptermA_0 {+} \constt{\varepsilon}\timevar {>} 0}$ is cut into the antecedents after which rule~\irref{dVcmpA} is used to obtain the premise of~\irref{dVcmp}.
Intuitively, this initial liveness assumption says that the solution exists for sufficiently long, so that $\ptermA$, which is provably bounded below by $\ptermA_0 {+} \constt{\varepsilon}\timevar$, becomes positive when starting from its initial value $\ptermA_0$.
The proof of this cut is abbreviated \textcircled{1} and proved below.
{\footnotesizeoff%
\begin{sequentdeduction}[array]
  \linfer[cut]{
  \linfer[dVcmpA]{
    \lsequent{\lnot{(\ptermA \cmp 0)}}{\lied[]{\genDE{x}}{\ptermA}\geq \constt{\varepsilon}}
  }
    {\lsequent{\Gamma, \ptermA=\ptermA_0, \timevar=0, \ddiamond{\pevolve{\D{x}=\genDE{x},\D{\timevar}=1}}{\ptermA_0 {+} \constt{\varepsilon}\timevar {>} 0} }{\ddiamond{\pevolve{\D{x}=\genDE{x},\D{\timevar}=1}}{\ptermA \cmp 0}}} \quad
    \textcircled{1}
  }
  {\lsequent{\Gamma, \constt{\varepsilon} > 0, \ptermA=\ptermA_0, i\varepsilon() = 1, \timevar=0}{\ddiamond{\pevolve{\D{x}=\genDE{x},\D{\timevar}=1}}{\ptermA \cmp 0}}}
\end{sequentdeduction}
}%

From premise~\textcircled{1}, a monotonicity step~\irref{MdW} equivalently rephrases the postcondition of the cut in real arithmetic.
The arithmetic rephrasing works using the constant assumption $\constt{\varepsilon} > 0$ and the choice of $i$ as the multiplicative inverse of $\constt{\varepsilon}$.
Axiom~\irref{GEx} finishes the derivation because the ODE $\D{x}=\genDE{x}$ is assumed to be globally Lipschitz continuous and $- i \ptermA_0$ is constant for the ODE.
{\footnotesizeoff%
\begin{sequentdeduction}[array]
  \linfer[qear+MdW]{
  \linfer[GEx]{
    \lclose
  }
    {\lsequent{}{\ddiamond{\pevolve{\D{x}=\genDE{x},\D{\timevar}=1}}{ \timevar > -i \ptermA_0}}}
  }
  {\lsequent{\constt{\varepsilon} > 0, i\varepsilon() = 1}{\ddiamond{\pevolve{\D{x}=\genDE{x},\D{\timevar}=1}}{\ptermA_0 + \constt{\varepsilon}\timevar > 0}}}
\\[-\normalbaselineskip]\tag*{\qedhere}
\end{sequentdeduction}
}%
\end{proofatend}

The premises of both rules require a constant (positive) lower bound on the Lie derivative $\lied[]{\genDE{x}}{\ptermA}$ which ensures that the value of $\ptermA$ strictly increases along solutions to the ODE, eventually becoming non-negative.
Soundness of both rules therefore crucially requires that ODE solutions exist for sufficiently long for $\ptermA$ to become non-negative.
This is usually left as a soundness-critical side condition in liveness proof rules~\cite{DBLP:journals/logcom/Platzer10,DBLP:conf/fm/SogokonJ15}, but such a side condition is antithetical to approaches for minimizing the soundness-critical core in implementations~\cite{DBLP:journals/jar/Platzer17} because it requires checking the (semantic) condition that solutions exist for sufficient duration.
The conclusion of rule~\irref{dVcmpA} formalizes this side condition as an assumption while rule~\irref{dVcmp} uses global Lipschitz continuity of the ODEs to show it.
All subsequent proof rules can also be presented with sufficient duration assumptions like~\irref{dVcmpA} but these are omitted for brevity.

\begin{example}
\label{ex:exlinproof}

Rule~\irref{dVcmp} enables a liveness proof for the linear ODE $\exlinear$ as suggested by~\rref{fig:odeexamples}.
The proof is shown on the left below and visualized on the right.
The first monotonicity step~\irref{MdW} strengthens the postcondition to the inner blue circle $u^2+v^2 = \frac{1}{4}$ which is contained within the green goal region.
Next, since solutions satisfy $u^2+v^2=1$ initially (black circle), the~\irref{Prog} step expresses an intermediate value property: to show that the \emph{continuous} solution eventually reaches $u^2+v^2 = \frac{1}{4}$, it suffices to show that it eventually reaches $u^2+v^2 \leq \frac{1}{4}$ (see \rref{cor:tt}).
The postcondition is rearranged before~\irref{dVcmp} is used with $\constt{\varepsilon} \mnodefeq \frac{1}{2}$.
Its premise proves with~\irref{qear} because the Lie derivative of $\frac{1}{4} - (u^2+v^2)$ with respect to $\exlinear$ is $2(u^2+v^2)$, which is bounded below by $\frac{1}{2}$ with assumption $\frac{1}{4} - (u^2+v^2) < 0$.

\noindent
\begin{minipage}[b]{0.66\textwidth}
{\footnotesizeoff\renewcommand{\arraystretch}{1.2}%
\centering
\begin{sequentdeduction}[array]
  \linfer[MdW]{
  \linfer[Prog]{
  \linfer[]{
  \linfer[dVcmp]{
  \linfer[]{
  \linfer[qear]{
    \lclose
  }
    {\lsequent{\frac{1}{4} < u^2+v^2}{2(u^2+v^2) \geq \frac{1}{2}}}
  }
    {\lsequent{\frac{1}{4} - (u^2+v^2) < 0}{2(u^2+v^2) \geq \frac{1}{2}}}
  }
    {\lsequent{u^2+v^2=1}{\ddiamond{\exlinear}{\frac{1}{4} - (u^2+v^2) \geq 0}}}
  }
    {\lsequent{u^2+v^2=1}{\ddiamond{\exlinear}{u^2+v^2 \leq \frac{1}{4}}}}
  }
    {\lsequent{u^2+v^2=1}{\ddiamond{\exlinear}{u^2+v^2 = \frac{1}{4}}}}
  }
  {\lsequent{u^2+v^2=1}{\ddiamond{\exlinear}{\big(\frac{1}{4} \leq \lnorm{(u,v)} \leq \frac{1}{2}\big)}}}
\end{sequentdeduction}
~\\
}%
\end{minipage}\hfill
\begin{minipage}[b]{0.33\textwidth}
\includegraphics[width=1\textwidth]{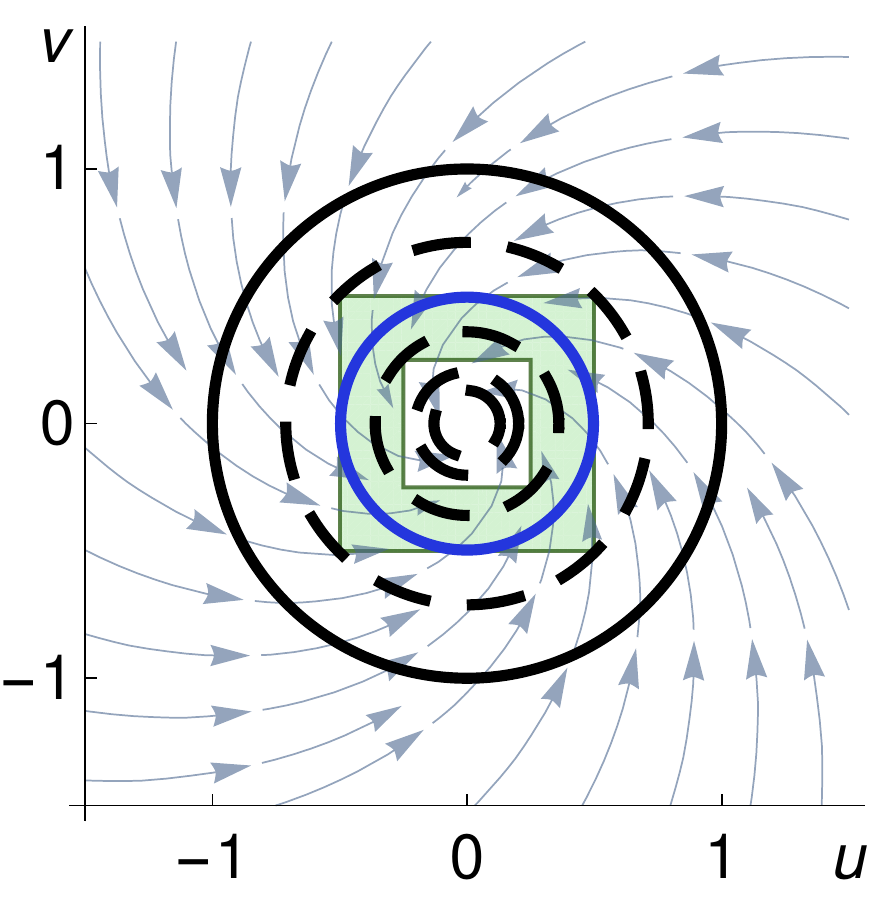}
\end{minipage}%

The Lie derivative calculation shows that the value of $u^2+v^2$ decreases along solutions of $\exlinear$, as visualized by the shrinking (dashed) circles.
However, the rate of shrinking converges to zero as solutions approach the origin, so solutions \emph{never} reach the origin in finite time!
This is why~\irref{dVcmpA+dVcmp} need a~\emph{constant} positive lower bound on the Lie derivative $\lied[]{\genDE{x}}{\ptermA}\geq \constt{\varepsilon}$ instead of merely requiring $\lied[]{\genDE{x}}{\ptermA} > 0$.

It is instructive to examine the chain of refinements~\rref{eq:refinementchain} underlying the proof.
The first~\irref{dVcmp} step refines the initial liveness property from~\irref{GEx}, i.e., that solutions exist globally (so, for at least $\frac{3}{4} \mathbin{/} \frac{1}{2} = \frac{3}{2}$ time), to the property $u^2+v^2 \leq \frac{1}{4}$.
Subsequent refinement steps can be read off from the proof steps above:
{\footnotesizeoff%
\begin{align*}
  \ddiamond{\pevolve{\exlinear,\D{\timevar}=1}}{\timevar {>} \frac{3}{2}}
  {\osetf{\irref{dVcmp}}{\limply}}
  \ddiamond{\exlinear}{u^2+v^2 {\leq} \frac{1}{4}}
  {\osetf{\irref{Prog}}{\limply}}
  \ddiamond{\exlinear}{u^2+v^2 {=} \frac{1}{4}}
  {\osetf{\irref{MdW}}{\limply}}
  \ddiamond{\exlinear}{\big(\frac{1}{4} {\leq} \lnorm{(u,v)} {\leq} \frac{1}{2}\big)}
\end{align*}
}%
\end{example}

The latter two steps illustrate the idea behind the next two surveyed proof rules.
In the original presentation~\cite{DBLP:conf/emsoft/TalyT10}, the ODE $\D{x}=\genDE{x}$ is only assumed to be \highlight{locally Lipschitz continuous}, which is insufficient for global existence of solutions, making the original rules unsound. See~\thereport{\rref{app:counterexamples}} for counterexamples.

\begin{corollary}[Equational differential variants~\cite{DBLP:conf/emsoft/TalyT10}]
\label{cor:tt}
The following proof rules are derivable in \dL.
Term $\constt{\varepsilon}$ is constant for ODE $\D{x}=\genDE{x}$ and the ODE is \highlight{globally Lipschitz continuous} for both rules.\\
{\renewcommand{\linferenceRuleNameSeparation}{\hspace{2pt}}%
\begin{calculuscollection}
\begin{calculus}
\dinferenceRule[dVeq|dV$_=$]{}
{\linferenceRule
  { \lsequent{\ptermA < 0}{\lied[]{\genDE{x}}{\ptermA}\geq \constt{\varepsilon}}
  }
  {\lsequent{\Gamma,\constt{\varepsilon} > 0, \ptermA \leq 0}{\ddiamond{\pevolve{\D{x}=\genDE{x}}}{\ptermA = 0}} }
}{}
\end{calculus}
\iflongversion
\qquad
\else
\;
\fi
\begin{calculus}
\dinferenceRule[TT|dV${_{=}^M}$]{}
{\linferenceRule
  { \lsequent{\ptermA = 0}{\rfvar} \quad
    \lsequent{\ptermA < 0}{\lied[]{\genDE{x}}{\ptermA}\geq \constt{\varepsilon}}
  }
  {\lsequent{\Gamma,\constt{\varepsilon} > 0, \ptermA \leq 0}{\ddiamond{\pevolve{\D{x}=\genDE{x}}}{\rfvar}} }
}{}
\end{calculus}
\end{calculuscollection}}%
\end{corollary}
\begin{proofatend}
Rule \irref{TT} derives directly from \irref{dVeq} with a~\irref{MdW} monotonicity step:
{\footnotesizeoff%
\begin{sequentdeduction}[array]
  \linfer[MdW]{
  \lsequent{\ptermA=0}{\rfvar}
  !
  \linfer[dVeq]{
    \lsequent{\ptermA < 0}{\lied[]{\genDE{x}}{\ptermA}\geq \constt{\varepsilon}}
  }
  {\lsequent{\Gamma,\constt{\varepsilon} > 0, \ptermA \leq 0}{\ddiamond{\pevolve{\D{x}=\genDE{x}}}{\ptermA = 0}}}
  }
  {\lsequent{\Gamma,\constt{\varepsilon} > 0, \ptermA \leq 0}{\ddiamond{\pevolve{\D{x}=\genDE{x}}}{\rfvar}}}
\end{sequentdeduction}
}%

The derivation of rule~\irref{dVeq} starts using axiom~\irref{Prog} with $\rgvar \mnodefequiv \ptermA \geq 0$ and rule \irref{dVcmp} (with $\cmp$ being $\geq$) on the resulting right premise, which yields the sole premise of~\irref{dVeq} (on the right):
{\footnotesizeoff%
\begin{sequentdeduction}[array]
  \linfer[Prog]{
  \lsequent{\ptermA \leq 0}{\dbox{\pevolvein{\D{x}=\genDE{x}}{\ptermA \neq 0}}{\ptermA< 0}}
   !
  \linfer[dVcmp]{
     \lsequent{\ptermA < 0}{\lied[]{\genDE{x}}{\ptermA}\geq \constt{\varepsilon}}
  }
    {\lsequent{\Gamma,\constt{\varepsilon} > 0}{\ddiamond{\pevolve{\D{x}=\genDE{x}}}{\ptermA \geq 0}}}
  }
  {\lsequent{\Gamma,\constt{\varepsilon} > 0, \ptermA \leq 0}{\ddiamond{\pevolve{\D{x}=\genDE{x}}}{\ptermA = 0}}}
\end{sequentdeduction}
}%

From the left premise after using~\irref{Prog}, axiom~\irref{DX} allows the domain constraint to be assumed true initially, which strengthens the assumption $\ptermA \leq 0$ to $\ptermA < 0$.
Rule~\irref{BC} completes the proof because the antecedents $\ptermA \neq 0, \ptermA = 0$ in its resulting premise are contradictory:
{\footnotesizeoff%
\begin{sequentdeduction}[array]
  \linfer[DX]{
  \linfer[BC]{
  \linfer[qear]{
    \lclose
  }
    {\lsequent{\ptermA \neq 0, \ptermA = 0}{\lied[]{\genDE{x}}{\ptermA} < 0}}
  }
    {\lsequent{\ptermA < 0}{\dbox{\pevolvein{\D{x}=\genDE{x}}{\ptermA \neq 0}}{\ptermA< 0}}}
  }
  {\lsequent{\ptermA \leq 0}{\dbox{\pevolvein{\D{x}=\genDE{x}}{\ptermA \neq 0}}{\ptermA< 0}}}
\\[-\normalbaselineskip]\tag*{\qedhere}
\end{sequentdeduction}
}%
\end{proofatend}

The view of~\irref{dVcmp} as a refinement of~\irref{GEx} immediately yields generalizations to higher Lie derivatives.
For example, it suffices that \emph{any} higher Lie derivative $\lied[k]{\genDE{x}}{\ptermA}$ is bounded below by a positive constant rather than just the first:
\begin{corollary}[Atomic higher differential variants]
\label{cor:higherdv}
The following proof rule (where $\cmp$ is either $\geq$ or $>$) is derivable in \dL.
Term $\constt{\varepsilon}$ is constant for ODE $\D{x}=\genDE{x}$ and the ODE is globally Lipschitz continuous.\\
\begin{calculuscollection}
\begin{calculus}
\dinferenceRule[dVcmpK|dV$_\cmp^k$]{}
{\linferenceRule
  { \lsequent{\lnot{(\ptermA \cmp 0)}}{\lied[k]{\genDE{x}}{\ptermA}\geq \constt{\varepsilon}}
  }
  {\lsequent{\Gamma, \constt{\varepsilon} > 0}{\ddiamond{\pevolve{\D{x}=\genDE{x}}}{\ptermA \cmp 0}} }
}{}
\end{calculus}
\end{calculuscollection}
\end{corollary}
\begin{proofsketch}[app:proofs]
Since $\lied[k]{\genDE{x}}{\ptermA}$ is strictly positive, the (lower) Lie derivatives of $\ptermA$ all eventually become positive.
This derives using a sequence of~\irref{dC+dIcmp} steps.
\end{proofsketch}
\begin{proofatend}
Rule~\irref{dVcmpK} can be derived in several ways.
For example, because $\lied[k]{\genDE{x}}{\ptermA}$ is strictly positive, we can prove that the solution successively reaches states where $\lied[k-1]{\genDE{x}}{\ptermA}$ is strictly positive, followed by $\lied[k-2]{\genDE{x}}{\ptermA}$ and so on.
The following derivation shows how \irref{dC} can be elegantly used for this argument.
The idea is to extend the derivation of rule~\irref{dVcmp} to higher Lie derivatives by (symbolically) integrating with respect to the time variable $\timevar$ using the following sequence of inequalities, where $\lied[i]{\genDE{x}}{\ptermA}_0$ is a symbolic constant that represents the initial value of the $i$-th Lie derivative of $\ptermA$ along $\D{x}=\genDE{x}$ for $i = 0,1,\dots,k-1$:
\begin{align}
\label{eq:integration}
\lied[k]{\genDE{x}}{\ptermA} &\geq \constt{\varepsilon} \nonumber \\
\lied[k-1]{\genDE{x}}{\ptermA} &\geq \lied[k-1]{\genDE{x}}{\ptermA}_0 + \constt{\varepsilon}\timevar \nonumber \\
\lied[k-2]{\genDE{x}}{\ptermA} &\geq \lied[k-2]{\genDE{x}}{\ptermA}_0 + \lied[k-1]{\genDE{x}}{\ptermA}_0\timevar + \constt{\varepsilon}\frac{\timevar^2}{2} \nonumber \\
&~\vdots \\
\lied[1]{\genDE{x}}{\ptermA} &\geq \lied[1]{\genDE{x}}{\ptermA}_0 + \dots + \lied[k-1]{\genDE{x}}{\ptermA}_0 \frac{\timevar^{k-2}}{(k-2)!} + \constt{\varepsilon} \frac{\timevar^{k-1}}{(k-1)!}  \nonumber \\
\ptermA &\geq \underbrace{\ptermA_0 + \lied[1]{\genDE{x}}{\ptermA}_0\timevar + \dots + \lied[k-1]{\genDE{x}}{\ptermA}_0 \frac{\timevar^{k-1}}{(k-1)!} + \constt{\varepsilon} \frac{\timevar^k}{k!}}_{\ptermB(\timevar)} \nonumber
\end{align}

The RHS of the final inequality in~\rref{eq:integration} is a polynomial $\ptermB(\timevar)$ in $\timevar$ which is positive for sufficiently large values of $\timevar$ because its leading coefficient $\constt{\varepsilon}$ is assumed to be strictly positive.
That is, with the assumption $\constt{\varepsilon} > 0$, the formula $\lexists{\timevar_1}{\lforall{\timevar > \timevar_1} {\ptermB(\timevar) > 0}}$ is provable in real arithmetic.

The derivation of~\irref{dVcmpK} starts by introducing fresh ghost variables that remember the initial values of $\ptermA$ and the (higher) Lie derivatives $\lied[1]{\genDE{x}}{\ptermA}, \dots, \lied[k-1]{\genDE{x}}{\ptermA}$ using~\irref{cut+qear+existsl}.
The resulting antecedents are abbreviated with $\Gamma_0 \mnodefequiv \big(\Gamma, \ptermA = \ptermA_0, \dots, \lied[k-1]{\genDE{x}}{\ptermA} = \lied[k-1]{\genDE{x}}{\ptermA}_0\big)$.
It also uses \irref{dGt} to introduce a fresh time variable $\timevar$ into the system.
Finally, the arithmetic fact that $\ptermB(\timevar)$ is eventually positive is introduced with~\irref{cut+qear+existsl}.
{\footnotesizeoff%
\begin{sequentdeduction}[array]
  \linfer[cut+qear+existsl]{
  \linfer[dGt]{
  \linfer[cut+qear+existsl]{
    \lsequent{\Gamma_0, \timevar=0, \lforall{\timevar > \timevar_1}{\ptermB(\timevar) > 0}}{\ddiamond{\pevolve{\D{x}=\genDE{x},\D{\timevar}=1}}{\ptermA \cmp 0}}
  }
    {\lsequent{\Gamma_0, \constt{\varepsilon} > 0, \timevar=0}{\ddiamond{\pevolve{\D{x}=\genDE{x},\D{\timevar}=1}}{\ptermA \cmp 0}}}
  }
    {\lsequent{\Gamma, \constt{\varepsilon} > 0, \ptermA=\ptermA_0, \dots, \lied[k-1]{\genDE{x}}{\ptermA} = \lied[k-1]{\genDE{x}}{\ptermA}_0}{\ddiamond{\pevolve{\D{x}=\genDE{x}}}{\ptermA \cmp 0}}}
  }
  {\lsequent{\Gamma, \constt{\varepsilon} > 0}{\ddiamond{\pevolve{\D{x}=\genDE{x}}}{\ptermA \cmp 0}} }
\end{sequentdeduction}
}%

Next, an initial liveness assumption, $\ddiamond{\pevolve{\D{x}=\genDE{x},\D{\timevar}=1}}{\ptermB(t) > 0}$, is cut into the assumptions.
Like the derivation of rule~\irref{dVcmp}, this initial liveness assumption says that the solution exists for sufficiently long for $p$ to become positive using the lower bound from~\rref{eq:integration}.
The cut premise is abbreviated \textcircled{1} and proved below.
The derivation continues from the remaining (unabbreviated) premise using~\irref{Prog}, with $\rgvar \mnodefequiv \ptermB(\timevar) > 0$:
{\footnotesizeoff%
\begin{sequentdeduction}[array]
  \linfer[cut]{
  \linfer[Prog]{
    \lsequent{\Gamma_0, \timevar=0}{\dbox{\pevolvein{\D{x}=\genDE{x},\D{\timevar}=1}{\lnot{(\ptermA \cmp 0)}}}{\ptermB(t) \leq 0}}
  }
    {\lsequent{\Gamma_0, \timevar=0,\ddiamond{\pevolve{\D{x}=\genDE{x},\D{\timevar}=1}}{\ptermB(t) > 0}}{\ddiamond{\pevolve{\D{x}=\genDE{x},\D{\timevar}=1}}{\ptermA \cmp 0}}} \quad
    \textcircled{1}
  }
  {\lsequent{\Gamma_0, \timevar=0, \lforall{\timevar > \timevar_1}{\ptermB(\timevar) > 0}}{\ddiamond{\pevolve{\D{x}=\genDE{x},\D{\timevar}=1}}{\ptermA \cmp 0}}}
\end{sequentdeduction}
}%

From the resulting open premise after~\irref{Prog}, monotonicity \irref{MbW} strengthens the postcondition to $\ptermA \geq \ptermB(t)$ using the domain constraint $\lnot{(\ptermA \cmp 0)}$.
Notice that the resulting postcondition $\ptermA \geq \ptermB(\timevar)$ is the final inequality from the sequence of inequalities~\rref{eq:integration}:
{\footnotesizeoff%
\begin{sequentdeduction}[array]
  \linfer[MbW]{
    \lsequent{\Gamma_0, \timevar=0}{\dbox{\pevolvein{\D{x}=\genDE{x},\D{\timevar}=1}{\lnot{(\ptermA \cmp 0)}}}{\ptermA \geq \ptermB(\timevar)}}
  }
  {\lsequent{\Gamma_0, \timevar=0}{\dbox{\pevolvein{\D{x}=\genDE{x},\D{\timevar}=1}{\lnot{(\ptermA \cmp 0)}}}{\ptermB(\timevar) \leq 0}}}
\end{sequentdeduction}
}%

The derivation continues by using~\irref{dC} to sequentially cut in the inequality bounds outlined in~\rref{eq:integration}.
The first differential cut~\irref{dC} step adds $\lied[k-1]{\genDE{x}}{\ptermA} \geq \lied[k-1]{\genDE{x}}{\ptermA}_0 + \constt{\varepsilon}\timevar $ to the domain constraint.
The proof of this cut yields the premise of~\irref{dVcmpK} after a~\irref{dIcmp} step, see the derivation labeled \textcircled{$\star$} immediately below:
{\footnotesizeoff%
\begin{sequentdeduction}
  \linfer[dC]{
    \lsequent{\Gamma_0, \timevar=0}{\dbox{\pevolvein{\D{x}=\genDE{x},\D{\timevar}=1}{\lnot{(\ptermA \cmp 0)} \land \lied[k-1]{\genDE{x}}{\ptermA} \geq \lied[k-1]{\genDE{x}}{\ptermA}_0 + \constt{\varepsilon}\timevar }}{\ptermA \geq \ptermB(\timevar)}}
    \quad \textcircled{$\star$}
  }
  {\lsequent{\Gamma_0, \timevar=0}{\dbox{\pevolvein{\D{x}=\genDE{x},\D{\timevar}=1}{\lnot{(\ptermA \cmp 0)}}}{\ptermA \geq \ptermB(\timevar)}}}
\end{sequentdeduction}
}%
From~\textcircled{$\star$}:
{\footnotesizeoff%
\begin{sequentdeduction}[array]
  \linfer[dIcmp]{
    \lsequent{\lnot{(\ptermA \cmp 0)}}{\lied[k]{\genDE{x}}{\ptermA} \geq \constt{\varepsilon} }
  }
  {\lsequent{\Gamma_0, \timevar=0}{\dbox{\pevolvein{\D{x}=\genDE{x},\D{\timevar}=1}{\lnot{(\ptermA \cmp 0)}}}{\lied[k-1]{\genDE{x}}{\ptermA} \geq \lied[k-1]{\genDE{x}}{\ptermA}_0 + \constt{\varepsilon}\timevar}}}
\end{sequentdeduction}
}%

Subsequent~\irref{dC+dIcmp} step progressively add the inequality bounds from~\rref{eq:integration} to the domain constraint until the last step where the postcondition is proved invariant with~\irref{dIcmp}:
{\footnotesizeoff\renewcommand{\arraystretch}{1.4}%
\begin{sequentdeduction}[array]
    \linfer[dC+dIcmp]{
    \linfer[dC+dIcmp]{
    \linfer[dC+dIcmp]{
    \linfer[dIcmp]{
      \lclose
    }
      {\lsequent{\Gamma_0, \timevar=0}{\dbox{\pevolvein{\D{x}=\genDE{x},\D{\timevar}=1}{\dots \land \lied[1]{\genDE{x}}{\ptermA} \geq \lied[1]{\genDE{x}}{\ptermA}_0 + \dots + \constt{\varepsilon} \frac{\timevar^{k-1}}{(k-1)!} }}{\ptermA \geq \ptermB(\timevar)}}}
    }
      {\vdots}
    }
    {\lsequent{\Gamma_0, \timevar=0}{\dbox{\pevolvein{\D{x}=\genDE{x},\D{\timevar}=1}{\dots \land \lied[k-2]{\genDE{x}}{\ptermA} \geq \lied[k-2]{\genDE{x}}{\ptermA}_0 + \lied[k-1]{\genDE{x}}{\ptermA}_0\timevar + \constt{\varepsilon}\frac{\timevar^2}{2}}}{\ptermA \geq \ptermB(\timevar)}}}
    }
    {\lsequent{\Gamma_0, \timevar=0}{\dbox{\pevolvein{\D{x}=\genDE{x},\D{\timevar}=1}{\lnot{(\ptermA \cmp 0)} \land \lied[k-1]{\genDE{x}}{\ptermA} \geq \lied[k-1]{\genDE{x}}{\ptermA}_0 + \constt{\varepsilon}\timevar }}{\ptermA \geq \ptermB(\timevar)}}}
\end{sequentdeduction}
}%

From premise~\textcircled{1}, a monotonicity step~\irref{MdW} rephrases the postcondition of the cut using the (constant) assumption $\lforall{\timevar > \timevar_1}{\ptermB(\timevar) > 0}$.
Axiom~\irref{GEx} finishes the derivation because the ODE $\D{x}=\genDE{x}$ is assumed to be globally Lipschitz continuous.
{\footnotesizeoff%
\begin{sequentdeduction}[array]
  \linfer[MdW]{
  \linfer[GEx]{
    \lclose
  }
    {\lsequent{}{\ddiamond{\pevolve{\D{x}=\genDE{x},\D{\timevar}=1}}{ \timevar > \timevar_1}}}
  }
  {\lsequent{\lforall{\timevar > \timevar_1}{\ptermB(\timevar) > 0}}{\ddiamond{\pevolve{\D{x}=\genDE{x},\D{\timevar}=1}}{\ptermB(\timevar) > 0}}}
\\[-\normalbaselineskip]\tag*{\qedhere}
\end{sequentdeduction}
}%
\end{proofatend}

\subsection{Staging Sets}
The idea behind \emph{staging sets}~\cite{DBLP:conf/fm/SogokonJ15} is to use an intermediary staging set formula $\rsfvar$ that \emph{can only be left by entering the goal region} $\rfvar$.
This staging property is expressed by the box modality formula $\dbox{\pevolvein{\D{x}=\genDE{x}}{\lnot{\rfvar}}}{\rsfvar}$ and is formally justified as a refinement using axiom~\irref{Prog} with $\rgvar \mnodefequiv \lnot{\rsfvar}$.

\begin{corollary}[Staging sets~\cite{DBLP:conf/fm/SogokonJ15}]
\label{cor:SP}
The following proof rule is derivable in \dL.
Term $\constt{\varepsilon}$ is constant for ODE $\D{x}=\genDE{x}$, which is globally Lipschitz continuous.\\
\begin{calculuscollection}
\begin{calculus}
\dinferenceRule[SP|SP]{}
{
\linferenceRule
  { \lsequent{\Gamma}{\dbox{\pevolvein{\D{x}=\genDE{x}}{\lnot{\rfvar}}}{\rsfvar}} \quad
    \lsequent{\rsfvar}{p \leq 0 \land \lied[]{\genDE{x}}{p}\geq \constt{\varepsilon}}
  }
  {\lsequent{\Gamma,\constt{\varepsilon}>0}{\ddiamond{\pevolve{\D{x}=\genDE{x}}}{\rfvar}} }
}{}
\end{calculus}
\end{calculuscollection}
\end{corollary}
\begin{proofatend}
The derivation of rule~\irref{SP} begins by using axiom~\irref{Prog} with $\rgvar \mnodefequiv \lnot{\rsfvar}$.
The resulting left premise is the left premise of~\irref{SP}, which is the staging property of the formula $\rsfvar$ that solutions can only leave $\rsfvar$ by entering $\rfvar$:
{\footnotesizeoff%
\begin{sequentdeduction}[array]
\linfer[Prog]{
  \lsequent{\Gamma}{\dbox{\pevolvein{\D{x}=\genDE{x}}{\lnot{\rfvar}}}{\rsfvar}} !
  \lsequent{\Gamma,\constt{\varepsilon}>0}{\ddiamond{\pevolve{\D{x}=\genDE{x}}}{\lnot{\rsfvar}}}
}
{\lsequent{\Gamma,\constt{\varepsilon}>0}{\ddiamond{\pevolve{\D{x}=\genDE{x}}}{\rfvar}}}
\end{sequentdeduction}
}%

The derivation continues on the right premise, similarly to~\irref{dVcmp}, by introducing fresh variables $\ptermA_0, i$ representing the initial value of $\ptermA$ and the multiplicative inverse of $\varepsilon()$ respectively using arithmetic cuts (\irref{cut+qear}).
It then uses \irref{dGt} to introduce a fresh time variable:
{\footnotesizeoff%
\begin{sequentdeduction}[array]
  \linfer[cut+qear]{
  \linfer[existsl]{
  \linfer[dGt]{
    \lsequent{\Gamma, \constt{\varepsilon} > 0, \ptermA=\ptermA_0, i\varepsilon() = 1, \timevar=0}{\ddiamond{\pevolve{\D{x}=\genDE{x},\D{\timevar}=1}}{\lnot{\rsfvar}}}
  }
    {\lsequent{\Gamma, \constt{\varepsilon} > 0, \ptermA=\ptermA_0, i\varepsilon() = 1}{\ddiamond{\pevolve{\D{x}=\genDE{x}}}{\lnot{\rsfvar}}}}
  }
  {\lsequent{\Gamma, \constt{\varepsilon} > 0, \lexists{\ptermA_0}{\ptermA=\ptermA_0}, \lexists{i}{i\varepsilon() = 1}}{\ddiamond{\pevolve{\D{x}=\genDE{x}}}{\lnot{\rsfvar}}}}
  }
  {\lsequent{\Gamma, \constt{\varepsilon} > 0}{\ddiamond{\pevolve{\D{x}=\genDE{x}}}{\lnot{\rsfvar}}} }
\end{sequentdeduction}
}%

The next cut introduces an initial liveness assumption (cut premise abbreviated \textcircled{1}).
The premise~\textcircled{1} is proved identically to the correspondingly abbreviated premise from the derivation of~\irref{dVcmp} using~\irref{GEx} because the ODE $\D{x}=\genDE{x}$ is assumed to be globally Lipschitz continuous.
{\footnotesizeoff%
\begin{sequentdeduction}[array]
  \linfer[cut]{
    \lsequent{\Gamma, \ptermA = \ptermA_0, \timevar = 0, \ddiamond{\pevolve{\D{x}=\genDE{x},\D{\timevar}=1}}{\ptermA_0 + \constt{\varepsilon}\timevar > 0} }{\ddiamond{\pevolve{\D{x}=\genDE{x}, \D{\timevar}=1}}{\lnot{\rsfvar}}} \quad\quad \textcircled{1}
  }
  {\lsequent{\Gamma, \constt{\varepsilon} > 0, \ptermA = \ptermA_0, i > 0, i\varepsilon() = 1, \timevar = 0}{\ddiamond{\pevolve{\D{x}=\genDE{x},\D{\timevar}=1}}{\lnot{\rsfvar}}}}
\end{sequentdeduction}
}%
From the remaining open premise, axiom \irref{Prog} is used with $\rgvar \mnodefequiv \ptermA_0 + \constt{\varepsilon} \timevar > 0$:
{\footnotesizeoff%
\begin{sequentdeduction}%
  \linfer[Prog]{
    \lsequent{\Gamma,\ptermA = \ptermA_0, \timevar = 0}{\dbox{\pevolvein{\D{x}=\genDE{x},\D{\timevar}=1}{\rsfvar}}{\ptermA_0 + \constt{\varepsilon}\timevar \leq 0}}
  }
  {\lsequent{\Gamma, \ptermA = \ptermA_0, \timevar = 0, \ddiamond{\pevolve{\D{x}=\genDE{x},\D{\timevar}=1}}{\ptermA_0 + \constt{\varepsilon}\timevar > 0} }{\ddiamond{\pevolve{\D{x}=\genDE{x},\D{\timevar}=1}}{\lnot{\rsfvar}}}}
\end{sequentdeduction}
}%

Finally, a monotonicity step~\irref{MbW} simplifies the postcondition using domain constraint $\rsfvar$, yielding the left conjunct of the right premise of rule~\irref{SP}.
The right premise after monotonicity is abbreviated~\textcircled{2} and continued below.
{\footnotesizeoff%
\begin{sequentdeduction}[array]
  \linfer[MbW]{
  \linfer[qear]{
    \lsequent{\rsfvar}{\ptermA \leq 0}
  }
    {\lsequent{\rsfvar,\ptermA \geq \ptermA_0 + \constt{\varepsilon}\timevar}{\ptermA_0 + \constt{\varepsilon}\timevar \leq 0}} !
    \textcircled{2}
    }
    {\lsequent{\Gamma,\ptermA = \ptermA_0,\timevar = 0}{\dbox{\pevolvein{\D{x}=\genDE{x},\D{\timevar}=1}{\rsfvar}}{\ptermA_0 + \constt{\varepsilon}\timevar \leq 0}}}
\end{sequentdeduction}
}%

From~\textcircled{2}, rule~\irref{dIcmp} yields the right conjunct of the right premise of rule~\irref{SP}.

{\footnotesizeoff%
\begin{sequentdeduction}[array]
\linfer[dIcmp]{
    \lsequent{\rsfvar}{\lied[]{\genDE{x}}{\ptermA}\geq \constt{\varepsilon}}
    }
    {\lsequent{\Gamma,\ptermA = \ptermA_0,\timevar = 0}{\dbox{\pevolvein{\D{x}=\genDE{x},\D{\timevar}=1}{\rsfvar}}{\ptermA \geq \ptermA_0 + \constt{\varepsilon}\timevar}}}
\\[-\normalbaselineskip]\tag*{\qedhere}
\end{sequentdeduction}
}%
\end{proofatend}

In rule~\irref{SP}, the staging set formula $\rsfvar$ provides a choice of intermediary between the differential variant $\ptermA$ and the desired postcondition $\rfvar$.
Proof rules can be significantly simplified by choosing $\rsfvar$ with desirable topological properties.
All proof rules derived so far either have an explicit sufficient duration assumption (like \irref{dVcmpA}) or use axiom~\irref{GEx} by assuming that ODEs are globally Lipschitz.
To make use of axiom~\irref{BEx}, an alternative is to choose staging set formulas $\rsfvar(x)$ that characterize a bounded (or even compact) set over the variables $x$.

\begin{corollary}[Bounded/compact staging sets]
\label{cor:boundedandcompact}
The following proof rules are derivable in \dL.
Term $\constt{\varepsilon}$ is constant for $\D{x}=\genDE{x}$.
In rule \irref{SPb}, formula $\rsfvar$ characterizes a bounded set over variables $x$.
In rule \irref{SPc}, it characterizes a compact, i.e., closed and bounded, set over those variables.\\
{\footnotesizeoff%
\renewcommand{\linferenceRuleNameSeparation}{\hspace{2pt}}%
\begin{calculuscollection}
\begin{calculus}
\dinferenceRule[SPb|SP$_b$]{}
{
\linferenceRule
  { \lsequent{\Gamma}{\dbox{\pevolvein{\D{x}=\genDE{x}}{\lnot{\rfvar}}}{\rsfvar}} \;\,
    \lsequent{\rsfvar}{\lied[]{\genDE{x}}{p} \geq \constt{\varepsilon}}
  }
  {\lsequent{\Gamma,\constt{\varepsilon} > 0}{\ddiamond{\pevolve{\D{x}=\genDE{x}}}{\rfvar}} }
}{}
\end{calculus}
\iflongversion
\qquad
\else
\;\;
\fi
\begin{calculus}
\dinferenceRule[SPc|SP$_c$]{}
{
\linferenceRule
  { \lsequent{\Gamma}{\dbox{\pevolvein{\D{x}=\genDE{x}}{\lnot{\rfvar}}}{\rsfvar}} \;\,
    \lsequent{\rsfvar}{\lied[]{\genDE{x}}{p} > 0}
  }
  {\lsequent{\Gamma}{\ddiamond{\pevolve{\D{x}=\genDE{x}}}{\rfvar}} }
}{}
\end{calculus}
\end{calculuscollection}
}%
\end{corollary}
\begin{proofsketch}[app:proofs]
Rule~\irref{SPb} derives using~\irref{BEx} and differential variant $\ptermA$ to establish a time bound.
Rule~\irref{SPc} is an arithmetical corollary of~\irref{SPb}, using the fact that continuous functions on a compact domain attain their extrema.
\end{proofsketch}
\begin{proofatend}
Rule~\irref{SPb} is derived first since rule~\irref{SPc} follows as a corollary.
Both proof rules make use of the fact that continuous functions on compact domains attain their extrema~\cite[Theorem 4.16]{MR0385023}.
Polynomial functions are continuous so this fact can be stated and proved as a formula of first-order real arithmetic~\cite{Bochnak1998}.
The derivation of~\irref{SPb} is essentially similar to~\irref{SP} except replacing the use of global existence axiom~\irref{GEx} with the bounded existence axiom~\irref{BEx}.
It starts by using axiom~\irref{Prog} with $\rgvar \mnodefequiv \lnot{\rsfvar}$, yielding the left premise of~\irref{SPb}:
{\footnotesizeoff%
\begin{sequentdeduction}[array]
\linfer[Prog]{
  \lsequent{\Gamma}{\dbox{\pevolvein{\D{x}=\genDE{x}}{\lnot{\rfvar}}}{\rsfvar}} !
  \lsequent{\Gamma,\constt{\varepsilon}>0}{\ddiamond{\pevolve{\D{x}=\genDE{x}}}{\lnot{\rsfvar}}}
}
{\lsequent{\Gamma,\constt{\varepsilon}>0}{\ddiamond{\pevolve{\D{x}=\genDE{x}}}{\rfvar}}}
\end{sequentdeduction}
}%

Continuing on the resulting right from~\irref{Prog} (similarly to~\irref{SP}), the derivation introduces fresh variables $\ptermA_0, i$ representing the initial value of $\ptermA$ and the multiplicative inverse of $\varepsilon()$ respectively using arithmetic cuts and Skolemizing (\irref{cut+qear+existsl}). Rule \irref{dGt} introduces a fresh time variable:
{\footnotesizeoff%
\begin{sequentdeduction}[array]
  \linfer[cut+qear+existsl+dGt]{
    \lsequent{\Gamma, \constt{\varepsilon} > 0, \ptermA=\ptermA_0, i\varepsilon() = 1, \timevar=0}{\ddiamond{\pevolve{\D{x}=\genDE{x},\D{\timevar}=1}}{\lnot{\rsfvar}}}
  }
  {\lsequent{\Gamma, \constt{\varepsilon} > 0}{\ddiamond{\pevolve{\D{x}=\genDE{x}}}{\lnot{\rsfvar}}} }
\end{sequentdeduction}
}%

The set characterized by formula $\rsfvar$ is bounded so its closure is compact (with respect to variables $x$).
On this compact closure, the continuous polynomial function $\ptermA$ attains its maximum value, which implies that the value of $\ptermA$ is bounded above in $\rsfvar$ and cannot increase without bound while staying in $\rsfvar$.
That is, the formula $\lexists{\ptermA_1}{\rrfvar(\ptermA_1)}$ where $\rrfvar(\ptermA_1) \mnodefequiv \lforall{x}{(\rsfvar(x) \limply \ptermA \leq \ptermA_1)}$ is valid in first-order real arithmetic and thus provable by \irref{qear}.
This formula is added to the assumptions next and the existential quantifier is Skolemized with \irref{existsl}.
The resulting symbolic constant $\ptermA_1$ represents the upper bound of $\ptermA$ on $\rsfvar$.
Note that $\rrfvar(\ptermA_1)$ is constant for the ODE $\D{x}=\genDE{x},\D{\timevar}=1$ because it does not mention any of the variables $x$ (nor $\timevar$) free:
{\footnotesizeoff%
\begin{sequentdeduction}[array]
  \linfer[cut+qear]{
  \linfer[existsl]{
    \lsequent{\Gamma, \constt{\varepsilon} > 0, \ptermA=\ptermA_0, i\varepsilon() = 1, \timevar=0, \rrfvar(\ptermA_1)}{\ddiamond{\pevolve{\D{x}=\genDE{x},\D{\timevar}=1}}{\lnot{\rsfvar}}}
  }
    {\lsequent{\Gamma, \constt{\varepsilon} > 0, \ptermA=\ptermA_0, i\varepsilon() = 1, \timevar=0, \lexists{\ptermA_1}{\rrfvar(\ptermA_1)} }{\ddiamond{\pevolve{\D{x}=\genDE{x},\D{\timevar}=1}}{\lnot{\rsfvar}}}}
  }
  {\lsequent{\Gamma, \constt{\varepsilon} > 0, \ptermA=\ptermA_0, i\varepsilon() = 1, \timevar=0}{\ddiamond{\pevolve{\D{x}=\genDE{x},\D{\timevar}=1}}{\lnot{\rsfvar}}}}
\end{sequentdeduction}
}%

Next, a~\irref{cut} introduces an initial liveness assumption saying that sufficient time exists for $\ptermA$ to become greater than its upper bound $\ptermA_1$ on $\rsfvar$, which implies that the solution must leave $\rsfvar$.
This assumption is abbreviated $\rtfvar \mnodefequiv \ddiamond{\pevolve{\D{x}=\genDE{x},\D{\timevar}=1}}{(\lnot{\rsfvar} \lor \ptermA_0 + \constt{\varepsilon}\timevar > \ptermA_1)}$.
The main difference from~\irref{SP} is that assumption $\rtfvar$ also adds a disjunction for the possibility of leaving $\rsfvar$ (which characterizes a bounded set).
This cut premise is abbreviated \textcircled{1} and proved below.
{\footnotesizeoff%
\begin{sequentdeduction}[array]
  \linfer[cut]{
    \lsequent{\Gamma, \ptermA=\ptermA_0, \timevar=0, \rrfvar(\ptermA_1), \rtfvar}{\ddiamond{\pevolve{\D{x}=\genDE{x},\D{\timevar}=1}}{\lnot{\rsfvar}}} \quad
    \textcircled{1}
  }
  {\lsequent{\Gamma, \constt{\varepsilon} > 0, \ptermA=\ptermA_0, i\varepsilon() = 1, \timevar=0, \rrfvar(\ptermA_1)}{\ddiamond{\pevolve{\D{x}=\genDE{x},\D{\timevar}=1}}{\lnot{\rsfvar}}}}
\end{sequentdeduction}
}%
Continuing from the open premise on the left, axiom \irref{Prog} is used with $\rgvar \mnodefequiv \lnot{\rsfvar} \lor \ptermA_0 + \constt{\varepsilon}\timevar > \ptermA_1$:
{\footnotesizeoff%
\begin{sequentdeduction}[array]
  \linfer[Prog]{
    \lsequent{\Gamma,\ptermA = \ptermA_0, \timevar=0, \rrfvar(\ptermA_1)}{\dbox{\pevolvein{\D{x}=\genDE{x},\D{\timevar}=1}{\rsfvar}}{(\rsfvar \land \ptermA_0 + \constt{\varepsilon}\timevar \leq \ptermA_1)}}
  }
  {\lsequent{\Gamma, \ptermA=\ptermA_0, \timevar=0, \rrfvar(\ptermA_1), \rtfvar}{\ddiamond{\pevolve{\D{x}=\genDE{x},\D{\timevar}=1}}{\lnot{\rsfvar}}}}
\end{sequentdeduction}
}%

The postcondition of the resulting box modality is simplified with a \irref{MbW} monotonicity step.
This crucially uses the assumption $\rrfvar(\ptermA_1)$ which is constant for the ODE.
A \irref{dIcmp} step yields the remaining premise of~\irref{SPb} on the right, see the derivation labeled \textcircled{$\star$} immediately below:
{\footnotesizeoff%
\begin{sequentdeduction}[array]
\linfer[MbW]{
    \linfer[qear]{
      \linfer[qear]{
        \lclose
      }
      {\lsequent{\rsfvar,\rrfvar(\ptermA_1)}{\ptermA \leq \ptermA_1}}
    }
      {\lsequent{\rsfvar,\rrfvar(\ptermA_1),\ptermA \geq \ptermA_0 + \constt{\varepsilon}\timevar}{\rsfvar \land \ptermA_0 + \constt{\varepsilon}\timevar \leq \ptermA_1}}
    !
    \textcircled{$\star$}
  }
  {\lsequent{\Gamma,\ptermA = \ptermA_0, \timevar=0, \rrfvar(\ptermA_1)}{\dbox{\pevolvein{\D{x}=\genDE{x},\D{\timevar}=1}{\rsfvar}}{(\rsfvar \land \ptermA_0 + \constt{\varepsilon}\timevar \leq \ptermA_1)}}
}
\end{sequentdeduction}
}%
From \textcircled{$\star$}:
{\footnotesizeoff%
\begin{sequentdeduction}[array]
\linfer[dIcmp]{
    \lsequent{\rsfvar}{\lied[]{\genDE{x}}{\ptermA}\geq \constt{\varepsilon}}
    }
    {\lsequent{\Gamma,\ptermA = \ptermA_0,\timevar=0}{\dbox{\pevolvein{\D{x}=\genDE{x},\D{\timevar}=1}{\rsfvar}}{\ptermA \geq \ptermA_0 + \constt{\varepsilon}\timevar}}}
\end{sequentdeduction}
}%

From premise~\textcircled{1}, a monotonicity step~\irref{MdW} equivalently rephrases the postcondition of the cut.
Axiom~\irref{BEx} finishes the proof because formula $\rsfvar(x)$ is assumed to be bounded over variables $x$.
{\footnotesizeoff%
\begin{sequentdeduction}[array]
  \linfer[qear+MdW]{
  \linfer[BEx]{
    \lclose
  }
    {\lsequent{}{\ddiamond{\pevolve{\D{x}=\genDE{x},\D{\timevar}=1}}{ (\lnot{\rsfvar} \lor \timevar > i(\ptermA_1-\ptermA_0))}}}
  }
  {\lsequent{\constt{\varepsilon} > 0, i\varepsilon() = 1}{\rtfvar}}
\end{sequentdeduction}
}%

Next, to derive rule~\irref{SPc} from~\irref{SPb}, the compactness of the set characterized by $\rsfvar(x)$ implies that the (abbreviated) formula
$\lexists{\varepsilon {>} 0}{A(\varepsilon)}$ where $A(\varepsilon) \mnodefequiv \lforall{x}{(\rsfvar(x) \limply \lied[]{\genDE{x}}{p} \geq \varepsilon)}$ and the (abbreviated) formula
$B \mnodefequiv \lforall{x}{(\rsfvar(x) \limply \lied[]{\genDE{x}}{p} > 0)}$ are provably equivalent in first-order real arithmetic.
Briefly, this provable equivalence follows from the fact that the continuous polynomial function $\lied[]{\genDE{x}}{p}$ is bounded below by its minima on the compact set characterized by $\rsfvar(x)$ and this minima is strictly positive.
The following derivation of~\irref{SPc} threads these two formulas through the use of rule~\irref{SPb}.
After Skolemizing $\lexists{\varepsilon {>} 0}{A(\varepsilon)}$ with \irref{existsl}, the resulting formula $A(\varepsilon)$ is constant for the ODE $\D{x}=\genDE{x}$ so it is kept as a constant assumption across the use of~\irref{SPb}, leaving only the two premises of rule~\irref{SPc}:
{\footnotesizeoff%
\renewcommand{\linferPremissSeparation}{\hspace{5pt}}%
\begin{sequentdeduction}[array]
  \linfer[cut]{
  \linfer[existsl]{
  \linfer[SPb]{
    \lsequent{\Gamma}{\dbox{\pevolvein{\D{x}=\genDE{x}}{\lnot{\rfvar}}}{\rsfvar}} !
    \linfer[qear]{\lclose}
    {\lsequent{\rsfvar,A(\varepsilon)}{ \lied[]{\genDE{x}}{p} \geq \varepsilon}}
  }
    {\lsequent{\Gamma,\varepsilon > 0, A(\varepsilon)}{\ddiamond{\pevolve{\D{x}=\genDE{x}}}{\rfvar}}}
  }
  {\lsequent{\Gamma,\lexists{\varepsilon {>} 0}{A(\varepsilon)}}{\ddiamond{\pevolve{\D{x}=\genDE{x}}}{\rfvar}}}
  !
    \linfer[qear]{
    \linfer[allr+implyr]{
      \lsequent{\rsfvar}{\lied[]{\genDE{x}}{p} > 0}
    }
      \lsequent{}{B}
    }
    {\lsequent{}{\lexists{\varepsilon {>} 0}{A(\varepsilon)}}}
  }
  {\lsequent{\Gamma}{\ddiamond{\pevolve{\D{x}=\genDE{x}}}{\rfvar}} }
\\[-\normalbaselineskip]\tag*{\qedhere}
\end{sequentdeduction}
}%
\end{proofatend}

\begin{example}
\label{ex:exnonlinproof}
Liveness for the non-linear ODE $\exnonlinear$ (as suggested by~\rref{fig:odeexamples}) is proved using rule~\irref{SPc} by choosing the staging set formula $\rsfvar \mnodefequiv 1 \leq u^2+v^2 \leq 2$ (blue annulus) and the differential variant $p \mnodefeq u^2+v^2$.
The Lie derivative $\lied[]{\genDE{x}}{p}$ with respect to $\exnonlinear$ is $2(u^2+v^2)(u^2+v^2-\frac{1}{4})$, which is bounded below by $\frac{3}{2}$ in $\rsfvar$.
Thus, the right premise of~\irref{SPc} closes trivially.
The left premise (abbreviated~\textcircled{1}) requires proving that $\rsfvar$ is an invariant within the domain constraint $\lnot{(u^2+v^2\geq 2)}$.
Intuitively, this is true because the blue annulus can only be left by entering $u^2+v^2\geq 2$.
Its (elided) invariance proof is easy~\cite{DBLP:conf/lics/PlatzerT18}.

\noindent
\begin{minipage}[b]{0.66\textwidth}
{\footnotesizeoff\renewcommand{\arraystretch}{1.2}%
\centering
\begin{sequentdeduction}[array]
  \linfer[SPc]{
  \textcircled{1} !
  \linfer[qear]{
    \lclose
  }
  {
  \lsequent{\rsfvar}{\lied[]{\genDE{x}}{p} > 0}}
  }
  {\lsequent{u^2+v^2=1}{\ddiamond{\exnonlinear}{u^2 + v^2 \geq 2}}}
\end{sequentdeduction}

\begin{sequentdeduction}[array]
  \linfer[cut+qear]{
  \linfer[]{
    \lclose
  }
  {\lsequent{\rsfvar}{\dbox{\pevolvein{\exnonlinear}{\lnot{(u^2+v^2\geq 2)}}}{\rsfvar}}}
  }
  {\textcircled{1}:\quad \lsequent{u^2+v^2=1}{\dbox{\pevolvein{\exnonlinear}{\lnot{(u^2+v^2\geq 2)}}}{\rsfvar}}}
\end{sequentdeduction}
}%
~\\
\end{minipage}\hfill
\begin{minipage}[b]{0.33\textwidth}
\includegraphics[width=1\textwidth]{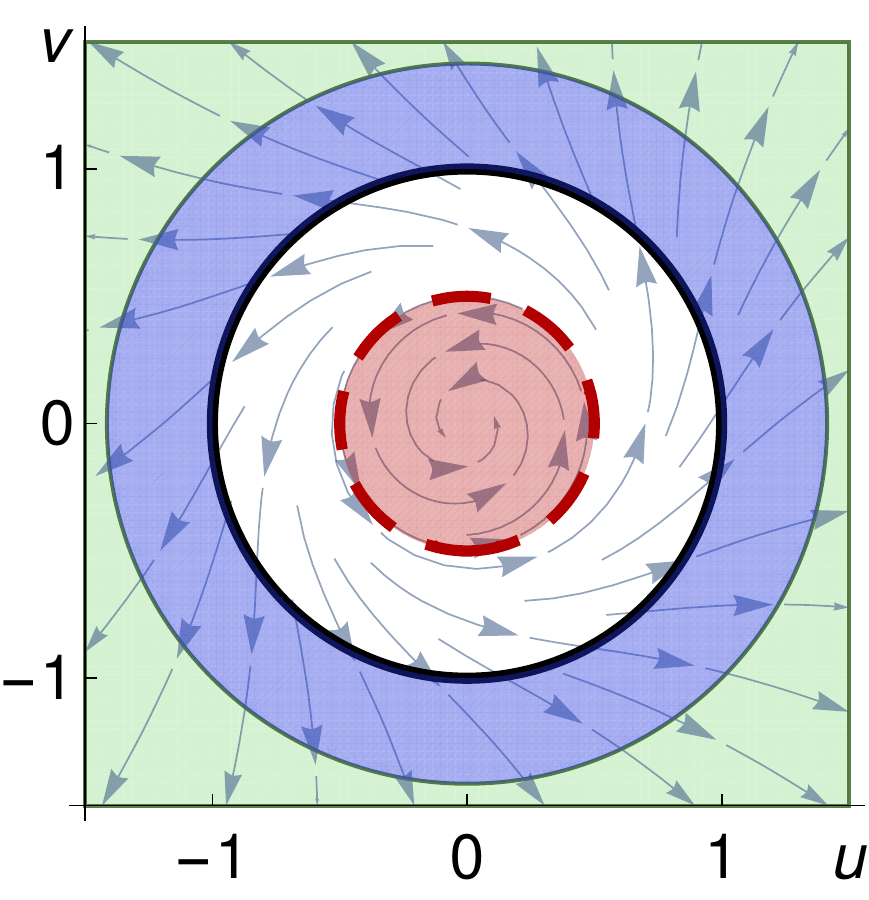}
\end{minipage}%

There are two subtleties to highlight in this proof.
First, $\rsfvar$ characterizes a compact, hence bounded, set (as required by rule~\irref{SPc}).
Solutions of $\exnonlinear$ can blow up in finite time which necessitates the use of~\irref{BEx} for proving its liveness properties.
Second, $\rsfvar$ is cleverly chosen to \emph{exclude} the red disk (dashed boundary) characterized by $u^2+v^2 \leq \frac{1}{4}$.
As mentioned earlier, solutions of $\exnonlinear$ behave differently in this region, e.g., the Lie derivative $\lied[]{\genDE{x}}{p}$ is \emph{non-positive} in this disk.
The chain of refinements~\rref{eq:refinementchain} behind this proof can be seen from the derivation of rules~\irref{SPb+SPc} in~\thereport{\rref{app:proofs}}.
It starts from the initial liveness property~\irref{BEx} (with time bound $1 \mathbin{/} \frac{3}{2} = \frac{2}{3}$) and uses two~\irref{Prog} refinement steps, first showing that the staging set is left ($\ddiamond{\pevolve{\exnonlinear}}{\lnot{\rsfvar}}
$), then showing the desired liveness property:
\[
  \ddiamond{\pevolve{\exnonlinear,\D{\timevar}=1}}{(\lnot{\rsfvar} \lor \timevar > \frac{2}{3})}
  {\osetf{\irref{Prog}}{\limply}}
  \ddiamond{\pevolve{\exnonlinear}}{\lnot{\rsfvar}}
  {\osetf{\irref{Prog}}{\limply}}
  \ddiamond{\pevolve{\exnonlinear}}{u^2 + v^2 \geq 2}
\]
\end{example}

The use of axiom \irref{BEx} is subtle and is sometimes overlooked in surveyed liveness arguments.
For example,~\cite[Remark 3.6]{DBLP:journals/siamco/PrajnaR07} incorrectly claims that their liveness argument works \highlight{without assuming that the relevant sets are bounded}.
The following proof rule derives from~\irref{SPc} and adapts ideas from~\cite[Theorem 2.4, Corollary 2.5]{DBLP:journals/siamco/RatschanS10}, but formula $K$ in the original presentation is only assumed to characterize a \highlight{closed rather than compact set}; the proofs (correctly) use the fact that the set is bounded but this assumption is not made explicit~\cite{DBLP:journals/siamco/RatschanS10}.

\begin{corollary}[Set Lyapunov functions~\cite{DBLP:journals/siamco/RatschanS10}]
\label{cor:rs}
The following proof rule is derivable in \dL.
Formula $K$ characterizes a \highlight{compact set} over variables $x$, while formula $\rfvar$ characterizes an open set over those variables.\\
\begin{calculuscollection}
\begin{calculus}
\dinferenceRule[RS|SLyap]{}
{
\linferenceRule
  {
     \lsequent{\ptermA \geq 0}{K}
     \quad
     \lsequent{\lnot{\rfvar},K}{\lied[]{\genDE{x}}{\ptermA} > 0}
  }
  {\lsequent{\Gamma, \ptermA \cmp 0}{\ddiamond{\pevolve{\D{x}=\genDE{x}}}{\rfvar}} }
}{}
\end{calculus}
\end{calculuscollection}
\end{corollary}
\begin{proofatend}
Rule \irref{RS} derives from \irref{SPc} with $\rsfvar \mnodefequiv \lnot{\rfvar} \land K$, since the intersection of a closed set with a compact set is compact.
The resulting right premise from using~\irref{SPc} is the right premise of~\irref{RS}:
{\footnotesizeoff%
\begin{sequentdeduction}[array]
\linfer[SPc]{
  \lsequent{\Gamma, \ptermA \cmp 0}{\dbox{\pevolvein{\D{x}=\genDE{x}}{\lnot{\rfvar}}}{(K \land \lnot{\rfvar})}} !
  \lsequent{\lnot{\rfvar},K}{\lied[]{\genDE{x}}{p} > 0}
}
{\lsequent{\Gamma, \ptermA \cmp 0}{\ddiamond{\pevolve{\D{x}=\genDE{x}}}{\rfvar}}}
\end{sequentdeduction}
}%

Continuing from the left premise, a monotonicity step with the premise $\lsequent{\ptermA \geq 0}{K}$ turns the postcondition to $\ptermA \cmp 0$.
Rule~\irref{BC} is used, which, along with the premise $\lsequent{\ptermA \geq 0}{K}$ results in the premise of rule~\irref{RS}:
{\footnotesizeoff%
\begin{sequentdeduction}[array]
\linfer[MbW]{
    \lsequent{\ptermA \geq 0}{K}
    !
    \linfer[BC]{
    \linfer[cut]{
      \lsequent{\lnot{\rfvar},K}{\lied[]{\genDE{x}}{\ptermA} > 0} !
      \linfer[qear]{
      \lsequent{\ptermA\geq 0}{K}
      }
      {\lsequent{\lnot{\rfvar},\ptermA=0}{K}}
    }
      {\lsequent{\lnot{\rfvar}, \ptermA = 0}{\lied[]{\genDE{x}}{\ptermA} > 0}}
    }
    {\lsequent{\ptermA \cmp 0}{\dbox{\pevolvein{\D{x}=\genDE{x}}{\lnot{\rfvar}}}{\ptermA \cmp 0}}}
  }
  {\lsequent{\Gamma, \ptermA \cmp 0}{\dbox{\pevolvein{\D{x}=\genDE{x}}{\lnot{\rfvar}}}{(K \land \lnot{\rfvar})}}}
\\[-\normalbaselineskip]\tag*{\qedhere}
\end{sequentdeduction}
}%
\end{proofatend}

\section{Liveness With Domain Constraints}
\label{sec:withdomconstraint}
This section presents proof rules for liveness properties $\pevolvein{\D{x}=\genDE{x}}{\ivr}$ with domain constraint $\ivr$.
Axiom~\irref{dDR} provides direct generalizations of the proof rules from~\rref{sec:nodomconstraint} with the following derivation choosing $\rrfvar \mnodefequiv \ltrue$:
{\footnotesizeoff%
\begin{sequentdeduction}[array]
\linfer[dDR]{
  \lsequent{\Gamma}{\dbox{\pevolve{\D{x}=\genDE{x}}}{\ivr}} !
  \lsequent{\Gamma}{\ddiamond{\pevolve{\D{x}=\genDE{x}}}{\rfvar}}
}
{\lsequent{\Gamma}{\ddiamond{\pevolvein{\D{x}=\genDE{x}}{\ivr}}{\rfvar}}}
\end{sequentdeduction}
}%
This extends all chains of refinements~\rref{eq:refinementchain} from~\rref{sec:nodomconstraint} with an additional step:
{\footnotesizeoff
\[
  \dots
  \limply
  \ddiamond{\pevolve{\D{x}=\genDE{x}}}{\rfvar}
  {\osetf{\irref{dDR}}{\limply}}
  \ddiamond{\pevolvein{\D{x}=\genDE{x}}{\ivr}}{\rfvar}
\]
}%

Liveness arguments become much more intricate when attempting to generalize beyond~\irref{dDR}, e.g., recall the unsound conjecture~\irref{badaxiom}.
Indeed, unlike the technical glitches of~\rref{sec:nodomconstraint}, our survey uncovers subtle soundness-critical errors here.
With our deductive approach, these intricacies are isolated to the topological axioms (\rref{lem:diatopaxioms}) which have been proved sound once and for all.
As before, errors and omissions in the surveyed techniques are \highlight{highlighted in blue}.

\subsection{Topological Proof Rules}

The first proof rule generalizes differential variants to handle domain constraints:

\begin{corollary}[Atomic differential variants with domains~\cite{DBLP:journals/logcom/Platzer10}]
\label{cor:atomicdvcmpQ}
The following proof rule (where $\cmp$ is either $\geq$ or $>$) is derivable in \dL.
Term $\constt{\varepsilon}$ is constant for the ODE $\D{x}=\genDE{x}$ and the ODE is globally Lipschitz continuous.
\highlight{Formula $\ivr$ characterizes a closed (resp. open) set when $\cmp$ is $\geq$ (resp. $>$).}\\
\begin{calculuscollection}
\begin{calculus}
\dinferenceRule[dVcmpQ|dV$_\cmp\&$]{}
{\linferenceRule
  {
    \lsequent{\Gamma}{\dbox{\pevolvein{\D{x}=\genDE{x}}{\lnot{(\ptermA \cmp 0)}}}{\ivr}} \quad
    \lsequent{\lnot{(\ptermA \cmp 0)}, \ivr}{\lied[]{\genDE{x}}{\ptermA}\geq \constt{\varepsilon}}
  }
  {\lsequent{\Gamma,\constt{\varepsilon} > 0,\highlight{\lnot{(\ptermA \cmp 0)}}}{\ddiamond{\pevolvein{\D{x}=\genDE{x}}{\ivr}}{p \cmp 0}} }
}{}
\end{calculus}
\end{calculuscollection}
\end{corollary}
\begin{proofsketch}[app:proofs]
The derivation uses axiom~\irref{CORef} choosing $\rrfvar \mnodefequiv \ltrue$, noting that $\ptermA \geq 0$ (resp. $\ptermA > 0$) characterizes a topologically closed (resp. open) set so the appropriate topological requirements of~\irref{CORef} are satisfied:
{\footnotesizeoff%
\begin{sequentdeduction}[array]
\linfer[CORef]{
  \lsequent{\Gamma}{\dbox{\pevolvein{\D{x}=\genDE{x}}{\lnot{(p \cmp 0)}}}{\ivr}} !
  \linfer[]{
  \linfer[]{
    \lsequent{\lnot{(\ptermA \cmp 0)}, \ivr}{\lied[]{\genDE{x}}{\ptermA}\geq \constt{\varepsilon}}
  }
    {\dots}
  }
  {\lsequent{\Gamma,\constt{\varepsilon} > 0}{\ddiamond{\pevolve{\D{x}=\genDE{x}}}{p\cmp 0}}}
}
{\lsequent{\Gamma,\constt{\varepsilon} > 0,\lnot{(\ptermA \cmp 0)}}{\ddiamond{\pevolvein{\D{x}=\genDE{x}}{\ivr}}{p \cmp 0}}}
\end{sequentdeduction}
}%
The right premise follows similarly to~\irref{dVcmp} although it uses an intervening~\irref{dC} step to add $\ivr$ to the antecedents.
\end{proofsketch}
\begin{proofatend}
The derivation uses axiom~\irref{CORef} choosing $\rrfvar \mnodefequiv \ltrue$, noting that $\ptermA \geq 0$ (resp. $\ptermA > 0$) characterizes a topologically closed (resp. open) set so the appropriate topological requirements of~\irref{CORef} are satisfied. The resulting left premise is the left premise of~\irref{dVcmpQ}:
{\footnotesizeoff%
\begin{sequentdeduction}[array]
\linfer[CORef]{
  \lsequent{\Gamma}{\dbox{\pevolvein{\D{x}=\genDE{x}}{\lnot{(p \cmp 0)}}}{\ivr}} !
  \lsequent{\Gamma,\constt{\varepsilon} > 0}{\ddiamond{\pevolve{\D{x}=\genDE{x}}}{p\cmp 0}}
}
{\lsequent{\Gamma,\constt{\varepsilon} > 0,\lnot{(\ptermA \cmp 0)}}{\ddiamond{\pevolvein{\D{x}=\genDE{x}}{\ivr}}{p \cmp 0}}}
\end{sequentdeduction}
}%
The proof continues from the resulting right premise identically to the derivation of~\irref{dVcmp} until the step where \irref{dVcmpA} is used.
The steps are repeated briefly here.

{\footnotesizeoff%
\begin{sequentdeduction}[array]
  \linfer[cut+qear+existsl]{
  \linfer[dGt]{
  \linfer[cut+GEx]{
    \lsequent{\Gamma, \ptermA = \ptermA_0, \timevar = 0, \ddiamond{\pevolve{\D{x}=\genDE{x},\D{\timevar}=1}}{\ptermA_0 + \constt{\varepsilon}\timevar > 0} }{\ddiamond{\pevolve{\D{x}=\genDE{x},\D{\timevar}=1}}{\ptermA \cmp 0}}
  }
  {\lsequent{\Gamma, \constt{\varepsilon} > 0, \ptermA = \ptermA_0, i\constt{\varepsilon}=1, \timevar = 0}{\ddiamond{\pevolve{\D{x}=\genDE{x},\D{\timevar}=1}}{\ptermA \cmp 0}} }
  }
  {\lsequent{\Gamma, \constt{\varepsilon} > 0, \ptermA = \ptermA_0, i\constt{\varepsilon}=1}{\ddiamond{\pevolve{\D{x}=\genDE{x}}}{\ptermA \cmp 0}} }
  }
  {\lsequent{\Gamma, \constt{\varepsilon} > 0}{\ddiamond{\pevolve{\D{x}=\genDE{x}}}{\ptermA \cmp 0}} }
\end{sequentdeduction}
}%

Like the derivation of~\irref{dVcmpA}, axiom \irref{Prog} is used with $\rgvar \mnodefequiv \constt{\ptermA_0} + \constt{\varepsilon} \timevar > 0$.
The key difference is an additional~\irref{dC} step, which adds $\ivr$ to the domain constraint.
The proof of this differential cut uses the left premise of~\irref{dVcmpQ}, it is labeled \textcircled{1} and shown below.
{\footnotesizeoff%
\begin{sequentdeduction}
  \linfer[Prog]{
  \linfer[dC]{
    \lsequent{\qquad\Gamma,\ptermA = \constt{\ptermA_0},\timevar=0}{\dbox{\pevolvein{\D{x}=\genDE{x},\D{\timevar}=1}{\lnot{(\ptermA \cmp 0) \land \ivr}}}{\constt{\ptermA_0} + \constt{\varepsilon} \timevar \leq 0}}
    \qquad \textcircled{1}
  }
    {\lsequent{\Gamma,\ptermA = \constt{\ptermA_0},\timevar=0}{\dbox{\pevolvein{\D{x}=\genDE{x},\D{\timevar}=1}{\lnot{(\ptermA \cmp 0)}}}{\constt{\ptermA_0} + \constt{\varepsilon} \timevar \leq 0}} \qquad\;\;\,\,}
  }
  {\lsequent{\Gamma, \ptermA=\ptermA_0, \timevar=0, \ddiamond{\pevolve{\D{x}=\genDE{x},\D{\timevar}=1}}{\ptermA_0 + \constt{\varepsilon}\timevar > 0} }{\ddiamond{\pevolve{\D{x}=\genDE{x},\D{\timevar}=1}}{\ptermA \cmp 0}}}
\end{sequentdeduction}
}%
The derivation from the resulting left premise (after the cut) continues similarly to~\irref{dVcmpA} using a monotonicity step~\irref{MbW} to rephrase the postcondition, followed by~\irref{dIcmp} which results in the right premise of~\irref{dVcmpQ}:
{\footnotesizeoff%
\begin{sequentdeduction}[array]
  \linfer[MbW]{
  \linfer[dIcmp]{
    \lsequent{\lnot{(\ptermA \cmp 0)}, \ivr}{\lied[]{\genDE{x}}{\ptermA}\geq \constt{\varepsilon}}
  }
    {\lsequent{\Gamma, \ptermA = \constt{\ptermA_0}, \timevar=0}{\dbox{\pevolvein{\D{x}=\genDE{x},\D{\timevar}=1}{\lnot{(\ptermA \cmp 0)}\land \ivr}}{\ptermA \geq \constt{\ptermA_0} + \constt{\varepsilon} \timevar}}}
  }
  {\lsequent{\Gamma,\ptermA = \constt{\ptermA_0}, \timevar=0}{\dbox{\pevolvein{\D{x}=\genDE{x},\D{\timevar}=1}{\lnot{(\ptermA \cmp 0)}\land \ivr}}{\constt{\ptermA_0} + \constt{\varepsilon} \timevar \leq 0}}}
\end{sequentdeduction}
}%

The derivation from~\textcircled{1} removes the time variable $t$ using the inverse direction of rule~\irref{dGt}~\cite{DBLP:journals/jar/Platzer17,Platzer18,DBLP:conf/lics/PlatzerT18}.
Just as rule~\irref{dGt} allows introducing a \emph{fresh} time variable $t$ for the sake of proof, its inverse direction simply removes the variable $t$ since it is irrelevant for the proof of the differential cut.
{\footnotesizeoff%
\begin{sequentdeduction}[array]
  \linfer[dGt]{
  \lsequent{\Gamma}{\dbox{\pevolvein{\D{x}=\genDE{x}}{\lnot{(\ptermA \cmp 0)}}}{\ivr}}
  }
  {\lsequent{\Gamma,\ptermA = \constt{\ptermA_0},\timevar=0}{\dbox{\pevolvein{\D{x}=\genDE{x},\D{\timevar}=1}{\lnot{(\ptermA \cmp 0)}}}{\ivr}}}
\\[-\normalbaselineskip]\tag*{\qedhere}
\end{sequentdeduction}
}%
\end{proofatend}

The original presentation of rule~\irref{dVcmpA}~\cite{DBLP:journals/logcom/Platzer10} omits the highlighted assumption $\highlight{\lnot{(\ptermA \cmp 0)}}$.
This premise is needed for the \irref{CORef} step and the rule is unsound without it.
In addition, it uses a form of syntactic weak negation~\cite{DBLP:journals/logcom/Platzer10}, which is also unsound for open postconditions, as pointed out earlier~\cite{DBLP:conf/fm/SogokonJ15}.
See~\thereport{\rref{app:counterexamples}} for counterexamples.
Our presentation of~\irref{dVcmpQ} recovers soundness by adding topological restrictions on the domain constraint $\ivr$.

The next two corollaries similarly make use of~\irref{CORef} to derive the proof rule~\irref{TTQ}~\cite{DBLP:conf/emsoft/TalyT10} and the adapted rule~\irref{RSQ}~\cite{DBLP:journals/siamco/RatschanS10}.
They respectively generalize~\irref{TT} and~\irref{RS} from~\rref{sec:nodomconstraint} to handle domain constraints.
The technical glitches in their original presentations~\cite{DBLP:journals/siamco/RatschanS10,DBLP:conf/emsoft/TalyT10}, which were identified in \rref{sec:nodomconstraint}, remain highlighted here:

\begin{corollary}[Equational differential variants with domains~\cite{DBLP:conf/emsoft/TalyT10}]
\label{cor:ttq}
The following proof rules are derivable in \dL.
Term $\constt{\varepsilon}$ is constant for the ODE $\D{x}=\genDE{x}$ and the ODE is \highlight{globally Lipschitz continuous} in both rules.
Formula $\ivr$ characterizes a closed set over variables $x$.\\
\begin{calculuscollection}
\begin{calculus}
\dinferenceRule[dVeqQ|dV$_=\&$]{}
{\linferenceRule
  { \lsequent{\Gamma}{\dbox{\pevolvein{\D{x}=\genDE{x}}{\ptermA < 0}}{\ivr}} \quad
    \lsequent{\ptermA < 0, \ivr}{\lied[]{\genDE{x}}{\ptermA}\geq \constt{\varepsilon}}
  }
  {\lsequent{\Gamma,\constt{\varepsilon} > 0, \ptermA \leq 0, \ivr}{\ddiamond{\pevolvein{\D{x}=\genDE{x}}{\ivr}}{\ptermA = 0}} }
}{}

\dinferenceRule[TTQ|dV${_{=}^M\&}$]{}
{\linferenceRule
  {
    \lsequent{\ivr, \ptermA = 0}{\rfvar} \quad
    \lsequent{\Gamma}{\dbox{\pevolvein{\D{x}=\genDE{x}}{\ptermA < 0}}{\ivr}} \quad
    \lsequent{\ptermA < 0, \ivr}{\lied[]{\genDE{x}}{\ptermA}\geq \constt{\varepsilon}}
  }
  {\lsequent{\Gamma,\constt{\varepsilon} > 0, \ptermA\leq 0, \ivr}{\ddiamond{\pevolvein{\D{x}=\genDE{x}}{\ivr}}{\rfvar}} }
}{}
\end{calculus}
\end{calculuscollection}
\end{corollary}
\begin{proofatend}
The derivations of rules~\irref{dVeqQ+TTQ} are similar to the derivations of rules~\irref{dVeq+TT} respectively.
Rule~\irref{TTQ} derives from~\irref{dVeqQ} by monotonicity:
{\footnotesizeoff%
\renewcommand{\linferPremissSeparation}{\hspace{4pt}}%
\begin{sequentdeduction}[array]
  \linfer[MdW]{
  \lsequent{\ivr,\ptermA = 0}{\rfvar}
  !
  \linfer[dVeqQ]{
     \lsequent{\Gamma}{\dbox{\pevolvein{\D{x}=\genDE{x}}{\ptermA < 0}}{\ivr}} !
     \lsequent{\ptermA < 0,\ivr}{\lied[]{\genDE{x}}{\ptermA}\geq \constt{\varepsilon}} !
  }
    {\lsequent{\Gamma,\constt{\varepsilon} > 0, \ptermA \leq 0, \ivr}{\ddiamond{\pevolvein{\D{x}=\genDE{x}}{\ivr}}{\ptermA = 0}}}
  }
  {\lsequent{\Gamma,\constt{\varepsilon} > 0, \ptermA \leq 0, \ivr}{\ddiamond{\pevolvein{\D{x}=\genDE{x}}{\ivr}}{\rfvar}}}
\end{sequentdeduction}
}%

The derivation of rule~\irref{dVeqQ} starts by using axiom~\irref{Prog} with $\rgvar \mnodefequiv \ptermA \geq 0$.
The resulting box modality (right) premise is abbreviated~\textcircled{1} and proved below.
On the resulting left premise, a~\irref{DX} step adds the negated postcondition $\ptermA < 0$ as an assumption to the antecedents since the domain constraint $\ivr$ is true initially.
Following that, rule~\irref{dVcmpQ} is used (with $\cmp$ being $\geq$, since $\ivr$ characterizes a closed set). This yields the two premises of~\irref{dVeqQ}:
{\footnotesizeoff%
\begin{sequentdeduction}[array]
  \linfer[Prog]{
  \linfer[DX]{
  \linfer[dVcmpQ]{
     \lsequent{\Gamma}{\dbox{\pevolvein{\D{x}=\genDE{x}}{\ptermA < 0}}{\ivr}} !
     \lsequent{\ptermA < 0,\ivr}{\lied[]{\genDE{x}}{\ptermA}\geq \constt{\varepsilon}}
  }
    {\lsequent{\Gamma,\constt{\varepsilon} > 0, \ptermA < 0}{\ddiamond{\pevolvein{\D{x}=\genDE{x}}{\ivr}}{\ptermA \geq 0}}}}
  {\lsequent{\Gamma,\constt{\varepsilon} > 0, \ivr}{\ddiamond{\pevolvein{\D{x}=\genDE{x}}{\ivr}}{\ptermA \geq 0}}  \qquad \textcircled{1}}
  }
  {\lsequent{\Gamma,\constt{\varepsilon} > 0, \ptermA \leq 0, \ivr}{\ddiamond{\pevolvein{\D{x}=\genDE{x}}{\ivr}}{\ptermA = 0}}}
\end{sequentdeduction}
}%
From premise \textcircled{1}, the derivation is closed similarly to~\irref{dVeq} using~\irref{DX} and \irref{BC}:
{\footnotesizeoff%
\begin{sequentdeduction}[array]
  \linfer[DX]{
  \linfer[BC]{
  \linfer[qear]{
    \lclose
  }
    {\lsequent{\ptermA \neq 0, \ptermA = 0}{\lied[]{\genDE{x}}{\ptermA} < 0}}
  }
    {\lsequent{\ptermA < 0}{\dbox{\pevolvein{\D{x}=\genDE{x}}{\ivr \land \ptermA \neq 0}}{\ptermA< 0}}}
  }
  {\lsequent{\ptermA \leq 0}{\dbox{\pevolvein{\D{x}=\genDE{x}}{\ivr \land \ptermA \neq 0}}{\ptermA< 0}}}
\\[-\normalbaselineskip]\tag*{\qedhere}
\end{sequentdeduction}
}%
\end{proofatend}

\begin{corollary}[Set Lyapunov functions with domains~\cite{DBLP:journals/siamco/RatschanS10}]
\label{cor:rsq}
The following proof rule is derivable in \dL.
Formula $K$ characterizes a \highlight{compact set} over variables $x$, while formula $\rfvar$ characterizes an open set over those variables.\\
\begin{calculuscollection}
\begin{calculus}
\dinferenceRule[RSQ|SLyap$\&$]{}
{
\linferenceRule
  {
     \lsequent{\ptermA \geq 0}{K}
     \quad
     \lsequent{\lnot{\rfvar},K}{\lied[]{\genDE{x}}{\ptermA} > 0}
  }
  {\lsequent{\Gamma, \ptermA > 0}{\ddiamond{\pevolvein{\D{x}=\genDE{x}}{\ptermA > 0}}{\rfvar}} }
}{}
\end{calculus}
\end{calculuscollection}
\end{corollary}
\begin{proofatend}
The derivation of rule~\irref{RSQ} starts by using~\irref{DX} to add assumption $\lnot{\rfvar}$ to the antecedents since the domain constraint $\ptermA > 0$ is in the antecedents.
Next, axiom~\irref{CORef} is used.
Its topological restrictions are met since both formulas $\rfvar$ and $\ptermA > 0$ characterize open sets.
From the resulting right premise, rule~\irref{RS} yields the corresponding two premises of~\irref{RSQ} because formula $K$ (resp. $\rfvar$) characterizes a compact set (resp. open set):
{\footnotesizeoff%
\begin{sequentdeduction}[array]
  \linfer[DX]{
  \linfer[CORef]{
    \lsequent{\Gamma, \ptermA > 0}{\dbox{\pevolvein{\D{x}=\genDE{x}}{\lnot{\rfvar}}}{\ptermA > 0}}
    !
    \linfer[RS]{
     \lsequent{\ptermA \geq 0}{K} !
     \lsequent{\lnot{\rfvar},K}{\lied[]{\genDE{x}}{\ptermA} > 0}
    }
    {\lsequent{\Gamma, \ptermA > 0}{\ddiamond{\pevolve{\D{x}=\genDE{x}}}{\rfvar}}}
  }
    {\lsequent{\Gamma, \ptermA > 0, \lnot{\rfvar}}{\ddiamond{\pevolvein{\D{x}=\genDE{x}}{\ptermA > 0}}{\rfvar}}}
  }
  {\lsequent{\Gamma, \ptermA > 0}{\ddiamond{\pevolvein{\D{x}=\genDE{x}}{\ptermA > 0}}{\rfvar}} }
\end{sequentdeduction}
}%

From the leftmost open premise after~\irref{CORef}, rule~\irref{BC} is used and the resulting $\ptermA = 0$ assumption is turned into $K$ using the left premise of~\irref{RSQ}.
The resulting open premises are the premises of~\irref{RSQ}:
{\footnotesizeoff%
\begin{sequentdeduction}[array]
  \linfer[BC]{
  \linfer[cut]{
    \lsequent{\lnot{\rfvar}, K}{ \lied[]{\genDE{x}}{\ptermA} > 0}
    !
    \linfer[qear]{
      \lsequent{\ptermA \geq 0}{K}
    }
    {\lsequent{\ptermA = 0}{K}}
  }
    {\lsequent{\lnot{\rfvar}, \ptermA = 0}{ \lied[]{\genDE{x}}{\ptermA} > 0}}
  }
  {\lsequent{\Gamma, \ptermA > 0}{\dbox{\pevolvein{\D{x}=\genDE{x}}{\lnot{\rfvar}}}{\ptermA > 0}}}
\\[-\normalbaselineskip]\tag*{\qedhere}
\end{sequentdeduction}
}%
\end{proofatend}

\noindent
The staging sets with domain constraints proof rule~\irref{SPQ}~\cite{DBLP:conf/fm/SogokonJ15} uses axiom \irref{SARef}:

\begin{corollary}[Staging sets with domains~\cite{DBLP:conf/fm/SogokonJ15}]
\label{cor:SPQ}
The following proof rule is derivable in \dL.
Term $\constt{\varepsilon}$ is constant for ODE $\D{x}=\genDE{x}$ and the ODE is globally Lipschitz continuous.\\
\begin{calculuscollection}
\begin{calculus}
\dinferenceRule[SPQ|SP$\&$]{}
{
\linferenceRule
  { \lsequent{\Gamma}{\dbox{\pevolvein{\D{x}=\genDE{x}}{\lnot{(\rfvar \land \ivr)}}}{\rsfvar}} \quad
    \lsequent{\rsfvar}{\ivr \land p \leq 0 \land \lied[]{\genDE{x}}{p}\geq \constt{\varepsilon}}
  }
  {\lsequent{\Gamma,\constt{\varepsilon}>0}{\ddiamond{\pevolvein{\D{x}=\genDE{x}}{\ivr}}{\rfvar}} }
}{}
\end{calculus}
\end{calculuscollection}
\end{corollary}
\begin{proofatend}
The derivation starts by using axiom \irref{SARef} which results in two premises:
{\footnotesizeoff%
\begin{sequentdeduction}[array]
\linfer[SARef]{
  \lsequent{\Gamma}{\dbox{\pevolvein{\D{x}=\genDE{x}}{\lnot{(\rfvar \land \ivr)}}}{\ivr}} !
  \lsequent{\Gamma}{\ddiamond{\pevolve{\D{x}=\genDE{x}}}{\rfvar}}
}
{\lsequent{\Gamma}{\ddiamond{\pevolvein{\D{x}=\genDE{x}}{\ivr}}{\rfvar}}}
\end{sequentdeduction}
}%
From the left premise after~\irref{SARef}, a monotonicity step turns the postcondition into $\rsfvar$, yielding the left premise and first conjunct of the right premise of~\irref{SPQ}.
{\footnotesizeoff%
\begin{sequentdeduction}[array]
\linfer[MbW]{
  \lsequent{\rsfvar}{\ivr} !
  \lsequent{\Gamma}{\dbox{\pevolvein{\D{x}=\genDE{x}}{\lnot{(\rfvar \land \ivr)}}}{\rsfvar}}
}
  {\lsequent{\Gamma}{\dbox{\pevolvein{\D{x}=\genDE{x}}{\lnot{(\rfvar \land \ivr)}}}{\ivr}}}
\end{sequentdeduction}
}%
From the right premise after using axiom~\irref{SARef}, rule~\irref{SP} yields the remaining two premises of~\irref{SPQ}:
{\footnotesizeoff%
\begin{sequentdeduction}[array]
\linfer[SP]{
  \linfer[dW+DMP]{
    \lsequent{\Gamma}{\dbox{\pevolvein{\D{x}=\genDE{x}}{\lnot{(\rfvar \land \ivr)}}}{\rsfvar}}
  }
  {\lsequent{\Gamma}{\dbox{\pevolvein{\D{x}=\genDE{x}}{\lnot{\rfvar}}}{\rsfvar}}} !
   \lsequent{\rsfvar}{p \leq 0 \land \lied[]{\genDE{x}}{p}\geq \constt{\varepsilon}}
}
  {\lsequent{\Gamma}{\ddiamond{\pevolve{\D{x}=\genDE{x}}}{\rfvar}}}
\end{sequentdeduction}
}%
The~\irref{dW+DMP} step uses the propositional tautology $\lnot{\rfvar} \limply \lnot{(\rfvar \land \ivr)}$ to weaken the domain constraint so that it matches the premise of rule~\irref{SPQ}.
\end{proofatend}

The rules derived in Corollaries~\ref{cor:atomicdvcmpQ}--\ref{cor:SPQ} demonstrate the flexibility of our refinement approach for deriving surveyed liveness arguments as proof rules.
Our approach is not limited to these surveyed arguments because refinement steps can be freely mixed-and-matched for specific liveness questions.

\begin{example}
The liveness property $u^2+v^2=1 \limply \ddiamond{\exnonlinear}{u^2 + v^2 \geq 2}$ was proved in~\rref{ex:exnonlinproof} using the staging set formula $\rsfvar \mnodefequiv 1 \leq u^2+v^2 \leq 2$.
Since $\rsfvar$ and $u^2 + v^2 \geq 2$ both characterize closed sets, axiom~\irref{CORef} extends the chain of refinements~\rref{eq:refinementchain} from~\rref{ex:exnonlinproof} to show a stronger liveness property for $\exnonlinear$:
\[
  \ddiamond{\pevolve{\exnonlinear,\D{\timevar}=1}}{(\lnot{\rsfvar} \lor \timevar > \frac{2}{3})}
  {\osetf{\irref{Prog}}{\limply}}
  \ddiamond{\pevolve{\exnonlinear}}{\lnot{\rsfvar}}
  {\osetf{\irref{Prog}}{\limply}}
  \ddiamond{\pevolve{\exnonlinear}}{u^2 + v^2 \geq 2}
  {\osetf{\irref{CORef}}{\limply}}
  \ddiamond{\pevolvein{\exnonlinear}{\rsfvar}}{u^2 + v^2 \geq 2}
\]

Formula $\rsfvarhat \mnodefequiv 1 \leq u^2+v^2 < 2$ also proves \rref{ex:exnonlinproof} but does \emph{not} characterize a closed set.
Thankfully, the careful topological restriction of \irref{CORef} prevents us from unsoundly concluding the property $u^2+v^2=1 \limply \ddiamond{\pevolvein{\exnonlinear}{\rsfvarhat}}{u^2 + v^2 \geq 2}$.
This latter property is unsatisfiable because $\rsfvarhat$ does not overlap with $u^2 + v^2 \geq 2$.
\end{example}

The refinement approach also enables discovery of new, general liveness proof rules by combining refinement steps in alternative ways.
As an example, the following chimeric proof rule combines ideas from Corollaries~\ref{cor:higherdv},~\ref{cor:boundedandcompact}, and~\ref{cor:SPQ}:

\begin{corollary}[Combination proof rule]
\label{cor:combination}
The following proof rule is derivable in \dL.
Formula $\rsfvar$ characterizes a compact set over variables $x$.\\
\begin{calculuscollection}
\begin{calculus}
\dinferenceRule[SPcQ|SP$_c^k\&$]{}
{
\linferenceRule
  { \lsequent{\Gamma}{\dbox{\pevolvein{\D{x}=\genDE{x}}{\lnot{(\rfvar \land \ivr)}}}{\rsfvar}} \quad
    \lsequent{\rsfvar}{\ivr \land \lied[k]{\genDE{x}}{p} > 0}
  }
  {\lsequent{\Gamma}{\ddiamond{\pevolvein{\D{x}=\genDE{x}}{\ivr}}{\rfvar}} }
}{}
\end{calculus}
\end{calculuscollection}
\end{corollary}
\begin{proofatend}
The chimeric proof rule~\irref{SPcQ} is an amalgamation of ideas behind the rules~\irref{SPQ+dVcmpK+SPc}.
It is therefore unsurprising that the derivation of~\irref{SPcQ} uses various steps from the derivations of those rules.
The derivation of~\irref{SPcQ} starts similarly to~\irref{SPQ} (following~\rref{cor:SPQ}) using axiom \irref{SARef}:
{\footnotesizeoff%
\begin{sequentdeduction}[array]
\linfer[SARef]{
  \lsequent{\Gamma}{\dbox{\pevolvein{\D{x}=\genDE{x}}{\lnot{(\rfvar \land \ivr)}}}{\ivr}} !
  \lsequent{\Gamma}{\ddiamond{\pevolve{\D{x}=\genDE{x}}}{\rfvar}}
}
{\lsequent{\Gamma}{\ddiamond{\pevolvein{\D{x}=\genDE{x}}{\ivr}}{\rfvar}}}
\end{sequentdeduction}
}%

From the left premise after~\irref{SARef}, a monotonicity step turns the postcondition into $\rsfvar$, yielding the left premise and first conjunct of the right premise of~\irref{SPcQ}.
{\footnotesizeoff%
\begin{sequentdeduction}[array]
  \linfer[MbW]{
  \lsequent{\Gamma}{\dbox{\pevolvein{\D{x}=\genDE{x}}{\lnot{(\rfvar \land \ivr)}}}{\rsfvar}} !
  \lsequent{\rsfvar}{\ivr}
  }
  {\lsequent{\Gamma}{\dbox{\pevolvein{\D{x}=\genDE{x}}{\lnot{(\rfvar \land \ivr)}}}{\ivr}}}
\end{sequentdeduction}
}%

From the right premise after~\irref{SARef}, the derivation continues using~\irref{Prog} with $\rgvar \mnodefequiv \lnot{\rsfvar}$, followed by~\irref{dW+DMP}.
The resulting left premise is (again) the left premise of~\irref{SPcQ}, while the resulting right premise is abbreviated \textcircled{1} and continued below:
{\footnotesizeoff%
\begin{sequentdeduction}[array]
\linfer[Prog]{
  \linfer[dW+DMP]{
    \lsequent{\Gamma}{\dbox{\pevolvein{\D{x}=\genDE{x}}{\lnot{(\rfvar \land \ivr)}}}{\rsfvar}}
  }
  {\lsequent{\Gamma}{\dbox{\pevolvein{\D{x}=\genDE{x}}{\lnot{\rfvar}}}{\rsfvar}}} !
  \textcircled{1}
}
  {\lsequent{\Gamma}{\ddiamond{\pevolve{\D{x}=\genDE{x}}}{\rfvar}}}
\end{sequentdeduction}
}%

The derivation continues from \textcircled{1} by intertwining proof ideas from~\rref{cor:higherdv} and~\rref{cor:boundedandcompact}.
First, compactness of the set characterized by $\rsfvar(x)$ implies that the formula
$\lexists{\varepsilon {>} 0}{A(\varepsilon)}$ where $A(\varepsilon) \mnodefequiv \lforall{x}{(\rsfvar(x) \limply \lied[k]{\genDE{x}}{p} \geq \varepsilon)}$ and the formula
$B \mnodefequiv \lforall{x}{(\rsfvar(x) \limply \lied[k]{\genDE{x}}{p} > 0)}$ are provably equivalent in first-order real arithmetic.
These facts are added to the assumptions similarly to the derivation of~\irref{SPc}.
The resulting right open premise is the right conjunct of the right premise of~\irref{SPcQ}:

{\footnotesizeoff%
\begin{sequentdeduction}[array]
  \linfer[cut]{
  \linfer[existsl]{
    \lsequent{\Gamma,\varepsilon > 0, A(\varepsilon)}{\ddiamond{\pevolve{\D{x}=\genDE{x}}}{\lnot{\rsfvar}}}
  }
  {\lsequent{\Gamma,\lexists{\varepsilon {>} 0}{A(\varepsilon)}}{\ddiamond{\pevolve{\D{x}=\genDE{x}}}{\lnot{\rsfvar}}}}
  !
    \linfer[qear]{
    \linfer[allr+implyr]{
      \lsequent{\rsfvar}{\lied[k]{\genDE{x}}{p} > 0}
    }
      \lsequent{}{B}
    }
    {\lsequent{}{\lexists{\varepsilon {>} 0}{A(\varepsilon)}}}
  }
  {\lsequent{\Gamma}{\ddiamond{\pevolve{\D{x}=\genDE{x}}}{\lnot{\rsfvar}}} }
\end{sequentdeduction}
}%

From the left premise, recall the derivation from~\rref{cor:higherdv} which introduces fresh variables for the initial values of the Lie derivatives with~\irref{cut+qear+existsl}.
The derivation continues similarly here, with the resulting antecedents abbreviated $\Gamma_0 \mnodefequiv \big(\Gamma,\ptermA=\ptermA_0, \dots, \lied[k-1]{\genDE{x}}{\ptermA} = \lied[k-1]{\genDE{x}}{\ptermA}_0\big)$.
Rule~\irref{dGt} is also used to add time variable $\timevar$ to the system of equations.

{\footnotesizeoff%
\begin{sequentdeduction}[array]
  \linfer[cut+qear+existsl]{
  \linfer[dGt]{
    \lsequent{\Gamma_0, \varepsilon > 0, A(\varepsilon),\timevar=0}{\ddiamond{\pevolve{\D{x}=\genDE{x},\D{\timevar}=1}}{\lnot{\rsfvar}}}
  }
    {\lsequent{\Gamma_0, \varepsilon > 0, A(\varepsilon)}{\ddiamond{\pevolve{\D{x}=\genDE{x}}}{\lnot{\rsfvar}}}}
  }
  {\lsequent{\Gamma,\varepsilon > 0, A(\varepsilon)}{\ddiamond{\pevolve{\D{x}=\genDE{x}}}{\lnot{\rsfvar}}}}
\end{sequentdeduction}
}%

Recall from~\rref{cor:boundedandcompact} that the formula $\rrfvar(\ptermA_1) \mnodefequiv \lforall{x}{(\rsfvar(x) \limply \ptermA \leq \ptermA_1)}$ can be added to the assumptions using~\irref{cut+qear+existsl}, for some fresh variable $\ptermA_1$ symbolically representing the maximum value of $\ptermA$ on the compact set characterized by $\rsfvar$:
{\footnotesizeoff%
\begin{sequentdeduction}[array]
  \linfer[cut+qear+existsl]{
    \lsequent{\Gamma_0, \varepsilon > 0, A(\varepsilon),\timevar=0,\rrfvar(\ptermA_1)}{\ddiamond{\pevolve{\D{x}=\genDE{x},\D{\timevar}=1}}{\lnot{\rsfvar}}}
  }
  {\lsequent{\Gamma_0, \varepsilon > 0, A(\varepsilon),\timevar=0}{\ddiamond{\pevolve{\D{x}=\genDE{x},\D{\timevar}=1}}{\lnot{\rsfvar}}}}
\end{sequentdeduction}
}%

One last arithmetic cut is needed to set up the sequence of differential cuts~\rref{eq:integration}.
Recall the polynomial $\ptermB(\timevar)$ from~\rref{eq:integration} is eventually positive for sufficiently large values of $\timevar$ because its leading coefficient is strictly positive.
The same applies to the polynomial $\ptermB(\timevar)-\ptermA_1$ so~\irref{cut+qear} (and Skolemizing with~\irref{existsl}) adds the formula $\lforall{\timevar > \timevar_1} {\ptermB(\timevar) - \ptermA_1 > 0}$ to the assumptions:
{\footnotesizeoff%
\begin{sequentdeduction}[array]
  \linfer[cut+qear+existsl]{
    \lsequent{\Gamma_0, \varepsilon > 0, A(\varepsilon),\timevar=0,\rrfvar(\ptermA_1),\lforall{\timevar > \timevar_1} {\ptermB(\timevar) - \ptermA_1 > 0}}{\ddiamond{\pevolve{\D{x}=\genDE{x},\D{\timevar}=1}}{\lnot{\rsfvar}}}
  }
  {\lsequent{\Gamma_0, \varepsilon > 0, A(\varepsilon),\timevar=0,\rrfvar(\ptermA_1)}{\ddiamond{\pevolve{\D{x}=\genDE{x},\D{\timevar}=1}}{\lnot{\rsfvar}}}
}
\end{sequentdeduction}
}%

Once all the arithmetic cuts are in place, an additional cut introduces a (bounded) sufficient duration assumption (antecedents temporarily abbreviated with $\dots$ for brevity).
The cut premise, abbreviated~\textcircled{1}, is proved below:
{\footnotesizeoff%
\begin{sequentdeduction}[array]
  \linfer[cut]{
    \lsequent{\Gamma_0, \dots, \ddiamond{\pevolve{\D{x}=\genDE{x},\D{\timevar}=1}}{(\lnot{\rsfvar} \lor \ptermB(t) - \ptermA_1 > 0)}}{\ddiamond{\pevolve{\D{x}=\genDE{x},\D{\timevar}=1}}{\lnot{\rsfvar}}} \quad
    \textcircled{1}
  }
  {\lsequent{\Gamma_0, \varepsilon > 0, A(\varepsilon),\timevar=0,\rrfvar(\ptermA_1),\lforall{\timevar > \timevar_1} {\ptermB(\timevar) - \ptermA_1 > 0}}{\ddiamond{\pevolve{\D{x}=\genDE{x},\D{\timevar}=1}}{\lnot{\rsfvar}}}}
\end{sequentdeduction}
}%

From the left, open premise, axiom~\irref{Prog} is used with $\rgvar \mnodefequiv \lnot{\rsfvar} \lor \ptermB(t) - \ptermA_1 > 0$:
{\footnotesizeoff%
\begin{sequentdeduction}%
  \linfer[Prog]{
    \lsequent{\Gamma_0, \varepsilon > 0, A(\varepsilon),\timevar=0,\rrfvar(\ptermA_1)}{\dbox{\pevolvein{\D{x}=\genDE{x},\D{\timevar}=1}{\rsfvar}}{(\rsfvar \land \ptermB(t) - \ptermA_1 \leq 0)}}
  }
  {\lsequent{\Gamma_0,\dots, \ddiamond{\pevolve{\D{x}=\genDE{x},\D{\timevar}=1}}{(\lnot{\rsfvar} \lor \ptermB(t) - \ptermA_1 > 0)}}{\ddiamond{\pevolve{\D{x}=\genDE{x},\D{\timevar}=1}}{\lnot{\rsfvar}}}}
\end{sequentdeduction}
}%

Next, a monotonicity step~\irref{MbW} simplifies the postcondition using the (constant) assumption $\rrfvar(\ptermA_1)$ and the domain constraint $\rsfvar$:

{\footnotesizeoff%
\begin{sequentdeduction}[array]
  \linfer[MbW]{
    \lsequent{\Gamma_0, \timevar=0, A(\varepsilon)}{\dbox{\pevolvein{\D{x}=\genDE{x},\D{\timevar}=1}{\rsfvar}}{\ptermA \geq \ptermB(t)}}
  }
    {\lsequent{\Gamma_0, \varepsilon > 0, A(\varepsilon),\timevar=0,\rrfvar(\ptermA_1)}{\dbox{\pevolvein{\D{x}=\genDE{x},\D{\timevar}=1}{\rsfvar}}{(\rsfvar \land \ptermB(t) - \ptermA_1 \leq 0)}}}
\end{sequentdeduction}
}%

The derivation closes using the chain of differential cuts from~\rref{eq:integration}.
In the first~\irref{dC} step, the (constant) assumption $A(\varepsilon)$ is used, see the derivation labeled \textcircled{$\star$} immediately below:
{\footnotesizeoff%
\begin{sequentdeduction}[array]
  \linfer[dC]{
    \lsequent{\Gamma_0, \timevar=0}{\dbox{\pevolvein{\D{x}=\genDE{x},\D{\timevar}=1}{\rsfvar \land \lied[k-1]{\genDE{x}}{\ptermA} \geq \lied[k-1]{\genDE{x}}{\ptermA}_0 + \constt{\varepsilon}\timevar}}{\ptermA \geq \ptermB(t)}} \quad \textcircled{$\star$}
  }
  {\lsequent{\Gamma_0, \timevar=0, A(\varepsilon)}{\dbox{\pevolvein{\D{x}=\genDE{x},\D{\timevar}=1}{\rsfvar}}{\ptermA \geq \ptermB(t)}}}
\end{sequentdeduction}
}%
From~\textcircled{$\star$}:
{\footnotesizeoff%
\begin{sequentdeduction}[array]
  \linfer[dIcmp]{
  \linfer[qear]{
    \lclose
  }
    {\lsequent{A(\varepsilon), \rsfvar}{\lied[k]{\genDE{x}}{\ptermA} \geq \constt{\varepsilon} }}
  }
  {\lsequent{\Gamma_0, \timevar=0,A(\varepsilon)}{\dbox{\pevolvein{\D{x}=\genDE{x},\D{\timevar}=1}{\rsfvar}}{\lied[k-1]{\genDE{x}}{\ptermA} \geq \lied[k-1]{\genDE{x}}{\ptermA}_0 + \constt{\varepsilon}\timevar}}}
\end{sequentdeduction}
}%

Subsequent~\irref{dC+dIcmp} steps are similar to the derivation in~\rref{cor:higherdv}:
{\footnotesizeoff\renewcommand{\arraystretch}{1.4}%
\begin{sequentdeduction}[array]
    \linfer[dC+dIcmp]{
    \linfer[dC+dIcmp]{
    \linfer[dIcmp]{
      \lclose
    }
      {\lsequent{\Gamma_0, \timevar=0}{\dbox{\pevolvein{\D{x}=\genDE{x},\D{\timevar}=1}{\dots \land \lied[1]{\genDE{x}}{\ptermA} \geq \lied[1]{\genDE{x}}{\ptermA}_0 + \dots + \constt{\varepsilon} \frac{\timevar^{k-1}}{(k-1)!} }}{\ptermA \geq \ptermB(\timevar)}}}
    }
      {\vdots}
    }
    {\lsequent{\Gamma_0, \timevar=0}{\dbox{\pevolvein{\D{x}=\genDE{x},\D{\timevar}=1}{\rsfvar \land \lied[k-1]{\genDE{x}}{\ptermA} \geq \lied[k-1]{\genDE{x}}{\ptermA}_0 + \constt{\varepsilon}\timevar }}{\ptermA \geq \ptermB(\timevar)}}}
\end{sequentdeduction}
}%

From premise \textcircled{1}, a monotonicity step~\irref{MdW} rephrases the postcondition of the cut using the assumption $\lforall{\timevar > \timevar_1} {\ptermB(\timevar) - \ptermA_1 > 0}$.
Axiom~\irref{BEx} finishes the derivation since $\rsfvar(x)$ characterizes a compact (hence bounded) set:
{\footnotesizeoff%
\begin{sequentdeduction}[array]
  \linfer[MdW]{
  \linfer[BEx]{
    \lclose
  }
    {\lsequent{}{\ddiamond{\pevolve{\D{x}=\genDE{x},\D{\timevar}=1}}{(\lnot{\rsfvar} \lor \timevar > \timevar_1)}}}
  }
  {\lsequent{\lforall{\timevar > \timevar_1} {\ptermB(\timevar) - \ptermA_1 > 0}}{\ddiamond{\pevolve{\D{x}=\genDE{x},\D{\timevar}=1}}{(\lnot{\rsfvar} \lor \ptermB(t) - \ptermA_1 > 0)}}}
\\[-\normalbaselineskip]\tag*{\qedhere}
\end{sequentdeduction}
}%
\end{proofatend}

Our logical approach derives even complicated proof rules like~\irref{SPcQ} from a small set of sound logical axioms, which ensures their correctness.
The proof rule~\irref{PRQ} below derives from~\irref{SPcQ} (for $k\mnodefeq1$) and is an adapted version of the liveness argument from~\cite[Theorem 3.5]{DBLP:journals/siamco/PrajnaR07}.
In the original presentation, additional restrictions are imposed on the sets characterized by $\Gamma,\rfvar,\ivr$, and different conditions are given compared to the left premise of~\irref{PRQ} (\highlight{highlighted below}).
These original conditions are overly permissive as they are checked on a smaller set than necessary for soundness. See~\thereport{\rref{app:counterexamples}} for counterexamples.

\begin{corollary}[Compact eventuality~\cite{DBLP:journals/siamco/PrajnaR07}]
\label{cor:prq}
The following proof rule is derivable in \dL.
Formula $\ivr \land \lnot{\rfvar}$ characterizes a compact set over variables $x$.\\
\begin{calculuscollection}
\begin{calculus}
\dinferenceRule[PRQ|E$_c\&$]{}
{
\linferenceRule
  { \highlight{\lsequent{\Gamma}{\dbox{\pevolvein{\D{x}=\genDE{x}}{\lnot{(\rfvar \land \ivr)}}}{\ivr}}} \quad
    \lsequent{\ivr,\lnot{\rfvar}}{\lied[]{\genDE{x}}{p} > 0}
  }
  {\lsequent{\Gamma}{\ddiamond{\pevolvein{\D{x}=\genDE{x}}{\ivr}}{\rfvar}} }
}{}
\end{calculus}
\end{calculuscollection}
\end{corollary}
\begin{proofatend}
Rule~\irref{PRQ} derives from~\irref{SPcQ} with (compact) $\rsfvar \mnodefequiv \ivr \land \lnot{\rfvar}$ and $k\mnodefeq1$:
{\footnotesizeoff%
\begin{sequentdeduction}[array]
\linfer[SPc]{
  \linfer[MbW]{
    \lsequent{\Gamma}{\dbox{\pevolvein{\D{x}=\genDE{x}}{\lnot{(\rfvar \land \ivr)}}}{\ivr}}
  }
  {\lsequent{\Gamma}{\dbox{\pevolvein{\D{x}=\genDE{x}}{\lnot{(\rfvar \land \ivr)}}}{(\ivr \land \lnot{\rfvar})}}} !
  \linfer[]{
    \lsequent{\ivr,\lnot{\rfvar}}{\lied[]{\genDE{x}}{p} > 0}
  }
  {\lsequent{\ivr,\lnot{\rfvar}}{\ivr \land \lied[]{\genDE{x}}{p} > 0}}
}
  {\lsequent{\Gamma}{\ddiamond{\pevolvein{\D{x}=\genDE{x}}{\ivr}}{\rfvar}} }
\end{sequentdeduction}
}%
The~\irref{MbW} step uses the propositional tautology $\lnot{(\rfvar \land \ivr)} \land \ivr \limply \ivr \land \lnot{\rfvar}$.
\end{proofatend}

\section{Related Work}
\label{sec:relatedwork}

\paragraph{Liveness Proof Rules.}
The liveness arguments surveyed in this paper were originally presented in various notations, ranging from proof rules~\cite{DBLP:journals/logcom/Platzer10,DBLP:conf/fm/SogokonJ15,DBLP:conf/emsoft/TalyT10} to other mathematical notation~\cite{DBLP:conf/hybrid/PrajnaR05,DBLP:journals/siamco/PrajnaR07,DBLP:journals/siamco/RatschanS10,DBLP:conf/fm/SogokonJ15}.
All of them were justified directly through semantical (or mathematical) means.
We unify (and correct) all of these arguments and present them as \dL proof rules which are syntactically derived with our refinement-based approach from \dL axioms.

\paragraph{Other Liveness Properties.}
The liveness property studied in this paper is the continuous analog of \emph{eventually}~\cite{DBLP:books/daglib/0077033} or \emph{eventuality}~\cite{DBLP:journals/siamco/PrajnaR07,DBLP:conf/fm/SogokonJ15} from temporal logics.
In discrete settings, temporal logic specifications give rise to a zoo of liveness properties~\cite{DBLP:books/daglib/0077033}.
In continuous settings, \emph{weak eventuality} (requiring \emph{almost all} initial states to reach the goal region) and \emph{eventuality-safety} have been studied~\cite{DBLP:conf/hybrid/PrajnaR05,DBLP:journals/siamco/PrajnaR07}. %
In (continuous) adversarial settings, \emph{differential game variants}~\cite{DBLP:journals/tocl/Platzer17} enable proofs of (Angelic) winning strategies for differential games.
In dynamical systems and controls, the study of \emph{asymptotic stability} requires both stability (an invariance property) with asymptotic attraction towards a fixed point or periodic orbit (an eventuality-like property)~\cite{Chicone2006,DBLP:journals/siamco/RatschanS10}.
For hybrid systems, various authors have proposed generalizations of classical asymptotic stability, such as \emph{persistence}~\cite{Sogokon2018}, \emph{stability}~\cite{DBLP:conf/hybrid/PodelskiW06}, and \emph{inevitability}~\cite{DBLP:conf/hybrid/DuggiralaM12}.
\emph{Controlled} versions of these properties are also of interest, e.g., \emph{(controlled) reachability and attractivity}~\cite{ABATE2009150,DBLP:conf/emsoft/TalyT10}.
Eventuality(-like) properties are fundamental to all of these advanced liveness properties.
The formal understanding of eventuality in this paper is therefore a key step towards enabling formal analysis of more advanced liveness properties.

\paragraph{Automated Liveness Proofs.}
Automated reachability analysis tools~\cite{DBLP:conf/cav/ChenAS13,DBLP:conf/cav/FrehseGDCRLRGDM11} can also be used for liveness verification.
For an ODE and initial set $\bigchi_0$, computing an over-approximation $\mathcal{O}$ of the reachable set $\bigchi_{t} \subseteq \mathcal{O}$ at time $t$ shows that \emph{all} states in $\bigchi_0$ reach $\mathcal{O}$ at time $t$~\cite{Sogokon2018} (if solutions do not blow up).
Similarly, an under-approximation $\mathcal{U} \subseteq \bigchi_{t}$ shows that \emph{some} state in $\bigchi_0$ eventually reaches $\mathcal{U}$~\cite{DBLP:conf/hybrid/GoubaultP17} (if $\mathcal{U}$ is non-empty).
Neither approach handles domain constraints directly~\cite{DBLP:conf/hybrid/GoubaultP17,Sogokon2018} and, unlike deductive approaches, the use of reachability tools limits them to concrete time bounds $t$ and bounded initial sets $\bigchi_0$.
Deductive liveness approaches can also be automated.
Lyapunov functions guaranteeing (asymptotic) stability can be found by sum-of-squares (SOS) optimization~\cite{1184414}.
Liveness arguments can be similarly combined with SOS optimization to find suitable differential variants~\cite{DBLP:conf/hybrid/PrajnaR05,DBLP:journals/siamco/PrajnaR07}.
Other approaches are possible, e.g., a constraint solving-based approach can be used for finding so-called \emph{set Lyapunov functions}~\cite{DBLP:journals/siamco/RatschanS10}.
Crucially, automated approaches must be based on sound liveness arguments.
The correct justification of these arguments is precisely what our approach enables.

\section{Conclusion}
This paper presents a refinement-based approach for proving liveness for ODEs.
Exploration of new ODE liveness proof rules is enabled by piecing together refinement steps identified through our approach.
Given its wide applicability and correctness guarantees, our approach is a suitable framework for justifying ODE liveness arguments, even for readers less interested in the logical aspects.

\iflongversion
\paragraph{Acknowledgments.} We thank Katherine Cordwell, Frank Pfenning, Andrew Sogokon, and the anonymous reviewers for their feedback on this paper.
This material is based upon work supported by the Alexander von Humboldt Foundation and the AFOSR under grant number FA9550-16-1-0288.
The first author was also supported by A*STAR, Singapore.

The views and conclusions contained in this document are those of the authors and should not be interpreted as representing the official policies, either expressed or implied, of any sponsoring institution, the U.S. government or any other entity.
\else
\subsubsection{Acknowledgments.} We thank Katherine Cordwell, Frank Pfenning, Andrew Sogokon, and the anonymous reviewers for their feedback on this paper.
This material is based upon work supported by the Alexander von Humboldt Foundation and the AFOSR under grant number FA9550-16-1-0288.
The first author was also supported by A*STAR, Singapore.
\fi

\iflongversion
\bibliographystyle{plainurl}
\bibliography{paperlong}
\else
\newpage
\bibliographystyle{splncs04}
\bibliography{paper}

\begin{thebibliography}{10}

\bibitem{ABATE2009150}
Alessandro Abate, Alessandro {{D{'}Innocenzo}}, Maria Domenica~Di Benedetto,
  and Shankar Sastry.
\newblock Understanding deadlock and livelock behaviors in hybrid control
  systems.
\newblock {\em Nonlinear Anal. Hybrid Syst.}, 3(2):150 -- 162, 2009.
\newblock \href {http://dx.doi.org/10.1016/j.nahs.2008.12.005}
  {\path{doi:10.1016/j.nahs.2008.12.005}}.

\bibitem{10.2307/j.ctt17kkb0d}
Rajeev Alur.
\newblock {\em Principles of Cyber-Physical Systems}.
\newblock MIT Press, 2015.

\bibitem{Bochnak1998}
Jacek Bochnak, Michel Coste, and Marie-Fran{\c{c}}oise Roy.
\newblock {\em Real Algebraic Geometry}.
\newblock Springer, Heidelberg, 1998.
\newblock \href {http://dx.doi.org/10.1007/978-3-662-03718-8}
  {\path{doi:10.1007/978-3-662-03718-8}}.

\bibitem{DBLP:conf/cav/ChenAS13}
Xin Chen, Erika {\'{A}}brah{\'{a}}m, and Sriram Sankaranarayanan.
\newblock Flow*: An analyzer for non-linear hybrid systems.
\newblock In Natasha Sharygina and Helmut Veith, editors, {\em CAV}, volume
  8044 of {\em LNCS}, pages 258--263, Heidelberg, 2013. Springer.
\newblock \href {http://dx.doi.org/10.1007/978-3-642-39799-8_18}
  {\path{doi:10.1007/978-3-642-39799-8_18}}.

\bibitem{Chicone2006}
Carmen Chicone.
\newblock {\em Ordinary Differential Equations with Applications}.
\newblock Springer, New York, second edition, 2006.
\newblock \href {http://dx.doi.org/10.1007/0-387-35794-7}
  {\path{doi:10.1007/0-387-35794-7}}.

\bibitem{DoyenFPP18}
Laurent Doyen, Goran Frehse, George~J. Pappas, and Andr{\'e} Platzer.
\newblock Verification of hybrid systems.
\newblock In Edmund~M. Clarke, Thomas~A. Henzinger, Helmut Veith, and Roderick
  Bloem, editors, {\em Handbook of Model Checking}, pages 1047--1110. Springer,
  Cham, 2018.
\newblock \href {http://dx.doi.org/10.1007/978-3-319-10575-8_30}
  {\path{doi:10.1007/978-3-319-10575-8_30}}.

\bibitem{DBLP:conf/hybrid/DuggiralaM12}
Parasara~Sridhar Duggirala and Sayan Mitra.
\newblock {Lyapunov} abstractions for inevitability of hybrid systems.
\newblock In Thao Dang and Ian~M. Mitchell, editors, {\em HSCC}, pages
  115--124, New York, 2012. {ACM}.
\newblock \href {http://dx.doi.org/10.1145/2185632.2185652}
  {\path{doi:10.1145/2185632.2185652}}.

\bibitem{DBLP:conf/cav/FrehseGDCRLRGDM11}
Goran Frehse, Colas~Le Guernic, Alexandre Donz{\'{e}}, Scott Cotton, Rajarshi
  Ray, Olivier Lebeltel, Rodolfo Ripado, Antoine Girard, Thao Dang, and Oded
  Maler.
\newblock {SpaceEx}: Scalable verification of hybrid systems.
\newblock In Ganesh Gopalakrishnan and Shaz Qadeer, editors, {\em CAV}, volume
  6806 of {\em LNCS}, pages 379--395, Heidelberg, 2011. Springer.
\newblock \href {http://dx.doi.org/10.1007/978-3-642-22110-1_30}
  {\path{doi:10.1007/978-3-642-22110-1_30}}.

\bibitem{DBLP:conf/tacas/GhorbalP14}
Khalil Ghorbal and Andr{\'{e}} Platzer.
\newblock Characterizing algebraic invariants by differential radical
  invariants.
\newblock In Erika {\'{A}}brah{\'{a}}m and Klaus Havelund, editors, {\em
  TACAS}, volume 8413 of {\em LNCS}, pages 279--294, Heidelberg, 2014.
  Springer.
\newblock \href {http://dx.doi.org/10.1007/978-3-642-54862-8_19}
  {\path{doi:10.1007/978-3-642-54862-8_19}}.

\bibitem{DBLP:conf/hybrid/GoubaultP17}
Eric Goubault and Sylvie Putot.
\newblock Forward inner-approximated reachability of non-linear continuous
  systems.
\newblock In Goran Frehse and Sayan Mitra, editors, {\em HSCC}, pages 1--10,
  New York, 2017. {ACM}.
\newblock \href {http://dx.doi.org/10.1145/3049797.3049811}
  {\path{doi:10.1145/3049797.3049811}}.

\bibitem{DBLP:conf/emsoft/LiuZZ11}
Jiang Liu, Naijun Zhan, and Hengjun Zhao.
\newblock Computing semi-algebraic invariants for polynomial dynamical systems.
\newblock In Samarjit Chakraborty, Ahmed Jerraya, Sanjoy~K. Baruah, and
  Sebastian Fischmeister, editors, {\em EMSOFT}, pages 97--106, New York, 2011.
  {ACM}.
\newblock \href {http://dx.doi.org/10.1145/2038642.2038659}
  {\path{doi:10.1145/2038642.2038659}}.

\bibitem{DBLP:books/daglib/0077033}
Zohar Manna and Amir Pnueli.
\newblock {\em The Temporal Logic of Reactive and Concurrent Systems -
  Specification}.
\newblock Springer, New York, 1992.
\newblock \href {http://dx.doi.org/10.1007/978-1-4612-0931-7}
  {\path{doi:10.1007/978-1-4612-0931-7}}.

\bibitem{DBLP:journals/toplas/OwickiL82}
Susan~S. Owicki and Leslie Lamport.
\newblock Proving liveness properties of concurrent programs.
\newblock {\em {ACM} Trans. Program. Lang. Syst.}, 4(3):455--495, 1982.
\newblock \href {http://dx.doi.org/10.1145/357172.357178}
  {\path{doi:10.1145/357172.357178}}.

\bibitem{1184414}
A.~{Papachristodoulou} and S.~{Prajna}.
\newblock On the construction of {Lyapunov} functions using the sum of squares
  decomposition.
\newblock In {\em CDC}, volume~3, pages 3482--3487. IEEE, 2002.
\newblock \href {http://dx.doi.org/10.1109/CDC.2002.1184414}
  {\path{doi:10.1109/CDC.2002.1184414}}.

\bibitem{DBLP:journals/logcom/Platzer10}
Andr{\'e} Platzer.
\newblock Differential-algebraic dynamic logic for differential-algebraic
  programs.
\newblock {\em J. Log. Comput.}, 20(1):309--352, 2010.
\newblock \href {http://dx.doi.org/10.1093/logcom/exn070}
  {\path{doi:10.1093/logcom/exn070}}.

\bibitem{DBLP:conf/lics/Platzer12a}
Andr{\'{e}} Platzer.
\newblock Logics of dynamical systems.
\newblock In {\em LICS}, pages 13--24. {IEEE}, 2012.
\newblock \href {http://dx.doi.org/10.1109/LICS.2012.13}
  {\path{doi:10.1109/LICS.2012.13}}.

\bibitem{DBLP:journals/jar/Platzer17}
Andr{\'e} Platzer.
\newblock A complete uniform substitution calculus for differential dynamic
  logic.
\newblock {\em J. Autom. Reas.}, 59(2):219--265, 2017.
\newblock \href {http://dx.doi.org/10.1007/s10817-016-9385-1}
  {\path{doi:10.1007/s10817-016-9385-1}}.

\bibitem{DBLP:journals/tocl/Platzer17}
Andr{\'{e}} Platzer.
\newblock Differential hybrid games.
\newblock {\em {ACM} Trans. Comput. Log.}, 18(3):19:1--19:44, 2017.
\newblock \href {http://dx.doi.org/10.1145/3091123}
  {\path{doi:10.1145/3091123}}.

\bibitem{Platzer18}
Andr{\'e} Platzer.
\newblock {\em Logical Foundations of Cyber-Physical Systems}.
\newblock Springer, Cham, 2018.
\newblock \href {http://dx.doi.org/10.1007/978-3-319-63588-0}
  {\path{doi:10.1007/978-3-319-63588-0}}.

\bibitem{DBLP:conf/lics/PlatzerT18}
Andr{\'{e}} Platzer and Yong~Kiam Tan.
\newblock Differential equation axiomatization: The impressive power of
  differential ghosts.
\newblock In Anuj Dawar and Erich Gr{\"{a}}del, editors, {\em LICS}, pages
  819--828, New York, 2018. ACM.
\newblock \href {http://dx.doi.org/10.1145/3209108.3209147}
  {\path{doi:10.1145/3209108.3209147}}.

\bibitem{DBLP:conf/hybrid/PodelskiW06}
Andreas Podelski and Silke Wagner.
\newblock Model checking of hybrid systems: From reachability towards
  stability.
\newblock In Jo{\~{a}}o~P. Hespanha and Ashish Tiwari, editors, {\em HSCC},
  volume 3927 of {\em LNCS}, pages 507--521, Heidelberg, 2006. Springer.
\newblock \href {http://dx.doi.org/10.1007/11730637_38}
  {\path{doi:10.1007/11730637_38}}.

\bibitem{DBLP:journals/tac/PrajnaJP07}
Stephen Prajna, Ali Jadbabaie, and George~J. Pappas.
\newblock A framework for worst-case and stochastic safety verification using
  barrier certificates.
\newblock {\em {IEEE} Trans. Automat. Contr.}, 52(8):1415--1428, 2007.
\newblock \href {http://dx.doi.org/10.1109/TAC.2007.902736}
  {\path{doi:10.1109/TAC.2007.902736}}.

\bibitem{DBLP:conf/hybrid/PrajnaR05}
Stephen Prajna and Anders Rantzer.
\newblock Primal-dual tests for safety and reachability.
\newblock In Manfred Morari and Lothar Thiele, editors, {\em HSCC}, volume 3414
  of {\em LNCS}, pages 542--556, Heidelberg, 2005. Springer.
\newblock \href {http://dx.doi.org/10.1007/978-3-540-31954-2_35}
  {\path{doi:10.1007/978-3-540-31954-2_35}}.

\bibitem{DBLP:journals/siamco/PrajnaR07}
Stephen Prajna and Anders Rantzer.
\newblock Convex programs for temporal verification of nonlinear dynamical
  systems.
\newblock {\em SIAM J. Control Optim.}, 46(3):999--1021, 2007.
\newblock \href {http://dx.doi.org/10.1137/050645178}
  {\path{doi:10.1137/050645178}}.

\bibitem{DBLP:journals/siamco/RatschanS10}
Stefan Ratschan and Zhikun She.
\newblock Providing a basin of attraction to a target region of polynomial
  systems by computation of {Lyapunov}-like functions.
\newblock {\em SIAM J. Control Optim.}, 48(7):4377--4394, 2010.
\newblock \href {http://dx.doi.org/10.1137/090749955}
  {\path{doi:10.1137/090749955}}.

\bibitem{MR0385023}
Walter Rudin.
\newblock {\em Principles of Mathematical Analysis}.
\newblock McGraw-Hill, third edition, 1976.

\bibitem{Sogokon16}
Andrew Sogokon.
\newblock {\em Direct methods for deductive verification of temporal properties
  in continuous dynamical systems}.
\newblock PhD thesis, Laboratory for Foundations of Computer Science, School of
  Informatics, University of Edinburgh, 2016.

\bibitem{DBLP:conf/fm/SogokonJ15}
Andrew Sogokon and Paul~B. Jackson.
\newblock Direct formal verification of liveness properties in continuous and
  hybrid dynamical systems.
\newblock In Nikolaj Bj{\o}rner and Frank~S. de~Boer, editors, {\em FM}, volume
  9109 of {\em LNCS}, pages 514--531, Cham, 2015. Springer.
\newblock \href {http://dx.doi.org/10.1007/978-3-319-19249-9_32}
  {\path{doi:10.1007/978-3-319-19249-9_32}}.

\bibitem{Sogokon2018}
Andrew Sogokon, Paul~B. Jackson, and Taylor~T. Johnson.
\newblock Verifying safety and persistence in hybrid systems using flowpipes
  and continuous invariants.
\newblock {\em J. Autom. Reas.}, to appear.
\newblock \href {http://dx.doi.org/10.1007/s10817-018-9497-x}
  {\path{doi:10.1007/s10817-018-9497-x}}.

\bibitem{DBLP:conf/emsoft/TalyT10}
Ankur Taly and Ashish Tiwari.
\newblock Switching logic synthesis for reachability.
\newblock In Luca~P. Carloni and Stavros Tripakis, editors, {\em EMSOFT}, pages
  19--28, New York, 2010. {ACM}.
\newblock \href {http://dx.doi.org/10.1145/1879021.1879025}
  {\path{doi:10.1145/1879021.1879025}}.

\bibitem{Walter1998}
Wolfgang Walter.
\newblock {\em Ordinary Differential Equations}.
\newblock Springer, New York, 1998.
\newblock \href {http://dx.doi.org/10.1007/978-1-4612-0601-9}
  {\path{doi:10.1007/978-1-4612-0601-9}}.

\end{thebibliography}
\fi

\iflongversion
\else
\end{document}
\fi
\clearpage
\appendix

\section{Proof Calculus}
\label{app:proofcalc}

For ease of reference, the following two lemmas list all of the (derived) \dL axioms and proof rules used in this paper.
Two additional axioms (\irref{DMP+DX}) and one additional derived proof rule (\irref{BC}) are listed in~\rref{lem:appdlaxioms} compared to~\rref{lem:dlaxioms}.
They are used in derivations in~\rref{app:proofs}.

\begin{lemma}[Axioms and proof rules of \dL~\cite{DBLP:journals/jar/Platzer17,Platzer18,DBLP:conf/lics/PlatzerT18}]
\label{lem:appdlaxioms}
The following are sound axioms and proof rules of \dL.\\
\begin{calculuscollection}
\begin{calculus}
\cinferenceRuleQuote{diamond}
\end{calculus}
\quad\quad\quad\quad
\begin{calculus}
\cinferenceRuleQuote{K}
\end{calculus}\\
\begin{calculus}
\dinferenceRuleQuote{dIcmp}
\dinferenceRuleQuote{dC}
\end{calculus}\\
\begin{calculus}
\dinferenceRuleQuote{dW}
\dinferenceRuleQuote{MbW}
\end{calculus}
\iflongversion
\qquad
\else
\hfill
\fi
\begin{calculus}
\dinferenceRuleQuote{dGt}
\dinferenceRuleQuote{MdW}
\end{calculus}\\
\begin{calculus}
\cinferenceRule[DMP|DMP]{differential modus ponens}
{\linferenceRule[impl]
  {\dbox{\pevolvein{\D{x}=\genDE{x}}{\ivr}}{(\ivr \limply \rrfvar)}}
  {(\dbox{\pevolvein{\D{x}=\genDE{x}}{\rrfvar}}{\rfvar} \limply \axkey{\dbox{\pevolvein{\D{x}=\genDE{x}}{\ivr}}{\rfvar}})}
}{}
\cinferenceRule[DX|DX]{}
{\linferenceRule[equiv]
  {(\ivr \limply \rfvar \land \dbox{\pevolvein{\D{x}=\genDE{x}}{\ivr}}{\rfvar})}
  {\axkey{\dbox{\pevolvein{\D{x}=\genDE{x}}{\ivr}}{\rfvar}}}
}{}
\dinferenceRule[BC|Barr]{}
{\linferenceRule
  { \lsequent{\ivr, \ptermA=0}{\lied[]{\genDE{x}}{p} > 0}
  }
  {\lsequent{\Gamma, \ptermA \cmp 0}{\dbox{\pevolvein{\D{x}=\genDE{x}}{\ivr}}{\ptermA \cmp 0}} }  \quad
}{where $\cmp$ is either $\geq$ or $>$}
\end{calculus}

\irlabel{sAIQ|sAI{$\&$}}
\end{calculuscollection}
\end{lemma}

Axiom~\irref{DMP} is the modus ponens principles for domain constraints.
The \emph{differential skip} axiom~\irref{DX} is a reflexivity property of differential equation solutions.
The ``$\lylpmi$'' direction says if domain constraint $\ivr$ is initially false, then the formula $\dbox{\pevolvein{\D{x}=\genDE{x}}{\ivr}}{\rfvar}$ is trivially true in that initial state because no solution of the ODE stays in the domain constraint.
Thus, this direction of~\irref{DX} allows domain constraint $\ivr$ to be assumed true initially when proving $\dbox{\pevolvein{\D{x}=\genDE{x}}{\ivr}}{\rfvar}$ (shown below, on the left).
The ``$\limply$'' direction has the following equivalent contrapositive reading using~\irref{diamond} and propositional simplification: $\ivr \land \rfvar \limply \ddiamond{\pevolvein{\D{x}=\genDE{x}}{\ivr}}{\rfvar}$, i.e., if the domain constraint $\ivr$ and postcondition $\rfvar$ were both true initially, then $\ddiamond{\pevolvein{\D{x}=\genDE{x}}{\ivr}}{\rfvar}$ is true because of the trivial solution of duration zero.
When proving the liveness property $\ddiamond{\pevolvein{\D{x}=\genDE{x}}{\ivr}}{\rfvar}$, this direction of~\irref{DX} says the formula $\ivr \land \rfvar$ can be assumed \emph{false} initially since there is nothing to prove otherwise (shown below, on the right).
\\~\\
\begin{minipage}[b]{0.5\textwidth}
{\footnotesizeoff%
\begin{sequentdeduction}[array]
  \linfer[DX]{
    \lsequent{\Gamma,\ivr}{\dbox{\pevolvein{\D{x}=\genDE{x}}{\ivr}}{\rfvar}}
  }
  {\lsequent{\Gamma}{\dbox{\pevolvein{\D{x}=\genDE{x}}{\ivr}}{\rfvar}} }
\end{sequentdeduction}
}%
\end{minipage}
\begin{minipage}[b]{0.5\textwidth}
{\footnotesizeoff%
\begin{sequentdeduction}[array]
  \linfer[DX]{
    \lsequent{\Gamma,\lnot{(\ivr \land \rfvar)}}{\ddiamond{\pevolvein{\D{x}=\genDE{x}}{\ivr}}{\rfvar}}
  }
  {\lsequent{\Gamma}{\ddiamond{\pevolvein{\D{x}=\genDE{x}}{\ivr}}{\rfvar}} }
\end{sequentdeduction}
}%
\end{minipage}
\\~\\
The proof rule~\irref{BC} is a \dL rendition of the strict barrier certificates proof rule~\cite{DoyenFPP18,DBLP:journals/tac/PrajnaJP07} for invariance of $\ptermA \cmp 0$.
Intuitively, the premise says that $\ptermA = 0$ is a \emph{barrier} along which the value of $\ptermA$ is increasing along solutions (succedent $\lied[]{\genDE{x}}{p} > 0$), so it is impossible for solutions starting from $\ptermA \cmp 0$ to cross this barrier into $p \pmc 0$.
In \dL, it derives as a special case of the general semialgebraic invariance proof rule~\irref{sAIQ} from its ODE invariance axiomatization~\cite{DBLP:conf/lics/PlatzerT18}.

\begin{lemma}[Diamond refinement and existence axioms from~\rref{sec:livenessaxioms}]
\label{lem:refexaxioms}
The following are sound axioms of \dL.
Term $\constt{\ptermA}$ is constant for ODE $\D{x}=\genDE{x},\D{\timevar}=1$.\\
In axiom~\irref{GEx}, the ODE $\D{x}=\genDE{x}$ is globally Lipschitz continuous.
In axiom~\irref{BEx}, the formula $\boundedf(x)$ characterizes a bounded set over variables $x$.
In axiom~\irref{CORef}, formulas $\rfvar,\ivr$ either both characterize topologically open or both characterize topologically closed sets over variables $x$.
In axiom~\irref{SARef}, formula $\ivr$ is a first-order formula of real arithmetic.\footnote{This condition on $\ivr$ is unnecessary because it is already guaranteed by our syntactic conventions.
It is stated here to highlight the fact that this property is crucially used in the soundness proof of~\irref{SARef}. See the proof of~\rref{lem:diatopaxioms}.}\\
\begin{calculuscollection}
\begin{calculus}
\dinferenceRuleQuote{dDR}
\dinferenceRuleQuote{Prog}
\cinferenceRuleQuote{GEx}
\cinferenceRuleQuote{BEx}
\cinferenceRuleQuote{CORef}
\cinferenceRuleQuote{SARef}
\end{calculus}
\end{calculuscollection}
\end{lemma}
\begin{proof}
See the respective proofs of Lemmas~\ref{lem:diarefaxioms},~\ref{lem:diainitaxioms}, and~\ref{lem:diatopaxioms} in~\rref{app:proofs}.
\end{proof}

The topological side conditions on formulas $\fvarA$, defined in~\rref{subsec:semantics}, generalize to the case where $\fvarA$ mentions additional parameters.
Let $x = (x_1,\dots,x_n)$, $(y_1,\dots,y_r) = \allvars \setminus \{x\}$ be parameters, and $\iget[state]{\I} \in \States$ be a state.
For brevity, we write $y=(y_1,\dots,y_r)$ for the parameters and $\iget[state]{\I}(y) = (\iget[state]{\I}(y_1), \dots, \iget[state]{\I}(y_r)) \in \reals^r$ for the component-wise projection, and similarly for $\iget[state]{\I}(x) \in \reals^n$.
Given the set $\imodel{\I}{\fvarA} \subseteq \States$ and $\gamma \in \reals^r$, define:
\[
  \imodel{\I}{\fvarA}_\gamma \mdefeq \{ \iget[state]{\I}(x) \in \reals^n~|~\iget[state]{\I} \in \imodel{\I}{\fvarA}, \iget[state]{\I}(y) = \gamma \}
\]

The set $\imodel{\I}{\fvarA}_\gamma \subseteq \reals^n$ is the projection onto variables $x$ of all states $\iget[state]{\I}$ that satisfy $\fvarA$ and having values $\gamma$ for the parameters $y$.
Formula $\fvarA$ \emph{characterizes} a (topologically) open (resp. closed, bounded, compact) set with respect to variables $x$ iff for all $\gamma \in \reals^r$, the set $\imodel{\I}{\fvarA}_\gamma \subseteq \reals^n$ is topologically open (resp. closed, bounded, compact) with respect to the Euclidean topology.

These topological side conditions are decidable~\cite{Bochnak1998} for first-order formulas of real arithmetic $\rfvar,\ivr$ because in Euclidean spaces they can be phrased as conditions using first-order real arithmetic.
The following conditions are standard~\cite{Bochnak1998}, although special care is taken to universally quantify over the parameters $y$.
Let $|\cdot|^2$ be the squared Euclidean norm and suppose $\rfvar(x,y)$ is a formula mentioning variables $x$ and parameters $y$, then it is (with respect to variables $x$):
\begin{itemize}
\item \emph{open} if the formula $\lforall{y}{ \lforall{x}{\Big(\rfvar(x,y) \limply \lexists{\varepsilon {>} 0}{ \lforall{z}{\big( |x-z|^2 < \varepsilon^2 \limply \rfvar(z,y) \big) } }\Big)} }$ is valid, where the variables $z = (z_1,\dots,z_n)$ are fresh for $\rfvar(x,y)$,
\item \emph{closed} if its complement formula $\lnot{\rfvar(x,y)}$ is open,
\item \emph{bounded} if the formula $\lforall{y}{ \lexists{r {>} 0}{ \lforall{x}{\big( \rfvar(x,y) \limply |x|^2 < r^2 \big)} }}$ is valid, where the variable $r$ is fresh for $\rfvar(x,y)$, and
\item \emph{compact} if it is closed and bounded, by the Heine-Borel theorem~\cite[Theorem 2.4.1]{MR0385023}.
\end{itemize}

There are simple syntactic criteria for checking whether a given formula satisfies these conditions, although these criteria are not complete.\footnote{If there are no parameters $y$, the finiteness theorem for semialgebraic sets~\cite[Theorem 2.7.2]{Bochnak1998} implies that the syntactic checks for formulas characterizing topologically open/closed sets are, in fact, complete.}
For example, the formula $\rfvar(x,y)$ is (with respect to variables $x$):
\begin{itemize}
\item \emph{open} if it is formed from finite conjunctions and disjunctions of strict inequalities $(\neq,>,<)$,
\item \emph{closed} if it is formed from finite conjunctions and disjunctions of non-strict inequalities $(=,\geq,\leq)$,
\item \emph{bounded} if it is of the form $|x|^2 \pmc \ptermA(y) \land \rrfvar(x,y)$, where $\ptermA(y)$ is a term depending only on parameters $y$ and $\rrfvar(x,y)$ is a formula. This syntactic criterion uses the fact that the intersection of a bounded set (characterized by $|x|^2 \pmc \ptermA(y)$) with any set (characterized by $\rrfvar(x,y)$) is bounded. The formula $\rfvar(x,y)$ is also \emph{compact} if $\pmc$ is $\leq$ and $\rrfvar(x,y)$ is closed.
\end{itemize}

\section{Proofs}
This appendix gives full proofs for the lemmas and corollaries in the paper.
The high-level intuition behind these proofs is available in the paper while motivation for important proof steps is given directly in the proofs.
Further motivation for the surveyed liveness arguments can also be found in their original presentations~\cite{DBLP:journals/logcom/Platzer10,DBLP:conf/hybrid/PrajnaR05,DBLP:journals/siamco/PrajnaR07,DBLP:journals/siamco/RatschanS10,DBLP:conf/fm/SogokonJ15,DBLP:conf/emsoft/TalyT10}.

\label{app:proofs}
\printproofs

\section{Counterexamples}
\newcommand{\uvar}{u}
\newcommand{\vvar}{v}
\label{app:counterexamples}
This appendix gives explicit counterexamples to illustrate the soundness errors identified in Sections~\ref{sec:nodomconstraint} and~\ref{sec:withdomconstraint}.

\subsection{Finite Time Blow Up}
The soundness errors identified in~\rref{sec:nodomconstraint} all arise because of incorrect handling of the fact that solutions may blow up in finite time.
This phenomenon is illustrated by $\exnonlinear$ (see~\rref{fig:odeexamples}), but occurs even for very simple non-linear ODEs.
The following is a counterexample for the original presentation of~\irref{dVeq} (and~\irref{TT+dVeqQ+TTQ})~\cite{DBLP:conf/emsoft/TalyT10}.
Similar counterexamples can be constructed for~\cite[Remark 3.6]{DBLP:journals/siamco/PrajnaR07} and for the original presentation of~\irref{RS+RSQ}~\cite{DBLP:journals/siamco/RatschanS10}.

\begin{counterexample}
\irlabel{dVeqbad|dV$_=$\usebox{\Lightningval}}
Consider rule~\irref{dVeq} \emph{without} the restriction of global Lipschitz continuity of the ODE.
This unrestricted rule is denoted~\irref{dVeqbad}.
It is unsound, as shown by the following derivation using rule~\irref{dVeqbad} with $\constt{\varepsilon} \mnodefeq 1$:
{\footnotesizeoff
\begin{sequentdeduction}[array]
  \linfer[dVeqbad]{
  \linfer[qear]{
    \lclose
  }
    {\lsequent{\vvar - 2 < 0}{ 1 \geq 1}}
  }
  {\lsequent{\vvar - 2 \leq 0}{\ddiamond{\pevolve{\D{\uvar}=\uvar^2,\D{\vvar}=1}}{\vvar - 2 = 0}}}
\end{sequentdeduction}
}%

The conclusion of this derivation is not valid.
Consider an initial state $\iget[state]{\I}$ satisfying the formula $\uvar=1 \land \vvar=0$.
The explicit solution of the ODE from $\iget[state]{\I}$ is given by $\uvar(t) = \frac{1}{1-t}, \vvar(t) = t$ for $t \in [0,1)$.
The solution \emph{does not exist} beyond the time interval $[0,1)$ because the $\uvar$-coordinate asymptotically approaches $\infty$, i.e., blows up, as time approaches $t=1$.
It is impossible to reach a state satisfying $\vvar-2=0$ from $\iget[state]{\I}$ along this solution since at least $2$ time units are required.

This counterexample further illustrates the difficulty in handling non-linear ODEs.
Neither the precondition ($\vvar-2 \leq 0$) nor postcondition ($\vvar-2=0$) mention the variable $\uvar$, and the ODEs $\D{\uvar}=\uvar^2, \D{\vvar}=1$ do not depend on variables $\vvar,\uvar$ respectively.
It is tempting to discard the variable $\uvar$ entirely: indeed, the liveness property $\vvar - 2 \leq 0 \limply \ddiamond{\pevolve{\D{\vvar}=1}}{\vvar - 2 = 0}$ is valid.
Yet, for liveness questions about the (original) ODE $\D{\uvar}=\uvar^2, \D{\vvar}=1$, the two variables are inextricably linked through the time axis of solutions to the ODE.
\end{counterexample}

\subsection{Topological Considerations}
The soundness errors identified in~\rref{sec:withdomconstraint} arise because of incorrect topological reasoning in subtle cases where the topological boundaries of the sets characterized by the domain constraint and desired liveness postcondition intersect.

The original presentation of~\irref{dVcmpQ}~\cite{DBLP:journals/logcom/Platzer10} gives the following proof rule for atomic inequalities $\ptermA \cmp 0$.
For simplicity, we assume that the ODE $\D{x}=\genDE{x}$ is globally Lipschitz continuous so that solutions exist for all time.

\[
\dinferenceRule[dVcmpQbad|dV$_\cmp\&$\usebox{\Lightningval}]{}
{\linferenceRule
  {
    \lsequent{\Gamma}{\dbox{\pevolvein{\D{x}=\genDE{x}}{\ptermA \leq 0}}{\ivr}} \quad
    \lsequent{\lnot{(\ptermA \cmp 0)}, \ivr}{\lied[]{\genDE{x}}{\ptermA}\geq \constt{\varepsilon}}
  }
  {\lsequent{\Gamma,\constt{\varepsilon} > 0}{\ddiamond{\pevolvein{\D{x}=\genDE{x}}{\ivr}}{p \cmp 0}} }
}{}
\]

Compared to~\irref{dVcmpQ}, this omits the assumption $\lnot{(\ptermA \cmp 0)}$ and makes no topological assumptions on the domain constraint $\ivr$.
The following two counterexamples show that these two assumptions are necessary.

\begin{counterexample}
Consider the following derivation using the unsound rule~\irref{dVcmpQbad} with $\constt{\varepsilon} \mnodefeq 1$:
{\footnotesizeoff
\begin{sequentdeduction}[array]
  \linfer[dVcmpQbad]{
    \linfer[dW+qear]{
      \lclose
    }
    {\lsequent{\uvar > 1}{\dbox{\pevolvein{\D{\uvar}=1}{\uvar \leq 0}}{\uvar \leq 1}}} !
    \linfer[qear]{
      \lclose
    }
    {\lsequent{\uvar < 0, \uvar \leq 1}{1 \geq 1}}
  }
  {\lsequent{\uvar > 1}{\ddiamond{\pevolvein{\D{\uvar}=1}{\uvar \leq 1}}{\uvar \geq 0}}}
\end{sequentdeduction}
}%

The conclusion of this derivation is not valid: in states where $\uvar > 1$ is true initially, the domain constraint is violated immediately so the diamond modality in the succedent is trivially false in these states.
\end{counterexample}

\begin{counterexample}[\cite{Sogokon16}]
This counterexample is adapted from~\cite[Example 142]{Sogokon16}, which has a minor typographical error (the sign of an inequality is flipped).
Consider the following derivation using the unsound rule~\irref{dVcmpQbad} with $\constt{\varepsilon} \mnodefeq 1$:
{\footnotesizeoff
\begin{sequentdeduction}[array]
  \linfer[dVcmpQbad]{
    \linfer[dW+qear]{
      \lclose
    }
    {\lsequent{}{\dbox{\pevolvein{\D{\uvar}=1}{\uvar \leq 1}}{\uvar \leq 1}}} !
    \linfer[qear]{
      \lclose
    }
    {\lsequent{\uvar \leq 1, \uvar \leq 1}{1 \geq 1}}
  }
  {\lsequent{}{\ddiamond{\pevolvein{\D{\uvar}=1}{\uvar \leq 1}}{\uvar > 1}}}
\end{sequentdeduction}
}%

The conclusion of this derivation is not valid: the domain constraint $\uvar \leq 1$ and postcondition $\uvar > 1$ are contradictory so no solution can reach a state satisfying both simultaneously.
The conclusion is, in fact, false in all states.
\end{counterexample}

The next two counterexamples are for the liveness arguments from~\cite[Corollary 1]{DBLP:conf/hybrid/PrajnaR05} and~\cite[Theorem 3.5]{DBLP:journals/siamco/PrajnaR07}.
For clarity, we use the original notation from~\cite[Theorem 3.5]{DBLP:journals/siamco/PrajnaR07}.
The following conjecture is quoted from~\cite[Theorem 3.5]{DBLP:journals/siamco/PrajnaR07}:

\begin{conjecture}
Consider the system $\D{x}=\genDE{x}$, with $f \in C(\reals^n,\reals^n)$. Let $\bigchi \subset \reals^n$, $\bigchi_0 \subseteq \bigchi$, and $\bigchi_r \subseteq \bigchi$ be bounded sets. If there exists a function $B \in C^1(\reals^n)$ satisfying:
\begin{align}
&B(x) \leq 0                              && \forall x \in \bigchi_0  \label{eq:init} \\
&B(x) > 0                                 && \forall x \in \closure{\bdr{\bigchi} \setminus \bdr{\bigchi_r}} \label{eq:unsoundbdr} \\
&\Dp[x]{B}f(x) < 0    && \forall x \in \closure{\bigchi \setminus \bigchi_r}
\end{align}
Then the eventuality property holds, i.e., for all initial conditions $x_0 \in \bigchi_0$, the trajectory $x(t)$ of the system starting at $x(0)=x_0$ satisfies $x(T) \in \bigchi_r$ and $x(t) \in \bigchi$ for all $t \in [0,T]$ for some $T \geq 0$.
The notation $\closure{\bigchi}$ (resp. $\bdr{\bigchi}$) denotes the topological closure (resp. boundary) of the set $\bigchi$.
\end{conjecture}

In~\cite[Corollary 1]{DBLP:conf/hybrid/PrajnaR05}, stronger conditions are required.
In particular, the sets $\bigchi_0,\bigchi_r,\bigchi$ are additionally required to be topologically open, and the inequality in~\rref{eq:init} is strict, i.e., $B(x) < 0$ instead of $B(x) \leq 0$.

The soundness errors in both of these liveness arguments stem from the condition~\rref{eq:unsoundbdr} being too permissive.
For example, notice that if the sets $\bdr{\bigchi}, \bdr{\bigchi_r}$ are equal then~\rref{eq:unsoundbdr} is vacuously true.
The first counterexample below applies for the requirements of~\cite[Theorem 3.5]{DBLP:journals/siamco/PrajnaR07}, while the second applies even for the more restrictive requirements of~\cite[Corollary 1]{DBLP:conf/hybrid/PrajnaR05}.

\begin{figure}[h]
\centering
\includegraphics[width=0.45\textwidth]{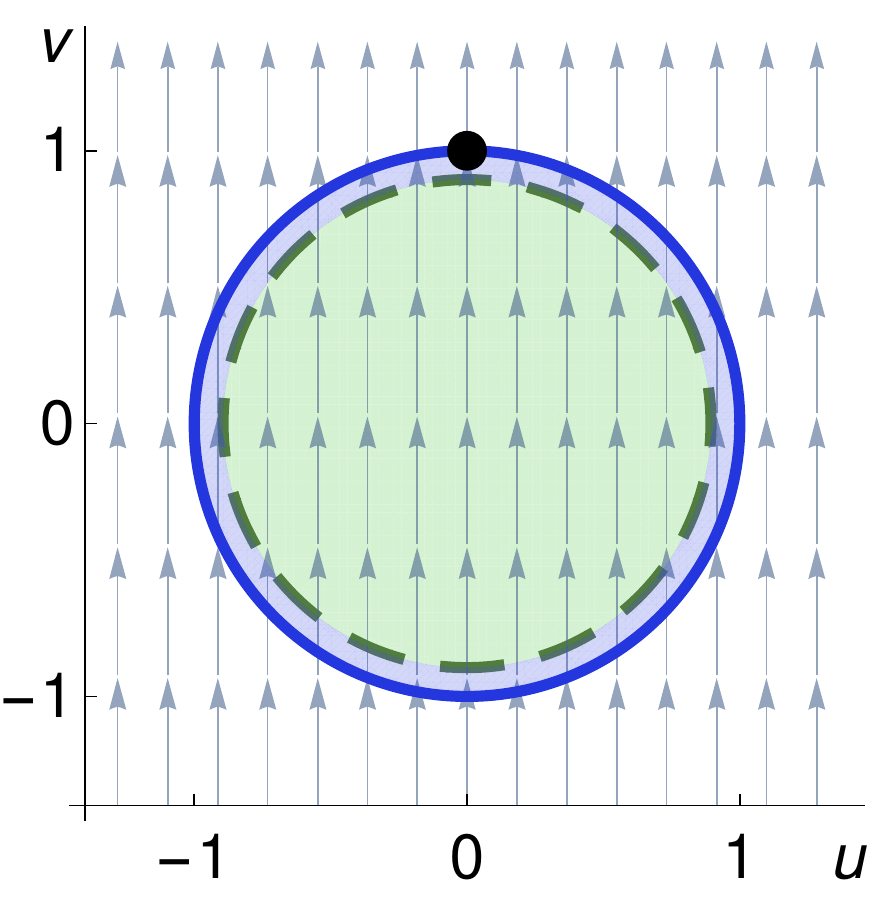}
\includegraphics[width=0.45\textwidth]{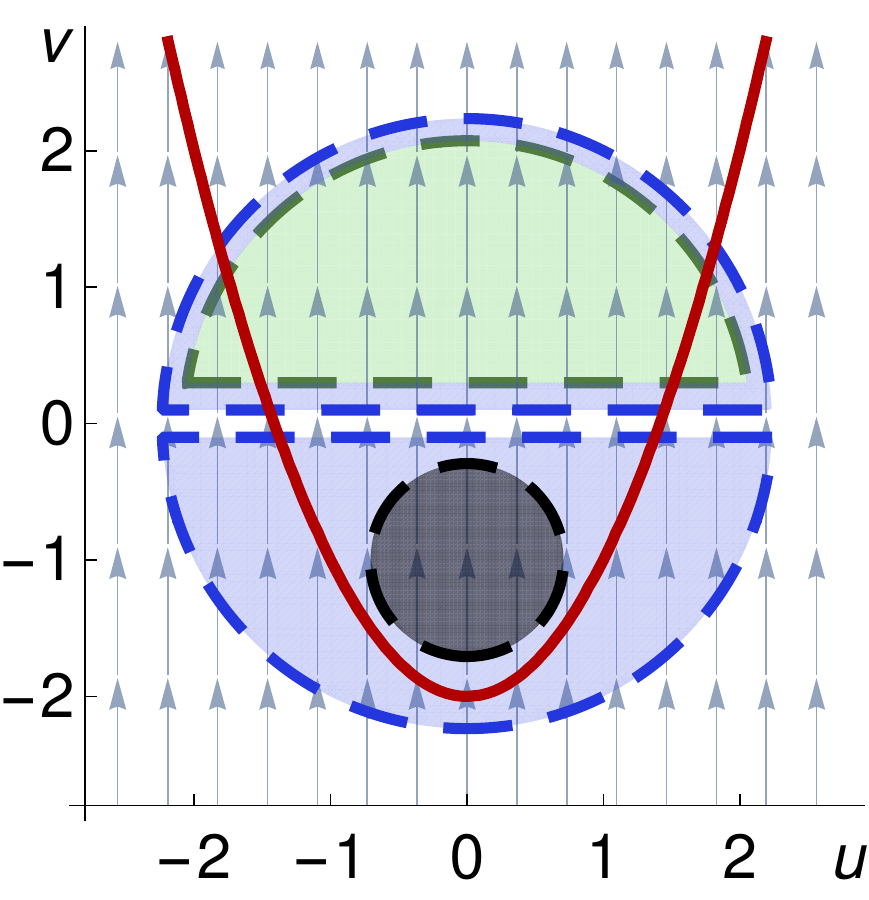}
\caption{\textbf{(Left)} Visualization of~\rref{cex:prajnarantzer1}. The solution from initial point $\uvar=0, \vvar=1$ ($\bigchi_0$, in black) leaves the domain unit disk ($\bigchi$, boundary in blue) immediately without ever reaching its interior ($\bigchi_r$, in green with dashed boundary). The interior is slightly shrunk for clarity in the visualization: the blue and green boundaries should actually overlap exactly.
\textbf{(Right)} Visualization of~\rref{cex:prajnarantzer2}.
Solutions from the initial set ($\bigchi_0$, in black with dashed boundary) eventually enter the goal region ($\bigchi_r$, in green with dashed boundary).
However, the domain ($\bigchi$, in blue with dashed boundary) shares an (open) boundary with $\bigchi_r$ at $\vvar=0$ which solutions are not allowed to cross.
As before, the sets are slightly shrunk for clarity in the visualization: the blue and green boundaries should actually overlap exactly.
The level curve $B = 0$ is plotted in red.
All points above the curve satisfy $B < 0$, while all points below it satisfy $B > 0$.}
\label{fig:cex}
\end{figure}

\begin{counterexample}
\label{cex:prajnarantzer1}
Let the system $\D{x}=\genDE{x}$ be $\D{\uvar}=0,\D{\vvar}=1$.
Let $\bigchi_r$ be the open unit disk characterized by $\uvar^2 + \vvar^2 < 1$, $\bigchi$ be the closed unit disk characterized by $\uvar^2+\vvar^2 \leq 1$, and $\bigchi_0$ be the single point characterized by $\uvar=0 \land \vvar=1$.
All of these sets are bounded.
Note that $\bdr{\bigchi} \setminus \bdr{\bigchi_r} = \emptyset$ since both topological boundaries are given by the unit circle $\uvar^2+\vvar^2=1$.
Let $B(\uvar,\vvar) \mnodefeq -\vvar$, so that $\Dp[x]{B}f(x) = \Dp[\uvar]{B}0 + \Dp[\vvar]{B}1 = -1 < 0$ and $B \leq 0$ on $\bigchi_0$.

All conditions of~\cite[Theorem 3.5]{DBLP:journals/siamco/PrajnaR07} are met but the eventuality property is not true.
The trajectory from $\bigchi_0$ leaves $\bigchi$ immediately and never enters $\bigchi_r$.
This is visualized in~\rref{fig:cex} (Left).
\end{counterexample}

\begin{counterexample}
\label{cex:prajnarantzer2}
Let the system $\D{x}=\genDE{x}$ be $\D{\uvar}=0,\D{\vvar}=1$.
Let $\bigchi_r$ be the set characterized by the formula $\uvar^2 + \vvar^2 < 5 \land \vvar > 0$, $\bigchi$ be the set characterized by the formula $\uvar^2+\vvar^2 < 5 \land \vvar \neq 0$, and $\bigchi_0$ be the set characterized by the formula $\uvar^2+(\vvar+1)^2 < \frac{1}{2}$.
All of these sets are bounded and topologically open.
Let $B(\uvar,\vvar) \mnodefeq -\vvar +\uvar^2 - 2$, so that $\Dp[x]{B}f(x) = \Dp[\uvar]{B}0 + \Dp[\vvar]{B}1 = -1 < 0$, and $B < 0$ on $\bigchi_0$.
The set $\closure{\bdr{\bigchi} \setminus \bdr{\bigchi_r}}$ is characterized by formula $\uvar^2+\vvar^2=5 \land \vvar \leq 0$ and $B$ is strictly positive on this set.
These claims can be checked arithmetically, see~\rref{fig:cex} (Right) for a plot of the curve $B=0$.

All conditions of~\cite[Corollary 1]{DBLP:conf/hybrid/PrajnaR05} are met but the eventuality property is not true.
Solutions starting in $\bigchi_0$ eventually enter $\bigchi_r$ but they can only do so by leaving the domain constraint $\bigchi$ at $\vvar=0$, see~\rref{fig:cex} (Right).

\end{counterexample}

\end{document}